\documentclass[12pt,a4paper]{article}
\usepackage{graphicx,amssymb}
\newcommand{\be}{\begin{equation}}
\newcommand{\ee}{\end{equation}}
\newcommand{\bea}{\begin{eqnarray}}
\newcommand{\eea}{\end{eqnarray}}

\catcode`@=12
\def\la{\mathrel{\mathchoice {\vcenter{\offinterlineskip\halign{\hfil
$\displaystyle##$\hfil\cr<\cr\sim\cr}}}
{\vcenter{\offinterlineskip\halign{\hfil$\textstyle##$\hfil\cr<\cr\sim\cr}}}
{\vcenter{\offinterlineskip\halign{\hfil$\scriptstyle##$\hfil\cr<\cr\sim\cr}}}
{\vcenter{\offinterlineskip\halign{\hfil$\scriptscriptstyle##$\hfil\cr<\cr\sim
\cr}}}}}
\def\ga{\mathrel{\mathchoice {\vcenter{\offinterlineskip\halign{\hfil
$\displaystyle##$\hfil\cr>\cr\sim\cr}}}
{\vcenter{\offinterlineskip\halign{\hfil$\textstyle##$\hfil\cr>\cr\sim\cr}}}
{\vcenter{\offinterlineskip\halign{\hfil$\scriptstyle##$\hfil\cr>\cr\sim\cr}}}
{\vcenter{\offinterlineskip\halign{\hfil$\scriptscriptstyle##$\hfil\cr>\cr\sim
\cr}}}}}
\begin{document}
\begin{flushright}
SINP-APC-13/01
\end{flushright}
\thispagestyle{empty}
\begin{center}
{\Large \bf
{Two Component Dark Matter : A Possible Explanation of 130 GeV $\gamma-$Ray Line from
the Galactic Centre }}\\
\vspace{0.25cm}
\begin{center}
{{\bf Anirban Biswas}$^{\dagger}$ \footnote{email: anirban.biswas@saha.ac.in},
{\bf Debasish Majumdar}$^{\dagger}$ \footnote{email: debasish.majumdar@saha.ac.in},
{\bf Arunansu Sil}$^{\ddagger}$ \footnote{email: sil.arunansu@gmail.com}, \\
{\bf Pijushpani Bhattacharjee}$^{\dagger}$ \footnote{email:
pijush.bhattacharjee@saha.ac.in}} \\
\vspace{0.5cm}
$^\dagger$ \it Astroparticle Physics and Cosmology Division, \\
\it Saha Institute of Nuclear Physics, Kolkata 700064, India \\
\vspace{0.25 cm}
$^\ddagger$\it Department of Physics,
 \it Indian Institute of Technology, Guwahati,\\ 
\it Guwahati 781039, India
\end{center}
\vspace{1cm}
{\bf ABSTRACT}
\end{center}
Recently there has been a hint of a gamma-ray line at 130 GeV 
originated from the galactic centre after the analysis of the 
Fermi-LAT satellite data. Being monochromatic in nature, it rules
out the possibility of having its astrophysical origin and 
there has been a speculation that this line could be originated 
from dark matter annihilation. In this work, we propose a two 
component dark matter scenario where an extension of the Standard 
model by an inert Higgs doublet and a gauge singlet scalar 
concocted with $\rm {Z_2 \times Z^{\prime}_2}$ symmetry, is considered. 
We find that our scenario can not only explain the 130 GeV gamma-ray
line through dark matter annihilation but also produce the correct dark
matter relic density. We have used the Standard Model Higgs mass around
125 GeV as intimated by the LHC data.
\newpage
\section {Introduction}
\label{intro}
Although the existence of dark matter (DM) is now established by various astronomical
measurements and observations where the gravitational effects of this huge amount of dark
matter is manifested, the particle nature of the dark matter still remains  unknown. The
particle nature of the dark matter can be probed if it is detected either by direct detection
process or by indirect detection. In the former the energy of recoil of a detector nucleus is
to be measured if a dark matter indeed scatters off such a nucleus or nucleon. On the other
hand the dark matter can be gravitationally trapped inside massive celestial objects 
such as sun, in regions of galactic centre etc. These trapped dark matter particles 
eventually annihilate to produce fermion-antifermion pairs or $\gamma$s. Study of dark matter
through the indirect detection whereby such annihilation products are detected and analyzed, 
may reveal the nature of the dark matter. 
Recently it was claimed in Ref. \cite{data-analysis} that there is a $4.6 \sigma$ local evidence of a
monochromatic gamma-ray line having an energy $\sim130$ GeV from the direction of galactic centre, obtained
from the analysis of Fermi-LAT publicly available data \cite{fermi-data}. Similar result was also followed
from the independent analysis of Ref. \cite{Tempel:2012ey}. Strong evidence for this gamma-ray
line from inner galaxy was also reported in Ref. \cite{Su:2012ft} where authors
show that this excess of gamma-ray may have a double peak structure with local significance 
$5\sigma$ ($5.4\sigma$) for one(two) line(s) case. Later works by \cite{hektor2, su1207, dan}
indicate the possibility of having double gamma-ray line emission from nearby galaxy clusters
and un-associated Fermi-LAT point sources. The analysis of Ref. \cite{su1209} was also in favour of
explaining the galactic centre gamma-ray emission at 130 GeV. There are some other works on this 130 GeV
gamma-line which took a critical observation on this line by looking at instrumental noise 
\cite{1205.4700,1208.3677}, statistical fluctuations, earth-limb magnification \cite{hektor3}
and emission from AGNs \cite{mirabal}. Recently in Ref. \cite{weniger2}, authors discuss current
observational situation of the 130 GeV gamma-line and propose a new modified survey strategy
which increases the data rate form the inner galaxy more than two times of the current
standard survey mode for the decisive measurement, by the end of 2014.  

A recent search for the spectral lines in the energy range 5-300 GeV using 3.7 years
of Fermi-LAT data has been carried out in Ref. \cite{fermi-lat-13} by Fermi-LAT collaboration. 
In this work they have used an updated instrumental calibration. They found that the most
significant fit occurred at an energy $\sim133$ GeV with a local significance of $3.3\sigma$.
The other claim was that the line like feature at energy 130 GeV has a weak significance compared to
the significance found in earlier analysis. However they also mentioned that at the present situation
133 GeV line can not be considered as a real signal, there may be some other systemic detector effects also.
More data and analyses are needed to clarify the origin of this ``line-like" gamma-ray feature. 
In this work, we adopt the 130 GeV gamma-ray line from the galactic centre. 
We propose a particle physics model for dark matter and attempt to explain
this gamma-ray line from such dark matter annihilation at the galactic centre. 
However in case the peak indeed exists in the vicinity of 130 GeV, our model can be minimally adjusted to
accommodate such deviation.

In the present work we propose a particle physics model for dark matter that can
explain the 130 GeV $\gamma$-line observed by Fermi-LAT. Our model is in fact a
two-component dark matter model in which a real scalar singlet and an inert doublet
are added to the Standard Model (SM) of particle physics. There are previous works where 
either the real scalar singlet model or the inert doublet model has been discussed as one 
component dark matter model. But, as it will be revealed later that any one of these two
models fail to explain individually (as one component dark matter) the observed 130 GeV 
$\gamma$-line from the galactic centre if it is produced due to dark matter annihilation.

In earlier works such as Ref. \cite{scalar-gamma}, the authors showed that 
the scalar dark matter can annihilate into $\gamma \gamma$ final state with 
the help of additional charged scalars in a model independent way. It would yield 
130 GeV $\gamma$s with the required annihilation cross section
($\langle \sigma {\rm{v}}_{\gamma\gamma} \rangle \sim 10^{-27} {\rm cm}^3/{\rm s}$ given by the 
analysis \cite{data-analysis, Tempel:2012ey} of the Fermi-LAT data mentioned above). Different
other possibilities involving new particles originated from different extensions 
of SM have been investigated \cite{gamma-papers,Buchmuller} to explain the 130 GeV 
$\gamma$-line through DM (single candidate) annihilation. However most of these 
endeavours are restricted in obtaining $\langle \sigma {\rm{v}}_{\gamma\gamma} \rangle \sim 10^{-27}
{\rm cm}^3/{\rm s}$ without going into detailed discussions on relic density 
calculations, calculation of scattering cross sections relevant for direct 
detection and their comparisons with experimental (direct detection experiments) 
or observational (WMAP \cite{wmap}) results. 
  
One of the simplest choice to accommodate a dark matter is to extend the 
SM with a gauge singlet real scalar field $S$ (we will call it real scalar 
singlet dark matter model, RSDM, from now on) \cite {RSDM, profumo, Biswas}, which couples to SM 
Higgs ($h$). The use of a ${\rm Z_2}$ symmetry ensures the stability of the 
dark matter candidate. In Ref. \cite{Biswas} it is shown that this type 
of model fails to explain the 130 GeV gamma-ray line from the Galactic centre. 
The reason of this failure is related with the relatively small annihilation cross 
section of two $S$ fields into two $\gamma$s considering the fact that $S$ should 
contribute to the correct amount of relic density as predicted by WMAP data \cite{wmap}. 
In \cite{profumo}, it was shown that with SM Higgs much heavier than 125 GeV could 
in principle lead to $\langle \sigma {\rm{v}}_{\gamma\gamma}\rangle \sim 10^{-27} {\rm cm}^3/{\rm s}$ with 
$m_S \sim$ 130 GeV, when the constraint on the $S$ field to produce right amount of 
dark matter abundance is relaxed. However once this constraint is applied, the 
$\langle \sigma {\rm{v}}_{\gamma\gamma} \rangle$ becomes few orders of magnitude less than required.
In addition to these findings, if we employ the recent constraint on the Higgs mass from LHC experiment
\cite{125higgs} on the SM Higgs, then we infer that strictly within RSDM picture, 
we can not accommodate both the relic density as well as the 130 GeV $\gamma$-line from DM annihilation.

Another well motivated dark matter model is the inert doublet model (IDM) \cite{IDM, LopezHonorez:2010tb, IDM1}
which requires an extension of the SM by a scalar Higgs (inert) doublet $\Phi$ having a 
${\rm Z_2}$. In Ref. \cite {LopezHonorez:2010tb}, it was 
shown that there exists an allowed region (consistent with the WMAP results of 
relic density) in IDM for dark matter\footnote{$\phi^0$ is the neutral 
component of the extra Higgs doublet.} $\phi^0$ having mass in the range between 80 GeV and 160 GeV provided 
the mass of the dark matter candidate ($m_{\phi^0}$)  is less than the mass of SM Higgs (and 
top quark), $m_{\phi^0} < m_{h, t}$. This condition was imposed to reduce the 
$\langle \sigma {\rm{v}} \rangle_{\rm total}$ (to get rid of the contributions 
like $\phi^0 \phi^0 \rightarrow hh$ ~(${\rm{and/or}}$ ~$tt$)) since otherwise the relic density 
would be small\footnote{since relic density $\Omega h^2 \propto 1/\langle \sigma {\rm{v}} \rangle$.}. 
Then they found that due to accidental cancellations of different Feynman diagrams for annihilation 
into gauge bosons, $\langle \sigma {\rm {v}} \rangle_{\rm total}$ can indeed be the right amount 
for a judicious choice of parameter space involved in the model. 

We are trying to find a resolution where a right amount of dark matter relic density 
could be obtained as well as an explanation of the 130 GeV gamma-ray line of Fermi-LAT 
can also be probed through DM annihilation. Now with the consideration that the SM Higgs 
boson has a mass $m_h \sim 125$ GeV \cite{125higgs} and mass of the DM candidate ~$130$ GeV 
(in order to explain the 130 GeV gamma-ray line from Fermi-LAT, the mass of the DM should be 
in this range), the above mentioned condition 
$m_{\phi^0} < m_{h, t}$ related to the IDM is evaded thereby the channel of annihilation, 
$\phi^0 \phi^0 \rightarrow hh$, opens up. Hence $\langle \sigma {\rm{v}} \rangle_{total}$ will 
increase and the final relic density in this sort of model \cite {LopezHonorez:2010tb} can accommodate only
(10-30)$\%$ of the observed DM relic density \cite {wmap}. However this result has an 
interesting consequence. We can compensate this deficit of the DM relic density by another 
candidate of DM while $\phi^0$ explains the 130 GeV gamma-line through its annihilation.    

Keeping in mind the above mentioned scenarios (particularly the RSDM and IDM), we propose 
that the dark matter can actually be composed of two fields, namely a scalar singlet ($S$) 
and an inert doublet ($\Phi$). The additional feature of this model would be that these two components 
possess an interaction between them via a term like $(\Phi^\dagger\Phi)SS$. The 130 GeV $\gamma$-line will
be produced by the annihilation of the component $\phi^0$ while the role of other 
component $S$, besides contributing to the overall relic density is to increase $\phi^0$ contribution 
to the combined relic density through $(\Phi^\dagger\Phi)SS$. A model with a scalar singlet and a doublet was 
presented in {\cite{0903.2475, 0907.1894}} in the context of GUT models. A multi-component dark matter 
was proposed in \cite{Ma} earlier where an additional fermion singlet and a scalar singlet 
were introduced, though not in the context of the possibility of having 130 GeV gamma-ray 
from DM annihilation. We have imposed a discrete ${\rm Z}_2 \times {\rm Z}^\prime_2$, 
under which $S$ and $\Phi$ transform non-trivially.

The calculation of the flux for 130 GeV $\gamma$-ray from galactic centre
region also requires the knowledge of dark matter density in the region of
the galactic centre. In the absence of a unique density profile in literature
we consider in the present work two dark matter density profiles namely
NFW (Navarro-Frenk-White) profile \cite{nfw} and Einasto profile \cite{einasto}
and compute the flux using the cross section $\langle {\sigma {\rm{v}}}_{\phi^0\phi^0
\rightarrow\gamma\gamma}\rangle$ calculated in this work from our model.    

The paper is arranged as follows. In Section \ref{model} we describe the structure of our two 
component dark matter model. Section \ref{relic} discusses the calculations of
the relic densities of each dark matter components and hence the combined relic
density by simultaneously solving the two Boltzmann's equations for the two components.
In Section \ref{calc} we discuss how to constrain the parameter space
of the model by comparing the relic density results with WMAP and the
scattering cross section results with the direct detection experiments
data. Section \ref{gamma} gives the cross section calculations for 130 GeV $\gamma$-ray
from $\phi^0 \phi^0$ annihilation and hence the calculations of $\gamma$-ray
flux using different dark matter halo models. Finally in Section \ref{disc}
we present some discussions and conclusions.
  
\section{The Two Component Dark Matter Model}
\label{model}
In the present work, the DM sector is composed of a real gauge singlet scalar 
field, ${S}$ and an extra (in addition to the usual Higgs doublet, $H$) scalar doublet 
field, $\Phi$ (doublet under $SU(2)_{\rm L}$). An exact (unbroken) discrete symmetry 
${\rm Z_2} \times {\rm Z'_2}$ is imposed under which all the SM particles are even, i.e. having 
${\rm Z_2 \times Z_2^{\prime}}$ charge as (1,1) and for $\Phi$, $S$ the 
${\rm Z_2 \times Z_2^{\prime}}$ charges are (1,-1) and (-1,1) respectively. Thereby
both $S$ and $\Phi$ are fermiophobic and do not develop 
any vacuum expectation value (VEV) and hence inert. The construction therefore 
ensures the stability of both $S$ and $\Phi$ in this two component dark matter scenario. 
The scalar doublet $\Phi$ can be written as
\begin{eqnarray}
\hspace{1cm} \Phi = \left(\begin{array}{cc}\phi^+ \\
\frac{\phi^0+iA^0}{\sqrt{2}}\end{array}\right)\,\,.
\end{eqnarray}

The Lagrangian of the construction can be read as,
${\cal L} = {\cal L}_{\rm SM} + {\cal L}_{\rm DM}$, 
where ${\cal L}_{\rm SM}$ is the Standard Model Lagrangian and ${\cal L}_{\rm DM}$ stands 
for the DM sector consistent with all the symmetries. ${\cal L}_{\rm DM}$ is then 
given by,
\begin{equation}
{\cal L}_{\rm DM} = {\cal L}_{\rm RSDM} + {\cal L}_{\rm IDM} + {\cal L}_{\rm INT},
\label{1}
\end{equation}
where ${\cal L}_{\rm RSDM}$ and ${\cal L}_{\rm IDM}$ refer to the individual Lagrangian of real 
singlet scalar and inert doublet dark matter respectively and ${\cal L}_{\rm INT}$ 
is the additional part representing the interaction between the two components of DM. 
So the most general form of them, consistent with the SM gauge group as well as 
the discrete symmetry imposed, is as follows, 
\begin{eqnarray} 
{\cal L}_{\rm RSDM} &=& \frac{1}{2} \partial_{\mu} {S} \partial^{\mu} {S}
- \frac{{\kappa}^2_1}{2}{S}^2 - \frac{{\kappa}_2}{4}{S}^4
-\lambda_6({\rm H}^{\dagger}{\rm H}){S}{S},
\label{RSDM}\\
{\cal L}_{\rm IDM}&=&(D_{\mu}\Phi)^{\dagger}(D^{\mu}\Phi)-
\mu_2^2(\Phi^{\dagger}\Phi)-\rho_{2}(\Phi^{\dagger}\Phi)^2 
-\lambda_1({\rm H}^{\dagger}{\rm H})(\Phi^{\dagger}\Phi) - \nonumber \\ 
&&\lambda_2(\Phi^{\dagger}{\rm H})({\rm H}^{\dagger}\Phi)-  
\lambda_3\left[(\Phi^{\dagger}{\rm H})^2 + h.c\right], 
\label{IDM}\\
{\cal L}_{\rm INT} &=& -{\lambda_5}\left(\Phi^{\dagger}\Phi\right){S}{S}.
\label{2}
\end{eqnarray}
The SM Higgs Lagrangian is included in the ${\cal L}_{\rm SM}$. In this model we have
five new particles, two charged scalar ($\phi^\pm$) and three neutral scalar particles
($\phi^0$, ${S}$, ${A}^0$). Due to the stability and electrical charge neutrality we consider 
$\phi^0$ and $S$ as two viable components of dark matter in this model.

After spontaneous breaking of the SM gauge symmetry, the
masses of these new particles and Higgs are given by,
\begin{eqnarray}
m_{{\phi}^\pm}^2 &=& \mu_2^2 + \frac{1}{2}\lambda_1v^2, 
\label{mphi+} \\
m_{\phi^0}^2 &=& \mu_2^2 + \alpha v^2, 
\label{mphi0}\\
m_{{A}^0}^2 &=& \mu_2^2 + \beta v^2, 
\label{ma0} \\
m_{S}^2 &=& {\kappa}_1^2 + \lambda_6 v^2, 
\label{ms}\\
m_{h}^2 &=& 2\rho_1 v^2,
\end{eqnarray}
where $v$ (= 246 GeV) is the Higgs VEV and $\rho_1$ 
is the coefficient of the quartic coupling of the SM Higgs (part of ${\cal L}_{\rm SM}$ here) potential. 
The parameters $\alpha$ and $\beta$ are defined in terms of $\lambda^{,}s$ as
\begin{eqnarray}
\alpha & = & \frac{1}{2}(\lambda_1 + \lambda_2 + 2\lambda_3)\, ,
\nonumber \\  
\beta & = &  \frac{1}{2}(\lambda_1 + \lambda_2 - 2\lambda_3).
\label{5}
\end{eqnarray}
As is evident from the above set up, the model involves 10 parameters in 
total, specifically $m_h$, $m_{\phi^0}$, $m_{A^0}$, $m_{\phi^+}$, 
$m_S$, $\alpha$, $\lambda_5$, $\lambda_6$, $\kappa_2$, $\rho_2$. Now fixing 
the Higgs mass $m_h$ at 125 GeV, we have altogether 
9 parameters, which are further restricted from theoretical bounds as 
well as from experimental results as discussed below.  

\begin{itemize}
\item {\bf Vacuum Stability} - The Lagrangian of this model (Eq. (\ref{1}))
must be bounded from below. This condition will be satisfied if 
\begin{eqnarray}
\rho_1,\,\rho_2,\,\kappa_2 &>& 0 \,\, ,
\label{cons1} \\
\alpha,\beta & > & -\sqrt{\rho_1\rho_2}\, ,
\label{cons2} \\
\lambda_1 & > & -2\sqrt{\rho_1\rho_2}\, ,
\label{cons2a}\\
\lambda_6 & > & -\sqrt{\rho_1{\kappa}_2}\, , 
\label{cons3} \\
\lambda_5 & > & -\sqrt{\rho_2{\kappa}_2}\, .
\label{cons4}
\end{eqnarray}

\item {\bf Zero VEV of $\Phi$ {\bf and} $S$} - Ground state of the Lagrangian Eq.
({\ref{1}}) must preserve ${\rm Z_2\times Z_2^{\prime}}$ symmetry for stability of the
dark matter candidates, this leads to the condition that the VEV of
both $\Phi$, $S$ is zero.

\item {\bf Perturbativity}- In order to be within the perturbative limit, the model
parameters cannot be too large. This can be ensured provided  
\begin{eqnarray}
|\rm model\,\,parameters|<4\pi
\label{cons5}.
\end{eqnarray}

\item {\bf Neutral Scalar Mass} - The LEP \cite {Abbiendi:2000hu}
measurement of the $Z$ boson decay width leads to the condition  
\begin{equation}
m_{A^0} + m_{\phi^0} > m_Z.
\label{cons6}
\end{equation}

\item {\bf WMAP Limit} - The Combined relic density of the dark matter components
must satisfy the WMAP limit \cite{wmap}, 
\begin{equation}
0.1053 < \Omega_{\rm DM} h^2 < 0.1165 {\rm{~~at ~~68}} \% {\rm{~~C.L.}}
\label{wmap}
\end{equation} 
for the dark matter in the Universe. This condition will further constrain the
parameter space of this model discussed above.

\item {\bf Direct detection limits of dark matter} - The results of the ongoing
experiments for direct detection of dark matter also impose additional limits on
the relevant parameters of the present two component dark matter model. 
\begin{figure}[h]
\centering
\includegraphics[width=5cm,height=3cm]{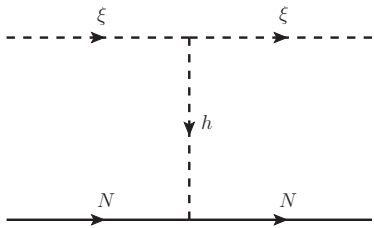}
\caption{Feynman diagram for the elastic scattering between dark matter particle $\xi$
(i,e. $S, \phi^0$) and nucleon $N$ of the detector material via Higgs exchange.}
\label{sigma-si-fig}
\end{figure}
Being inert, the elastic 
scattering between the dark matter candidate $\xi$ (here $\phi^0$ and/or $S$) 
and nucleons ($N$) can take place only with the SM Higgs exchange as in Fig 
\ref{sigma-si-fig}. The relevant term in the Lagrangian which describes 
the interaction between $\xi$ and Higgs is given by 
\begin{equation} 
{\cal{L}} = - k \xi^2 h, 
\end{equation}  
where the coupling $k= \alpha v$ ($\lambda_6 v$) for $\xi = \phi^0$ ($S$).
The spin independent scattering cross section for this 
process, $\xi N \rightarrow \xi N$, 
is given by
\cite{scatter-cross},
\begin{eqnarray}
\sigma^{\rm SI} = \frac{f^2}{\pi}\left(\frac{k}{v}\right)^2
\frac{\mu^2 \,\,m^2_N}{m^2_{\xi}\,\,m^4_h}\,\,,
\label{sigma-si}
\end{eqnarray}
where $m_{\xi}$ and $m_N$ are the masses of the DM candidate and nucleons respectively,  
$\mu = \frac{m_{\xi}m_N}{m_{\xi}+m_N}$ is the reduced mass and $f$ represents the 
strength of the effective interaction which depends upon the number of heavy quarks 
involved \cite {goran}. In this work, we have considered the value of $f = 0.3$ \cite{IDM1, f-value}.  

An upper limit on $\sigma^{\rm SI}$ for a particular mass of dark matter
particle would automatically sets an upper bound on the absolute value
of the couplings $\alpha$ and $\lambda_6$ through Eq. (\ref{sigma-si}) (note that
we are considering a two-component DM case). Now to find out this limit,
we need to know the masses of our DM candidates, which should not only
explain the correct DM relic abundance, but also can explain the 130 GeV
gamma-line obtained from Fermi-LAT data. It is discussed in the rest of our paper and we
infer that both masses should be similar and $\sim$ 130 GeV. With this two-component DM model,
the direct detection limit implies \cite{Ma},

\begin{equation}
\epsilon_{\phi^0} \sigma_{\phi^0 N} + \epsilon_{S} \sigma_{SN} < \sigma_0,
\label{xenon-Ma}
\end{equation}

where $\sigma_0$ is the upper limit of the DM-nucleon scattering cross section
(for one dark matter model) obtained from the direct search experiment.
In case of XENON 100 (2012) \cite {xenon2012}, it is $3 \times 10^{-45}$ cm$^2$ at $90\%$ C.L.
Using Eq.(\ref{sigma-si}), it then translates into the
following inequality,

\begin{equation}
\epsilon_{\phi^0} \alpha^2 + \epsilon_{S} \lambda^2_6 < (0.038)^2,
\label{xenon-number}
\end{equation}

where $\epsilon_{\phi^0, S} = \frac{\Omega_{\phi^0, S}h^2}{\Omega_c h^2}$. Here ${\Omega_{i}h^2}$
corresponds to the relic density of the $i$th type DM relic and $\Omega_c h^2$ is the total relic density
of DM in the universe (see Eq. (\ref{totalDM})).     
\end{itemize} 
\section{Combined Relic Density Calculation for the two dark matter Candidates 
$\phi^0$ and $S$}
\label{relic}
In the present two component dark matter model, the total relic density of the 
dark matter in the universe will have contributions from both the components $S$ 
and $\phi^0$. While both are annihilating into the SM particles, the heavier component, $S$ 
can annihilate into the lighter component, $\phi^0$ too. In order to obtain the correct 
combined relic density we have to solve Boltzmann's equation for each components 
simultaneously. The coupled Boltzmann's equations \cite{Belanger2011} to study the 
evolution of the number densities of the two dark matter candidates ($n_S$ and $n_{\phi^0}$) 
are given by,
\begin{eqnarray}
\frac{dn_{S}}{dt} + 3n_{S}{H} &=& -{\langle{\sigma {\rm{v}}}
_{SS\rightarrow \chi{\bar\chi}}\rangle} \left(n_{S}^2 - (n_{S}^{eq})^2\right) 
- {\langle{\sigma {\rm{v}}}_{SS\rightarrow\phi^0\phi^0}\rangle} \left(n_{S}^2 - 
\frac{(n_{S}^{eq})^2}{(n_{\phi^0}^{eq})^2}n_{\phi^0}^2\right) \, ,
\label{eq1}
\end{eqnarray}
\begin{eqnarray}
\frac{dn_{\phi^0}}{dt} + 3n_{\phi^0}{H} &=& -\langle{\sigma {\rm{v}}}
_{\phi^0\phi^0\rightarrow {\chi {\bar \chi}}}\rangle \left(n_{\phi^0}^2 -
(n_{\phi^0}^{eq})^2\right) 
 + {\langle{\sigma {\rm{v}}}_{SS\rightarrow\phi^0\phi^0}\rangle} \left(n_{S}^2 - 
\frac{(n_{S}^{eq})^2}{(n_{\phi^0}^{eq})^2}n_{\phi^0}^2\right).
\label{eq2}
\end{eqnarray}
Here $n_{\phi^0}^{eq}$ and $n_{S}^{eq}$ are the equilibrium values of $n_{\phi^0}$ and
$n_{S}$ respectively, $H$ is the Hubble's constant. $\chi$ represents any SM particle 
such as leptons, quarks, gauge bosons, Higgs boson. The annihilation of $SS$ into 
$\phi^0 \phi^0$ is included through the annihilation cross section 
$\langle{\sigma {\rm{v}}}_{SS\rightarrow\phi^0\phi^0}\rangle$, the expression 
of which in our scenario is discussed later in this section explicitly along with the total 
annihilation cross section.
The possible inclusion of co-annihilation terms for the
channels $\phi^0\phi^{\pm}\rightarrow\chi\chi^{\prime}$ and $\phi^0A^{0}\rightarrow\chi\chi^{\prime}$
will be discussed later.

Introducing two dimensionless variables $Y_i = \frac{n_i}{s}$ and $x_i = \frac{m_i}{T}$ 
with $i = S, \phi^0$, where $s$ and $T$ are the entropy density and temperature of 
the universe respectively, Eqs. (\ref{eq1}, \ref{eq2}) can be written as 
\begin{eqnarray}
\frac{dY_{S}}{dx_S} &=&-\left(\frac{45G}{\pi}\right)^{-\frac{1}{2}}
\frac{m_S}{x_S^2}{\sqrt{g_\star}}\left({\langle{\sigma {\rm{v}}}_
{SS\rightarrow {\chi \bar{\chi}}}\rangle}\left(Y_{S}^2-(Y_{S}^{eq})^2\right) 
+ {\langle{\sigma {\rm{v}}}_{SS\rightarrow\phi^0\phi^0}\rangle}\left(Y_{S}^2 - 
\frac{(Y_{S}^{eq})^2}{(Y_{\phi^0}^{eq})^2}Y_{\phi^0}^2\right)\right)
\,\, , \nonumber \\
\label{eq3}
\end{eqnarray}
\begin{eqnarray}
\frac{dY_{\phi^0}}{dx_{\phi^0}} &=&-\left(\frac{45G}{\pi}\right)^{-\frac{1}{2}}
\frac{m_{\phi^0}}{x_{\phi^0}^2}\sqrt{g_\star}\left({\langle{\sigma {\rm{v}}}_
{\phi^0\phi^0\rightarrow {\chi \bar{\chi}}}\rangle}
\left(Y_{\phi^0}^2-(Y_{\phi^0}^{eq})^2\right) 
 - {\langle{\sigma {\rm{v}}}_{SS\rightarrow\phi^0\phi^0}\rangle} \left(Y_{S}^2 - 
\frac{(Y_{S}^{eq})^2}{(Y_{\phi^0}^{eq})^2}Y_{\phi^0}^2\right)\right)
\,\, . \nonumber \\
\label{eq4}
\end{eqnarray}
Here $G$ is the Gravitation constant and $g_{\star}$ is defined as,
\begin{eqnarray}
\sqrt{g_\star} = \frac{h_{{\rm{eff}}}(T)}{\sqrt{g_{{\rm{eff}}}(T)}}
\left(1 + \frac{1}{3}\frac{d\,{\rm ln}(h_{{\rm{eff}}}(T))}{d\,{\rm ln}(T)}\right) \,\, ,
\end{eqnarray}
with $g_{{\rm{eff}}}(T)$ and $h_{{\rm{eff}}}(T)$ are the effective degrees of freedom related to 
the energy and entropy densities through
$\rho = g_{{\rm{eff}}}(T)\frac{\pi^2}{30}T^4, \, s = h_{{\rm{eff}}}(T)\frac{2\pi^2}{45}T^3$.

Once we get the values of $Y_{\phi^0}$ and  $Y_{S}$ at the present temperature $T_0$ 
after solving the coupled Eqs. (\ref{eq3}, \ref{eq4}), we will be able to calculate the 
individual contributions $\Omega_{\phi^0}$ and  $\Omega_{S}$ from  \cite{gondolo},
\begin{eqnarray}
{\Omega_i h^2} = 2.755\times10^8 \left ( \frac{m_i}{\rm GeV} \right ) Y_i(T_0)\,\, ,
\label{omega_individual}
\end{eqnarray}
(using the present values of $s$ and $h$). In this work we have solved the coupled Boltzmann 
equations numerically to get the values of $Y_i (T_0)$. After having these estimates, the 
total relic density of the universe can be obtained through 
\begin{eqnarray}
{\Omega_c h^2} = {\Omega_{\phi^0} h^2} + {\Omega_{S} h^2}\,\,.
\label{totalDM}
\end{eqnarray}

As we mentioned before, ${\langle{\sigma {\rm{v}}}_{SS\rightarrow {\chi \bar{\chi}}}\rangle}$
in Eq.(\ref{eq1}) represents the total annihilation cross section of two $S$ particles into
SM particles such as leptons and quarks ($f \bar{f}$), gauge bosons ($W^+ W^-, Z Z$) and Higgs boson 
($h$). 
\begin{figure}[h]
\centering
\includegraphics[width=5cm,height=3cm]{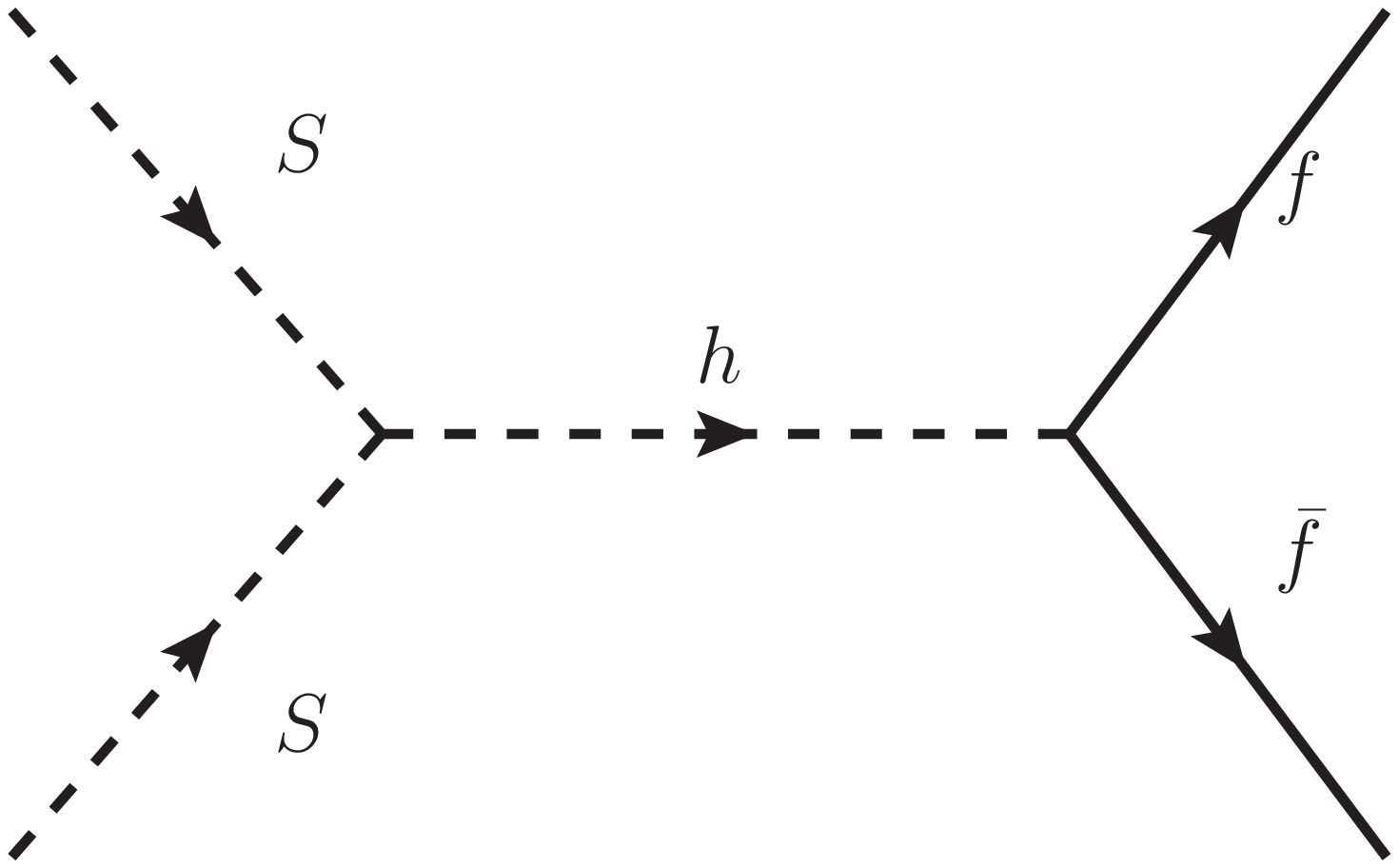}
\includegraphics[width=5cm,height=3cm]{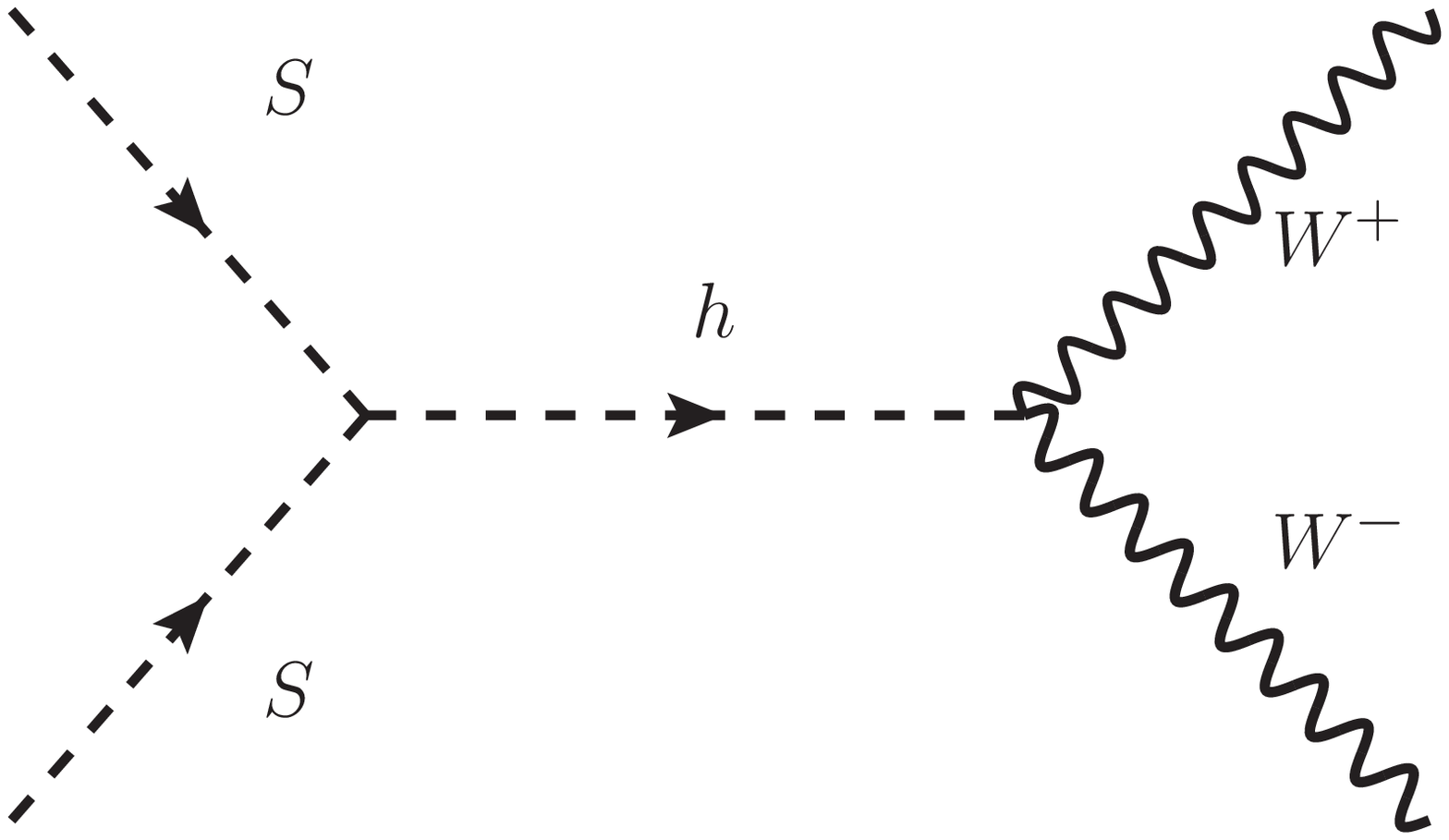}
\includegraphics[width=5cm,height=3cm]{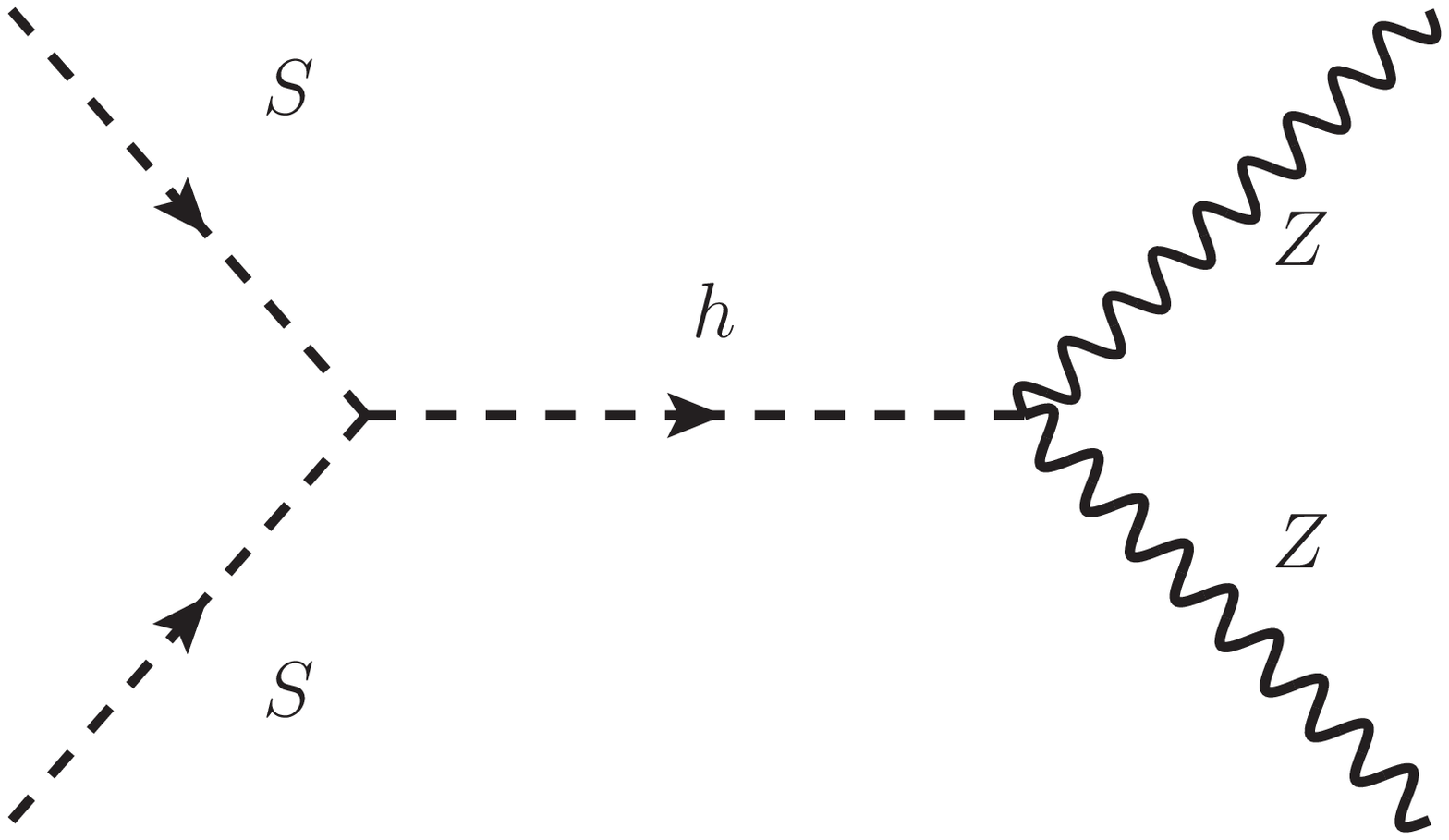}
\includegraphics[width=5cm,height=3cm]{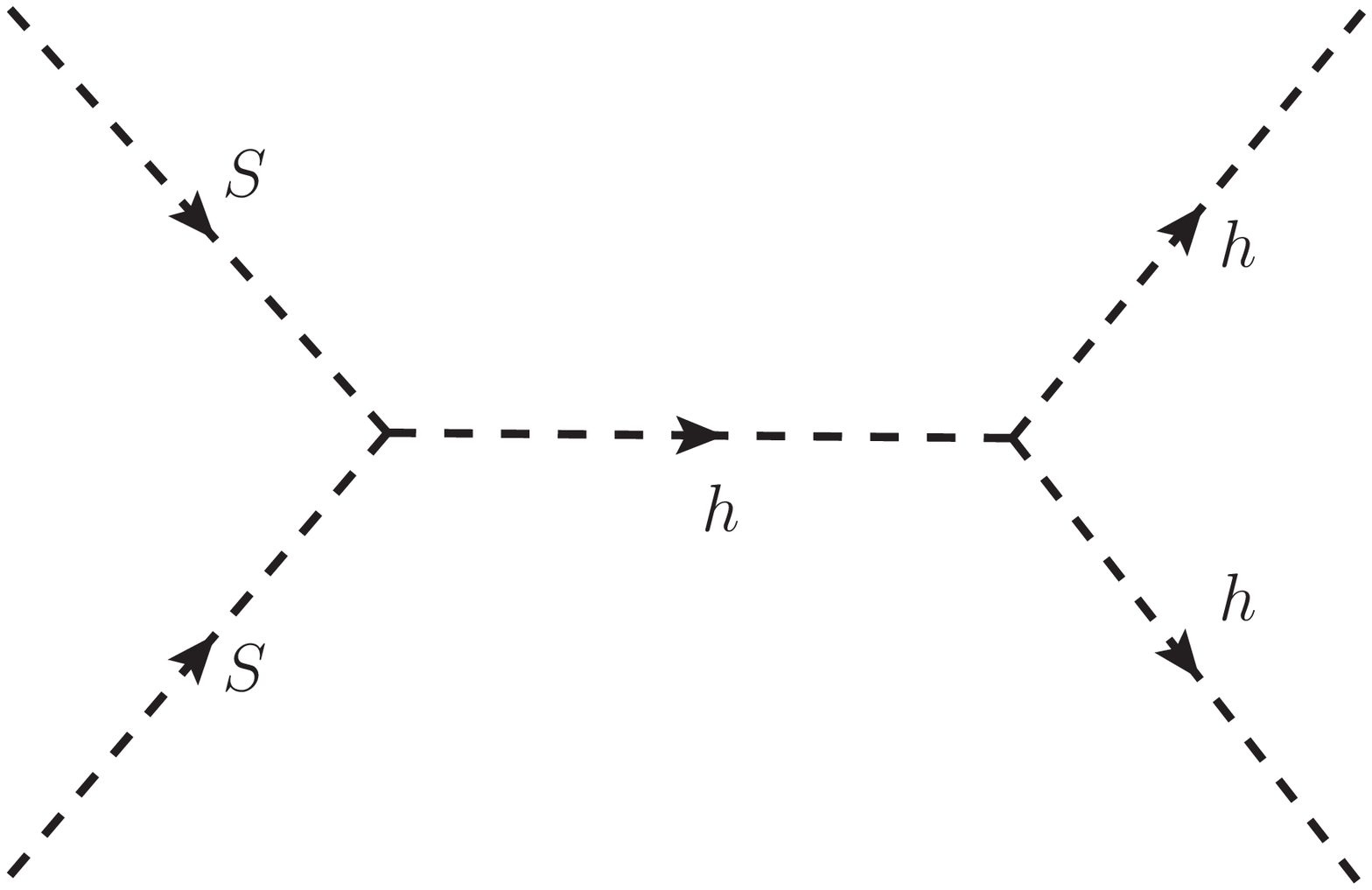}
\includegraphics[width=5cm,height=3.4cm]{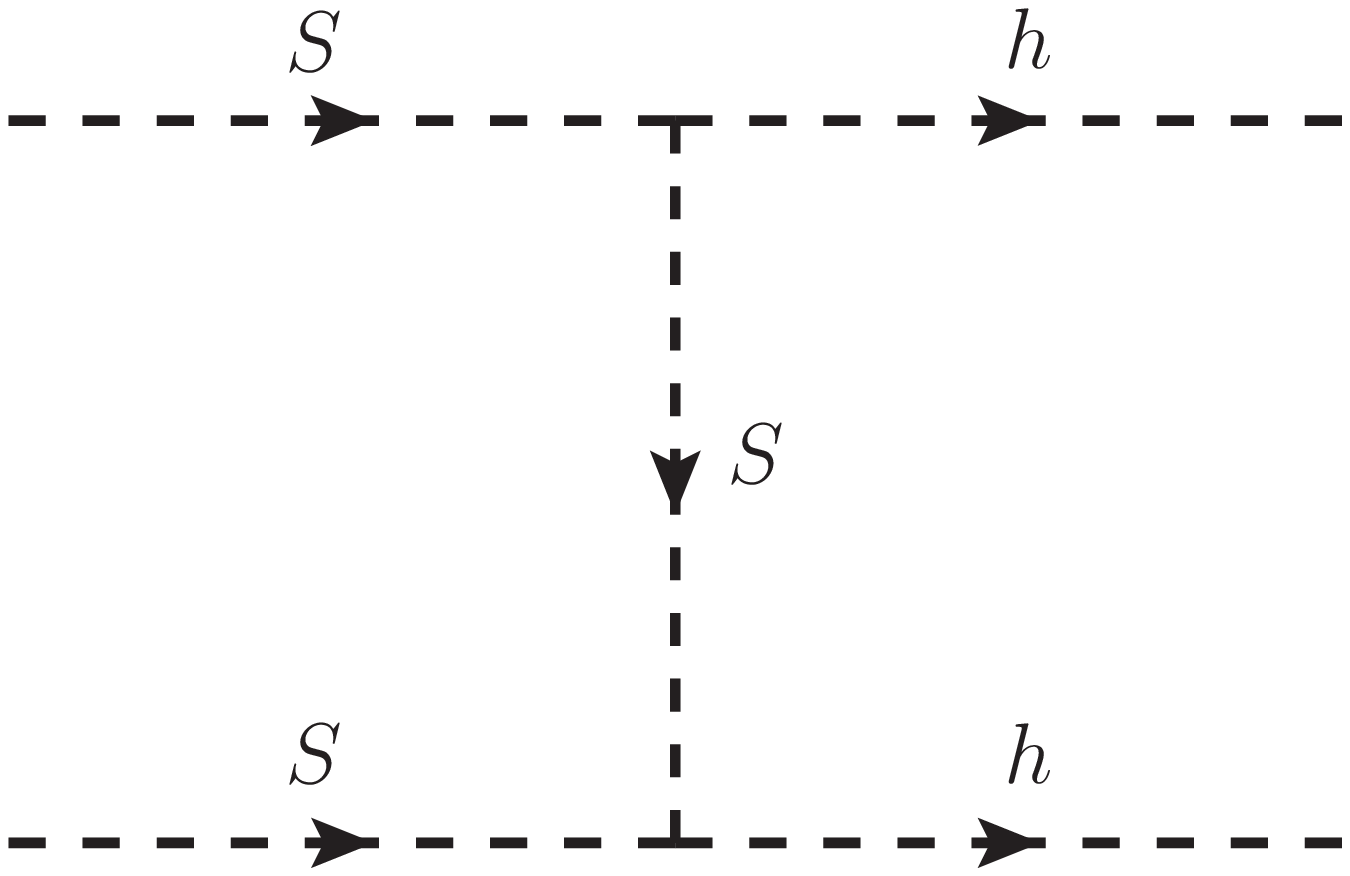}
\includegraphics[width=5cm,height=3cm]{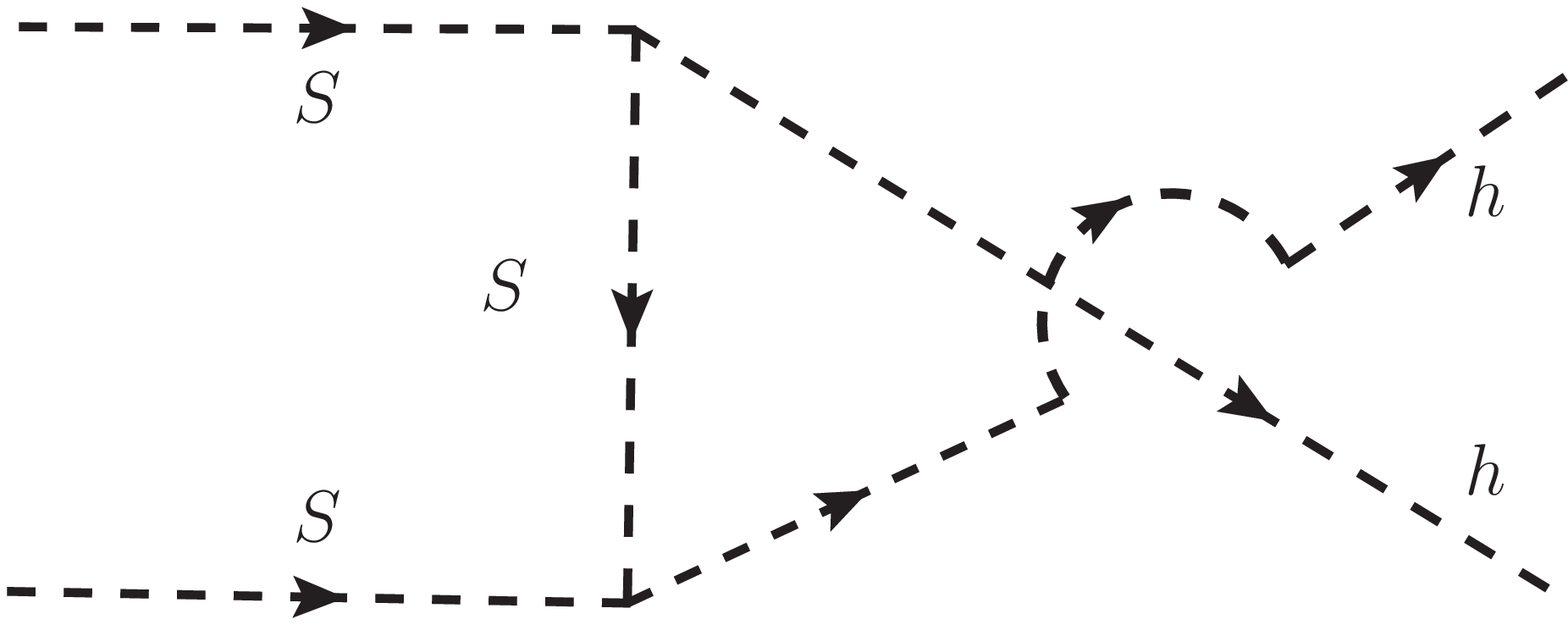}
\includegraphics[width=5cm,height=3cm]{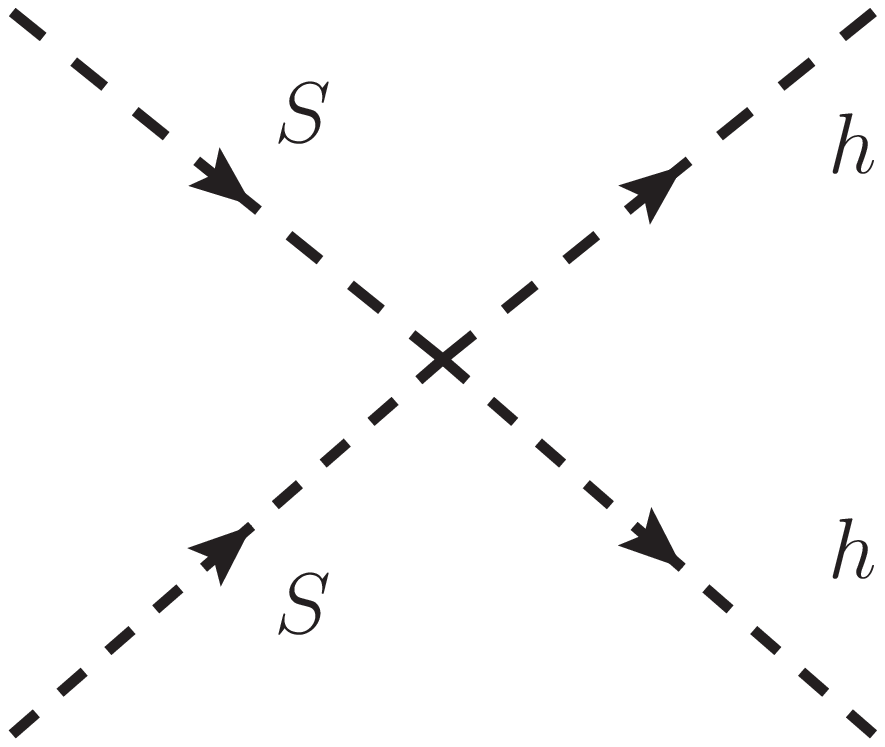}
\caption{Lowest order Feynman diagrams of two S annihilate into a pair
of fermion and anti-fermion, $ W^+ W^-$, $ZZ$ and Higgs.}
\label {feyn_dia_s}
\end{figure}
The Feynman diagrams for all the processes are shown in Fig. \ref{feyn_dia_s}. The 
expressions of annihilation cross sections of two $S$ particles for these final states are 
given below \cite{lei}.

\begin{eqnarray}
\langle{\sigma {\rm{v}}}_{SS\rightarrow f\bar f}\rangle &=& \left(\frac{g_{SSh}}{v}\right)^2 
\frac{{m^2_f}}{\pi}\frac{\left(1-\frac{{m^2_f}}{{m^2_S}}\right)^{3/2}}
{\left[(4{m^2_S}-{m^2_h})^2+(\Gamma_h m_h)^2\right]}\,\, , \\
\langle{\sigma {\rm{v}}}_{SS\rightarrow W^+W^-}\rangle &=& 2\left(\frac{g_{SSh}}{v}\right)^2\frac{{m^2_S}}{\pi}
\frac{\left(1-\frac{{m^2_W}}{{m^2_S}}\right)^{1/2}}{\left[(4{m^2_S}-{m^2_h})^2+
(\Gamma_h m_h)^2\right]} 
\left(1-\frac{{m^2_W}}{{m^2_S}} + \frac{3}{4}\frac{{m^4_W}}{{m^4_S}}\right)\,\, , \\
\langle{\sigma {\rm{v}}}_{SS\rightarrow ZZ}\rangle &=& \left(\frac{g_{SSh}}{v}\right)^2\frac{{m^2_S}}{\pi}
\frac{\left(1-\frac{{m^2_Z}}{{m^2_S}}\right)^{1/2}}{\left[(4{m^2_S}-{m^2_h})^2+
(\Gamma_h m_h)^2\right]} 
\left(1-\frac{{m^2_Z}}{{m^2_S}} + \frac{3}{4}\frac{{m^4_Z}}{{m^4_S}}\right)\,\, , \\
\langle{\sigma {\rm{v}}}_{SS\rightarrow hh}\rangle &=& \frac{1}{4\pi
{m^2_S}}\left(1-\frac{{m^2_h}}{{m^2_S}}\right)^{1/2}
\left[\left(\frac{3\,\,g_{SSh}            \,\,m^2_h}{2v(4{m^2_S}-{m^2_h})}\right)^2 
+ \frac{3\,\,g_{SSh}\,\,\,g_{SShh}\,\,\,m^2_h}{v(4{m^2_S}-{m^2_h})} \right.\nonumber\\
&&\left. + {g^2_{SShh}}  
+\frac{4\,\,{g^2_{SSh}} g_{SShh}}{2{m^2_S}-{m^2_h}}+\frac{6\,\,{g^3_{SSh}}m^2_h}
{v(2{m^2_S}-{m^2_h})(4{m^2_S}-{m^2_h})}\right.\nonumber\\&&\left.+
\left(\frac{2\,\,{g^2_{SSh}}}{2{m^2_S}-{m^2_h}}\right)^2 \right].
\end{eqnarray}
In writing the expression for the cross sections, we have introduced the 
notation $g_{xyz}$ (or $g_{xyzp}$) which are related to the coupling 
constants of the corresponding interaction Lagrangian through 
${\cal{L}} = g_{xyz} XYZ$ (or ${\cal{L}} = g_{xyzp} XYZP$). Here $X,Y,Z,P$ are the 
fields involved in a particular process. All the $g_{XYZ}$ and $g_{XYZP}$ are listed 
in Table {\ref{tab2}}. The masses for fermion $f$, $W$ boson, 
$Z$ boson and Higgs boson are denoted by $m_f, m_W, m_Z, m_h$ respectively and
$\Gamma_h$ represents the Higgs decay width. 

\begin{table}[t]
\begin{center}
\begin{tabular} {|c|c|}
\hline
{Interactions involving $XYZ(P)$} & {\bf $g_{xyz(p)}$ } \\
\hline
$S S h$&$g_{SSh} = -{\lambda_6}v$\\
\hline
$S S h h$& $g_{SShh} = -\frac{1}{2}{\lambda_6}$\\
\hline
$\phi^0 \phi^0 h$&$g_{\phi^0\phi^0h} = -{\alpha}v$\\
\hline
$\phi^0 \phi^0 h h$&$g_{\phi^0\phi^0hh} = -\frac{1}{2}{\alpha}$\\
\hline
$\phi^0 \phi^0 W^+ W^-$&$g_{\phi^0\phi^0W^+W^-} = \frac{m^2_W}{v^2}$\\
\hline
$\phi^0 \phi^0 Z Z$&$g_{\phi^0\phi^0ZZ} = \frac{m^2_Z}{2v^2}$\\
\hline
$\phi^0 \phi^0 \phi^+ \phi^-$ & $g_{\phi^0 \phi^0 \phi^+ \phi^-} = -\rho_2$\\
\hline
$S S \phi^0 \phi^0$& $-\frac{1}{2}{\lambda_5}$\\
\hline
\end{tabular}
\end{center}
\caption{Interactions and the corresponding couplings.}
\label{tab2}
\end{table}

\begin{figure}[h]
\centering
\includegraphics[width=5cm,height=3cm]{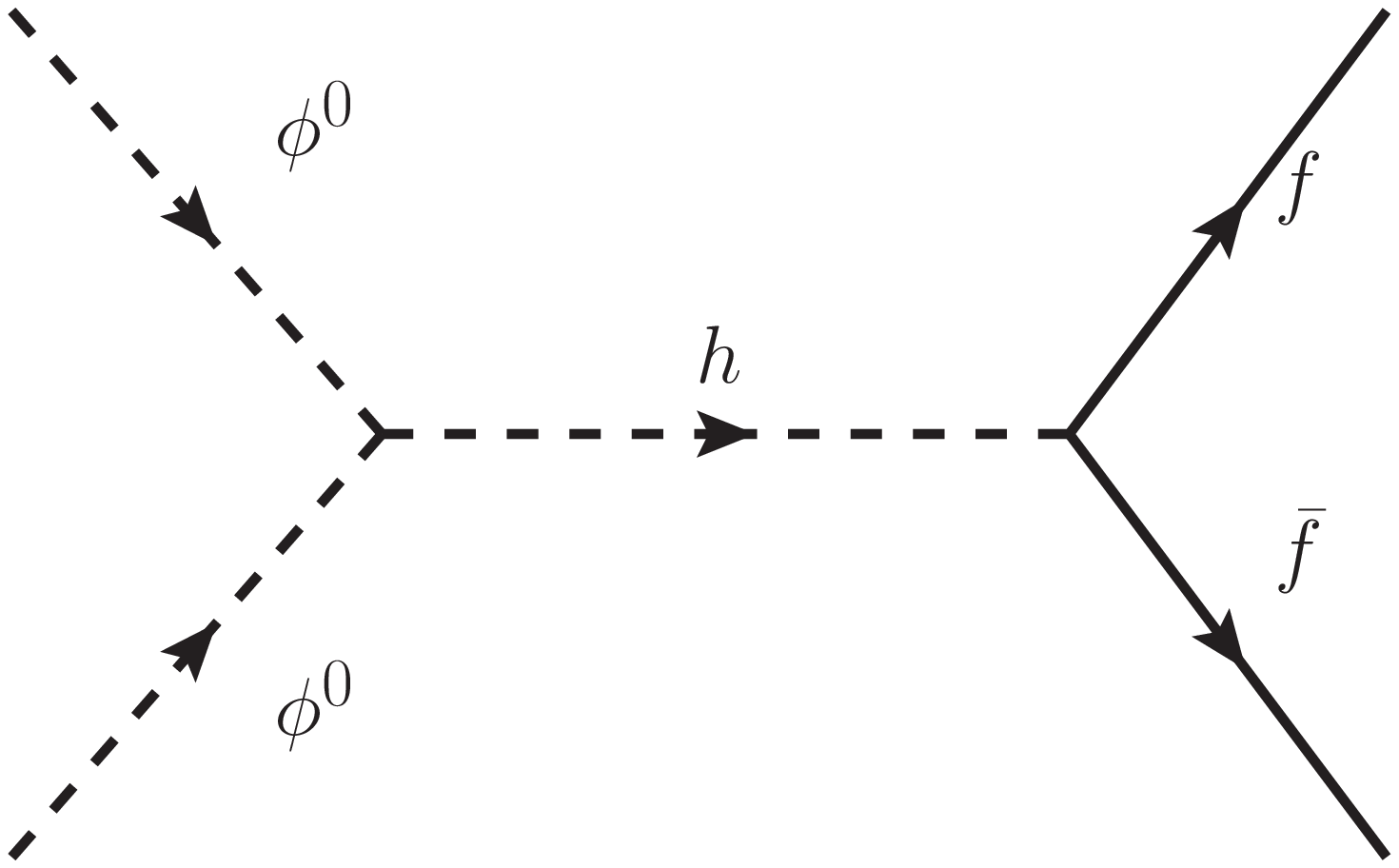}
\includegraphics[width=3.5cm,height=3cm]{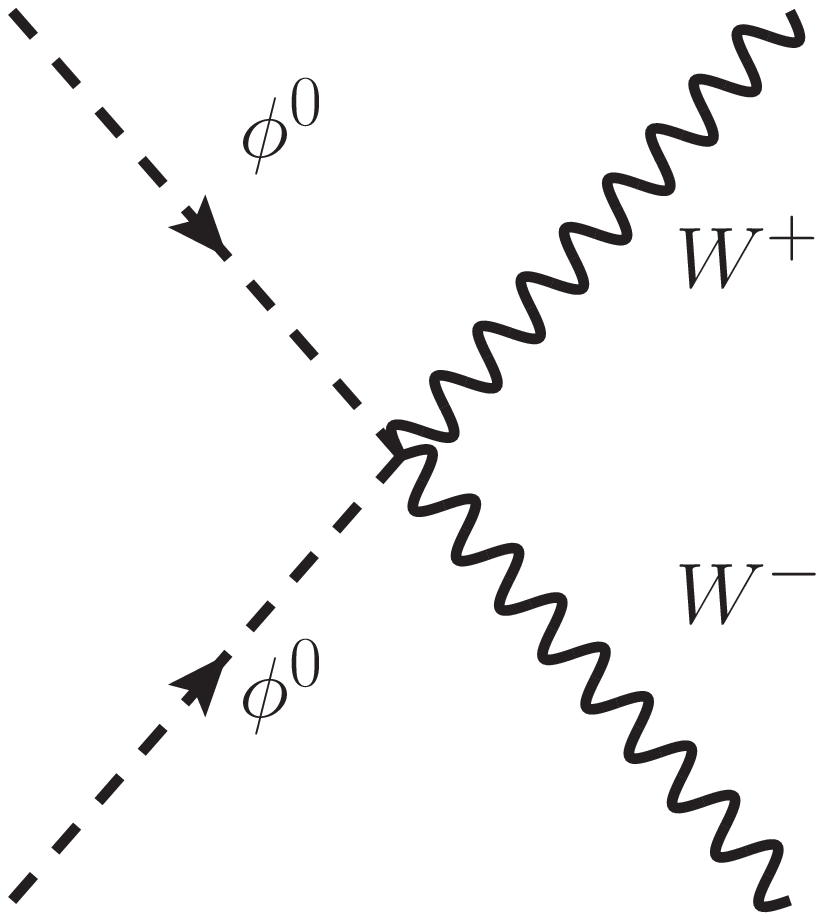}
\includegraphics[width=3.5cm,height=3cm]{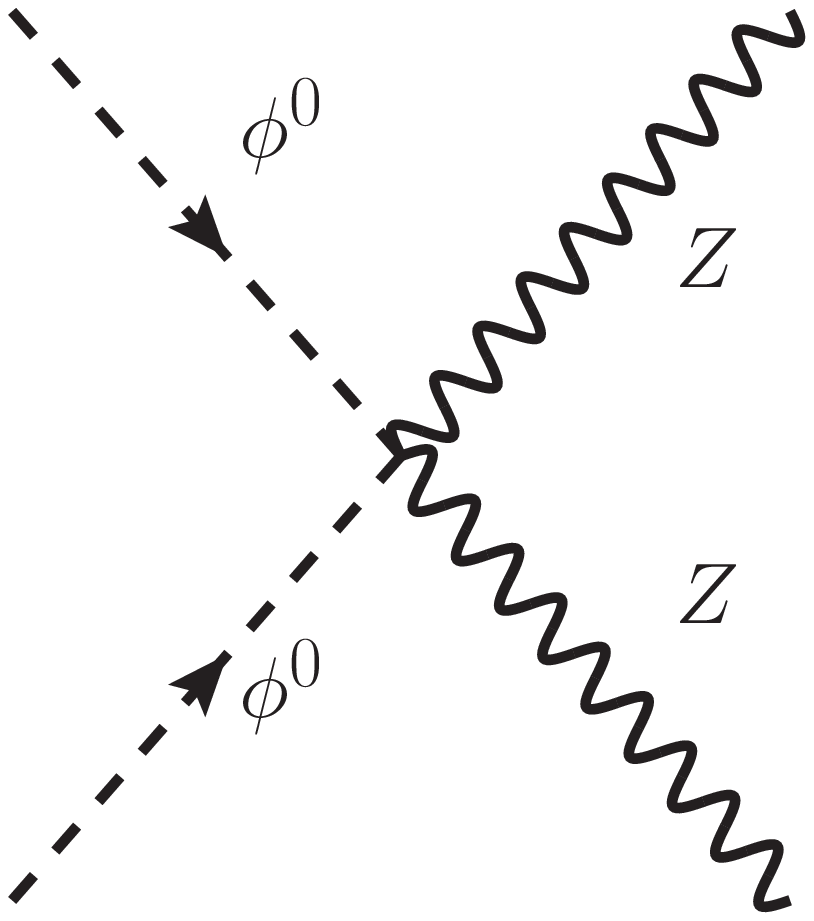}
\includegraphics[width=3.5cm,height=3cm]{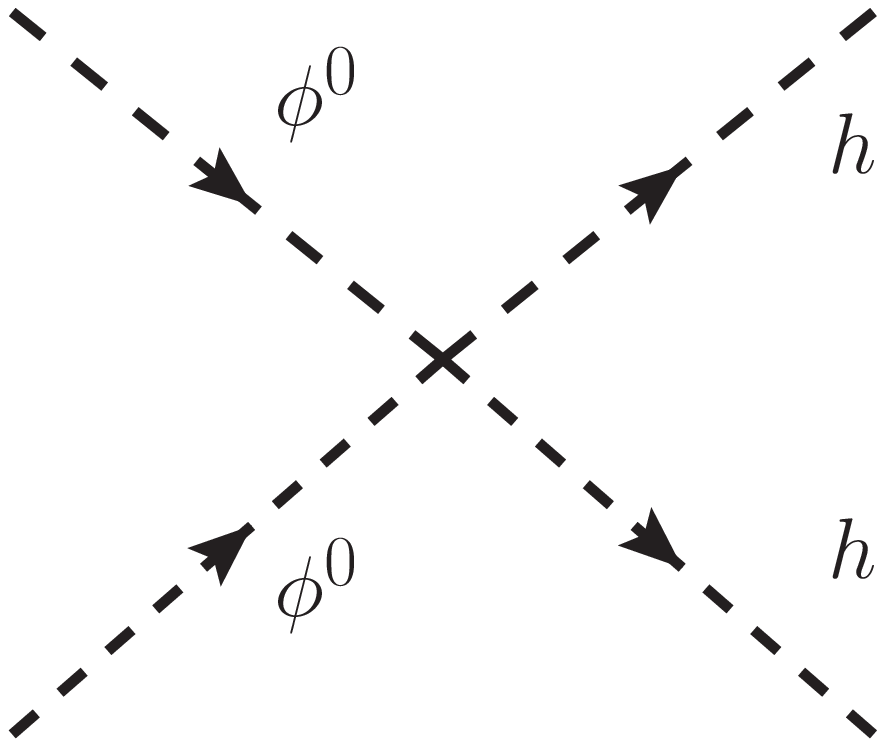}
\includegraphics[width=5cm,height=3cm]{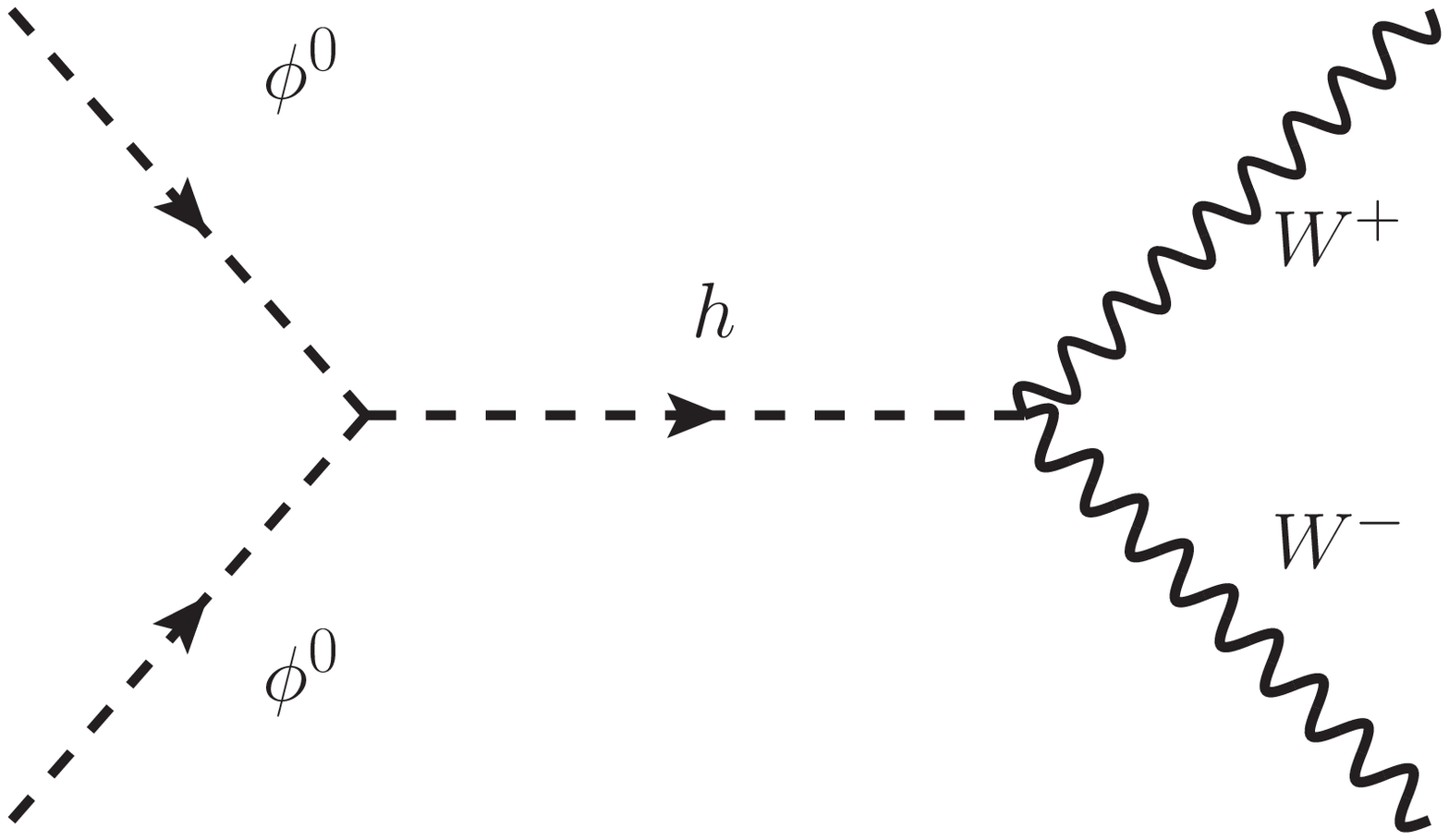}
\includegraphics[width=5cm,height=3.4cm]{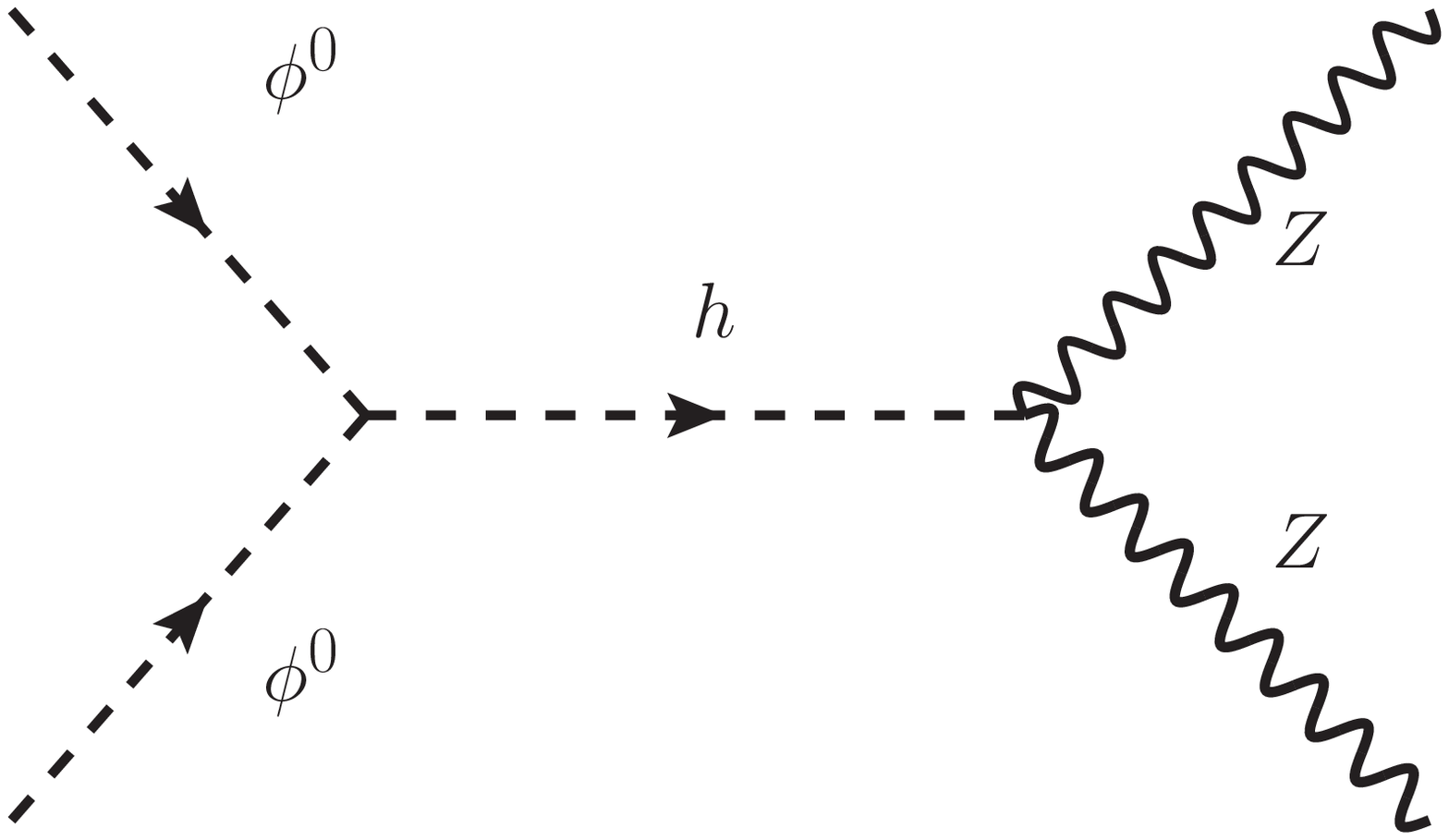}
\includegraphics[width=5cm,height=3cm]{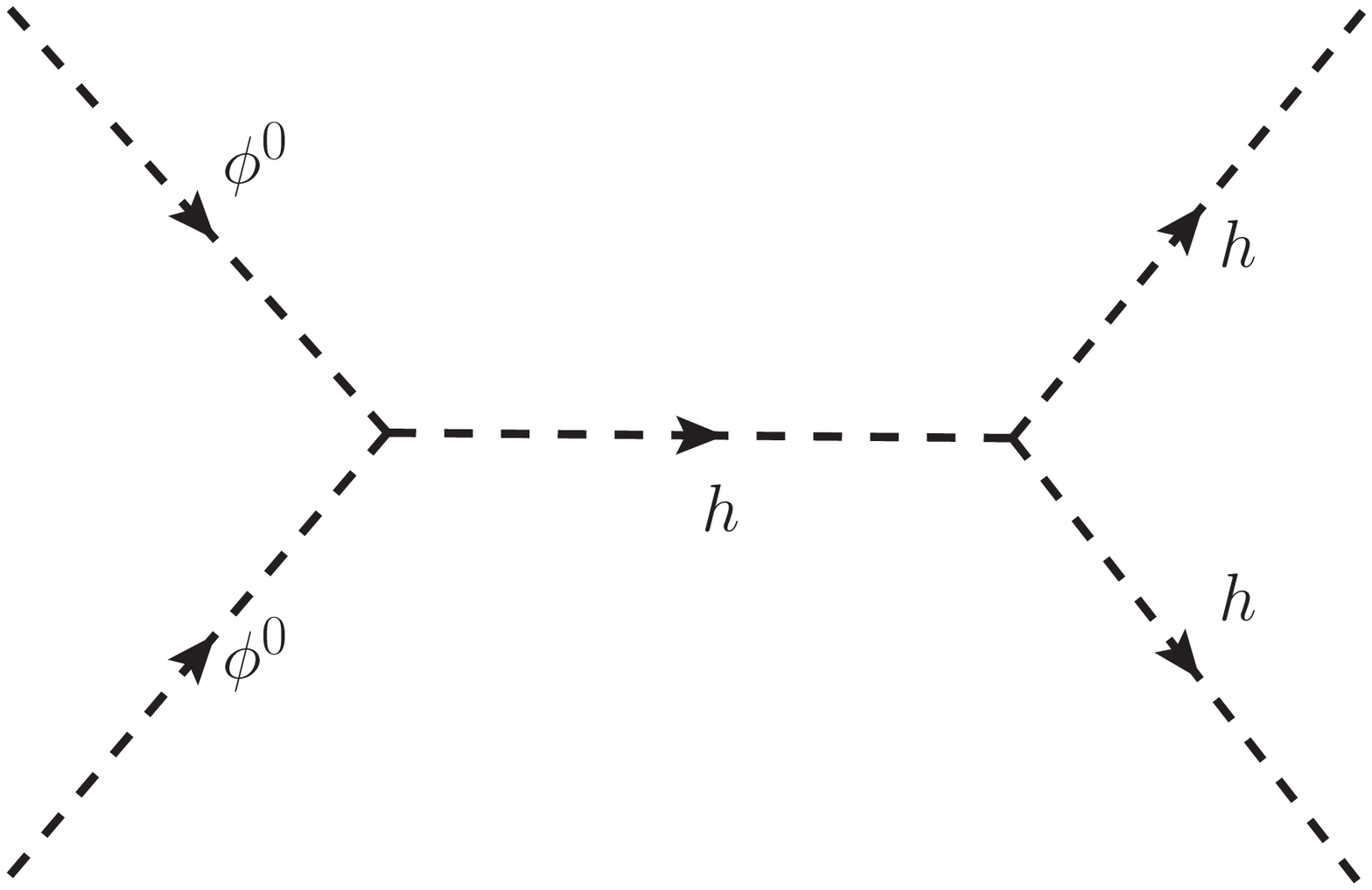}
\includegraphics[width=5cm,height=3cm]{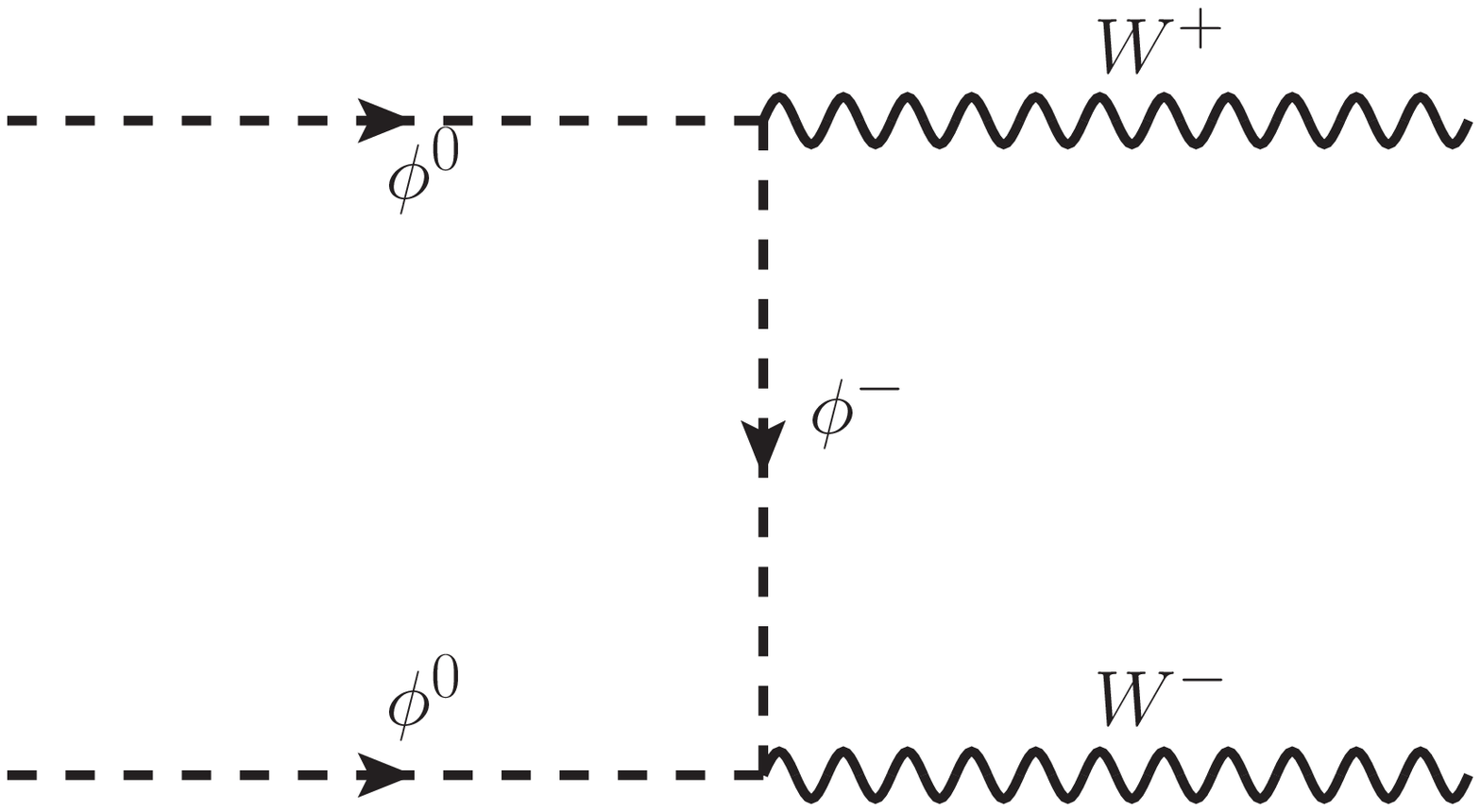}
\includegraphics[width=5cm,height=3cm]{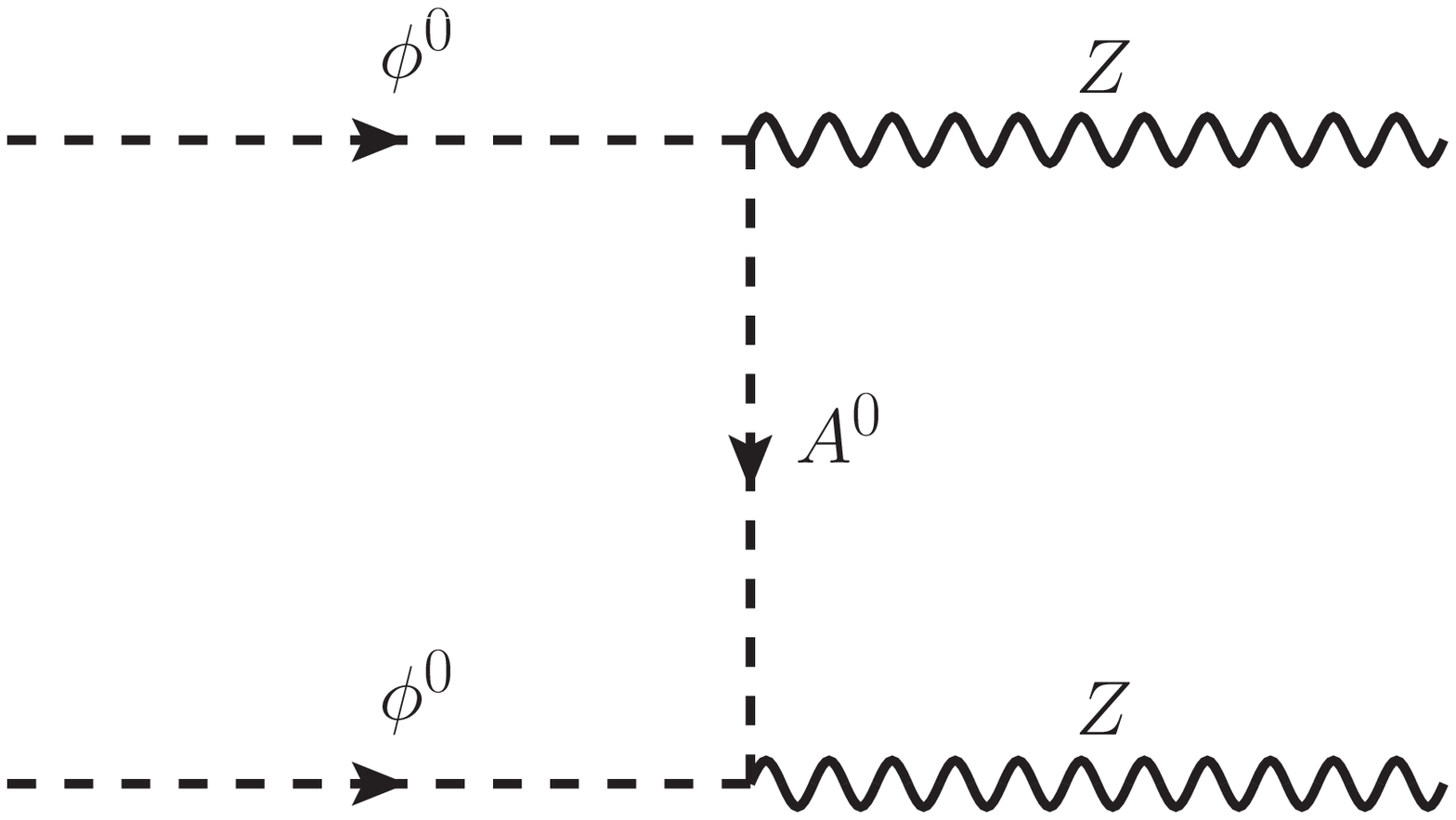}
\includegraphics[width=5cm,height=3cm]{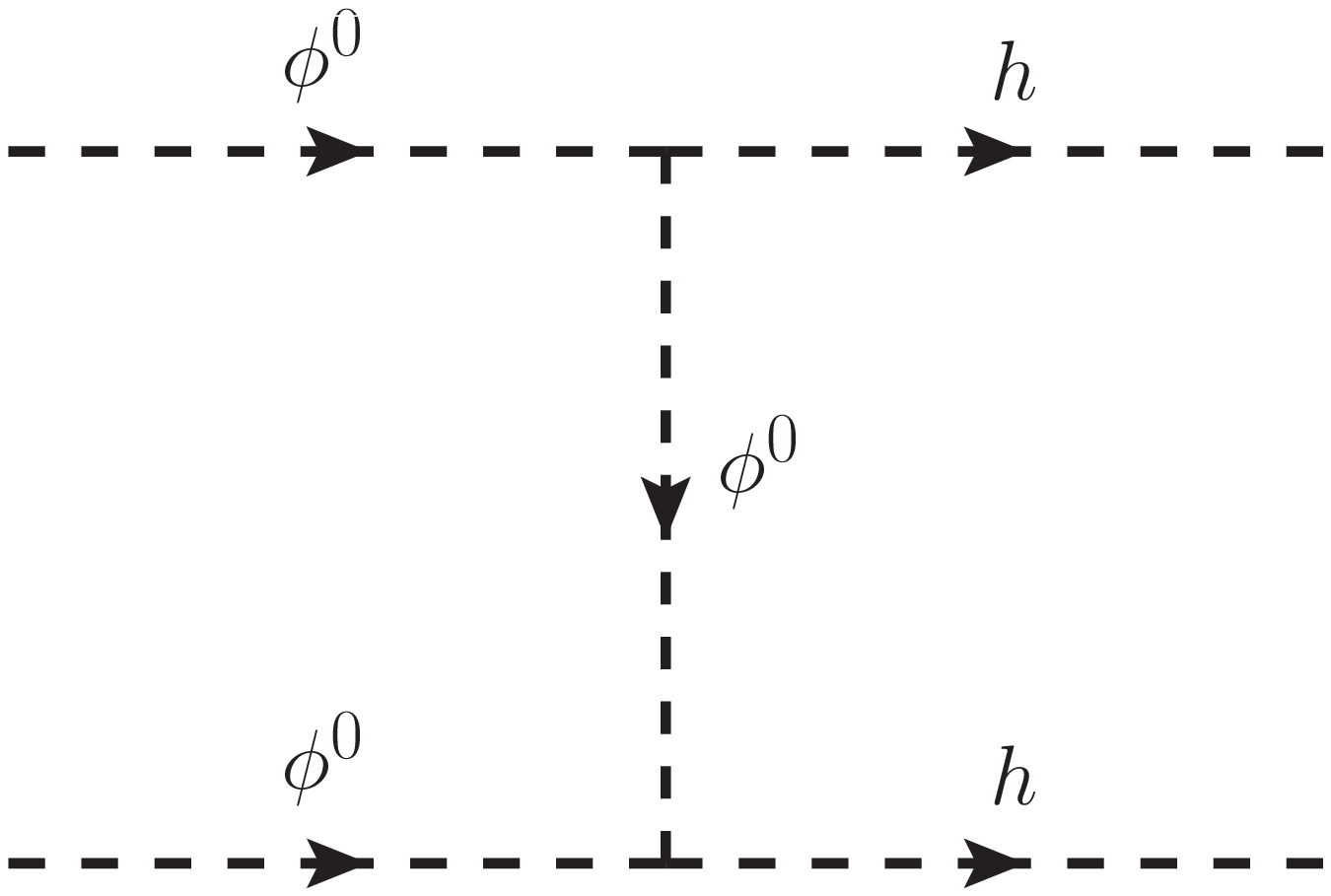}
\includegraphics[width=5cm,height=3cm]{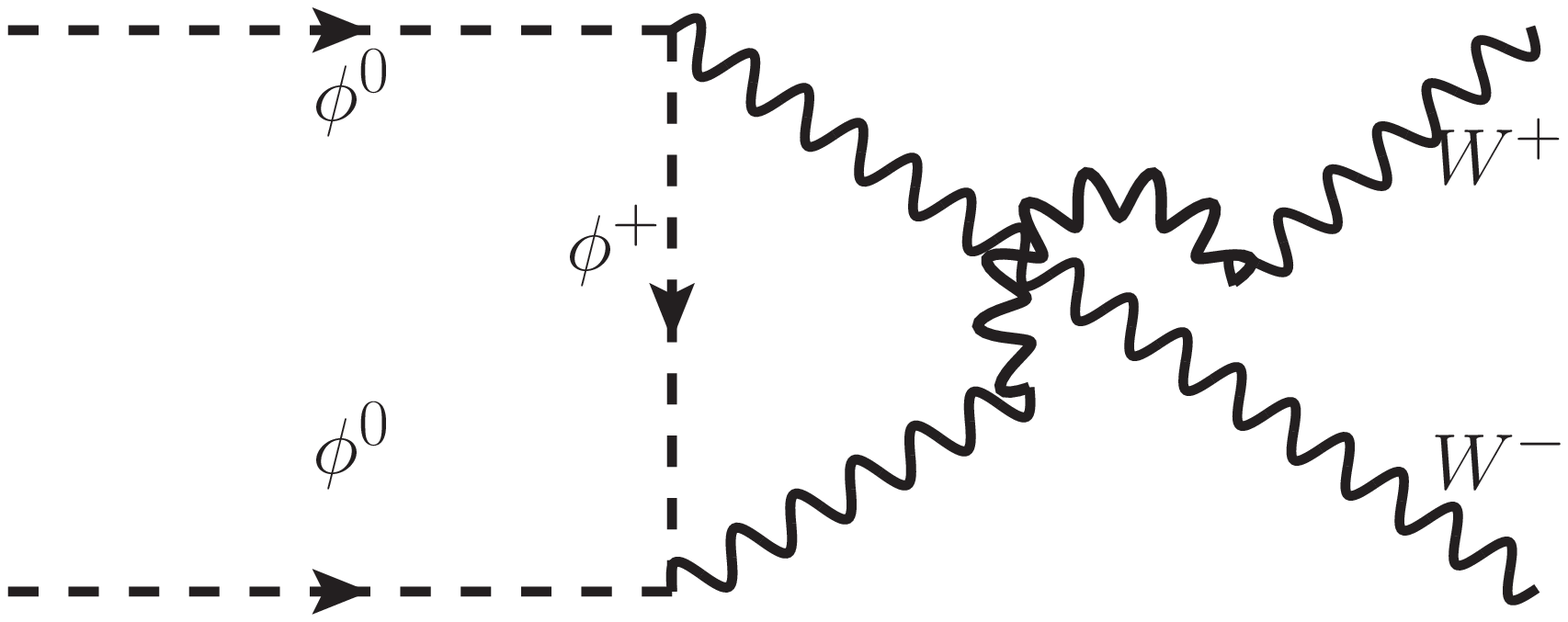}
\includegraphics[width=5cm,height=3cm]{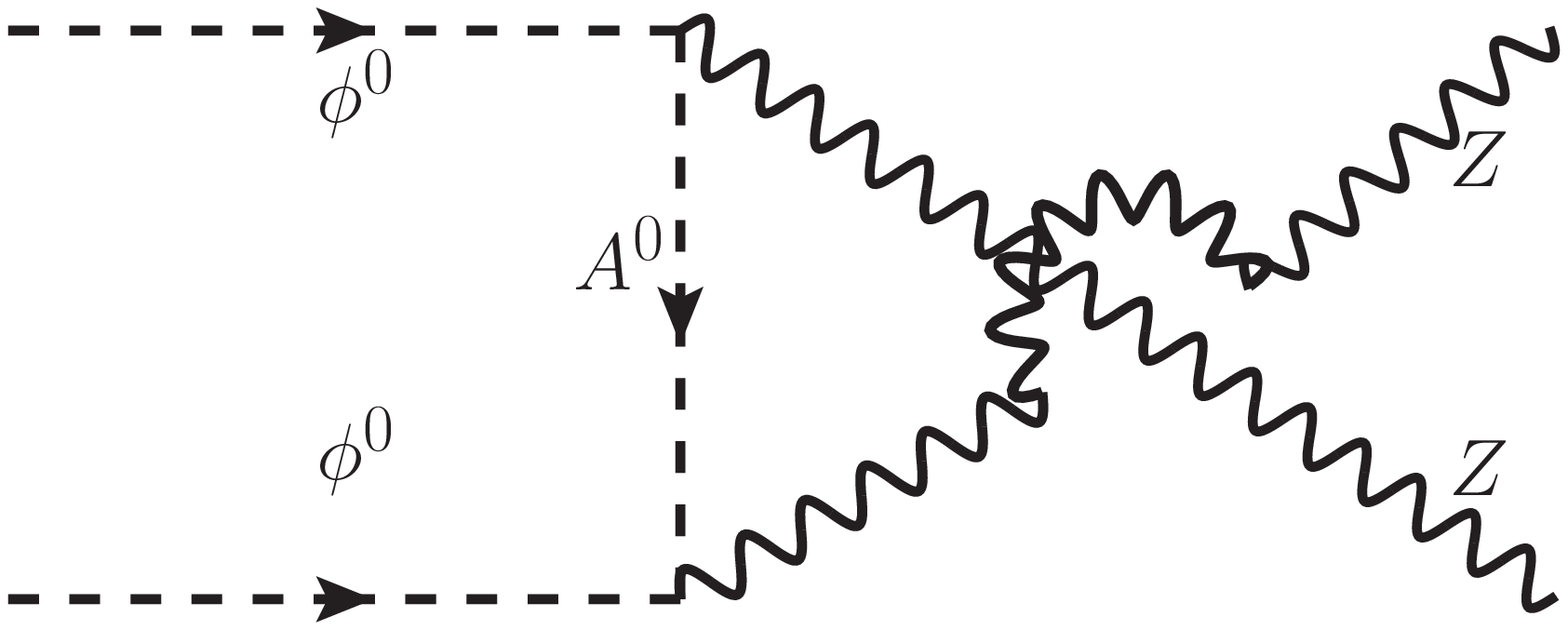}
\includegraphics[width=5cm,height=3cm]{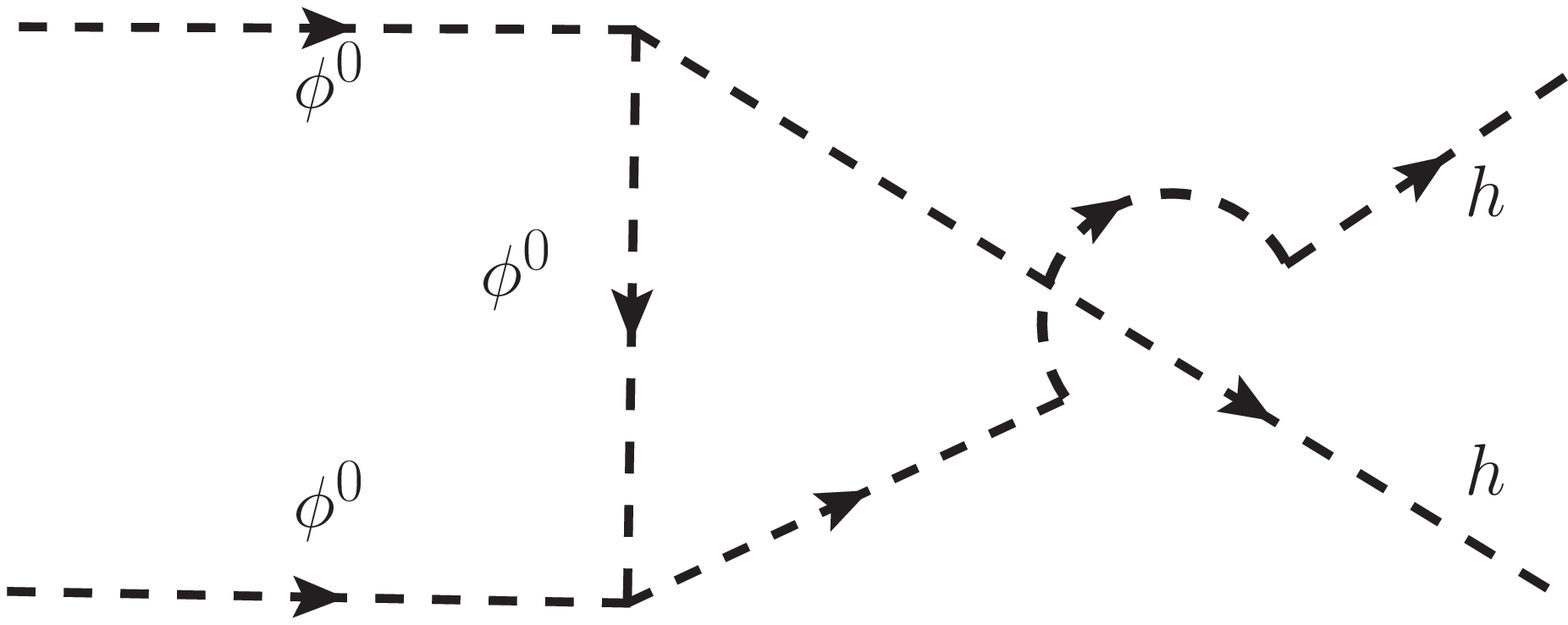}
\caption{Lowest order Feynman diagrams of two $\phi^0$ annihilate into a pair
of fermion and anti-fermion, $ W^+ W^-$, $ZZ$ and Higgs.}
\label{feyn_dia_phi0}
\end{figure}

Similarly in Eq. (\ref {eq2}),  
$\langle{\sigma {\rm{v}}}_{\phi^0\phi^0\rightarrow {\chi \bar{\chi}}}\rangle$ represents
the total annihilation cross section of two $\phi^0$ particles into SM particles. 
The Feynman diagrams of individual processes are shown in Fig.
\ref{feyn_dia_phi0} 
and the expressions of annihilation cross sections for these channels
are given below.
\begin{eqnarray}
\langle{\sigma {\rm{v}}}_{\phi^0\phi^0\rightarrow f\bar f}\rangle &=& \left(\frac{g_{\phi^0\phi^0h}}
{v}\right)^2 \frac{{m^2_f}}{\pi}\frac{\left(1-\frac{{m^2_f}}{{m^2_{\phi^0}}}\right)^{3/2}}
{\left[(4{m^2_{\phi^0}}-{m^2_h})^2+(\Gamma_h m_h)^2\right]} \,\, ,\\
\langle{\sigma {\rm{v}}}_{\phi^0\phi^0\rightarrow W^+ W^-}\rangle &=& \frac{1}{2\pi}
\left(1-\frac{{m^2_W}}{{m^2_{\phi^0}}}\right)^{1/2}\left[{m^2_{\phi^0}}
\left(1-\frac{{m^2_W}}{m^2_{\phi^0}}+\frac{3{m^4_W}}{4{m^4_{\phi^0}}}\right) 
\left\{\left(\frac{{g_{\phi^0\phi^0W^+ W^-}}}{m^2_W}\right)^2 +
\right.\right.\nonumber \\&&\left.\left.
 \frac{\left(\frac{2\,\,g_{\phi^0\phi^0h}}{v}\right)^2}{\left[(4{m^2_{\phi^0}}-{m^2_h})^2+
(\Gamma_h m_h)^2\right]} - \frac{\left(\frac{4\,\,g_{\phi^0\phi^0h} \,\,\,
g_{\phi^0\phi^0W^+W^-}}{{m^2_W}v}\right)(4{m^2_{\phi^0}}-{m^2_h})}
{\left[(4{m^2_{\phi^0}}-{m^2_h})^2+(\Gamma_h m_h)^2\right]}\right\}\right.\nonumber \\
&&\left.+ \frac{g^2}{4}\left\{
\left(g_{\phi^0\phi^0W^+W^-}-\frac{2\,\,g_{\phi^0\phi^0h}
(4{m^2_{\phi^0}}-{m^2_h})\,{m^2_W}}
{v{\left[(4{m^2_{\phi^0}}-{m^2_h})^2+(\Gamma_h m_h)^2\right]}}\right)\times
\right.\right.\nonumber\\&&\left.\left.
2\left(\frac{({m^4_W}-3{m^2_W}{m^2_{\phi^0}} +
2{m^4_{\phi^0}})}{{m^4_W}({m^2_W}-{m^2_{\phi^+}}-{m^2_{\phi^0}})}\right) 
\right.\right.\nonumber \\&&\left.\left.+ 
g^2\left(\frac{m_{\phi^0}
({m^2_W}-{m^2_{\phi^0}})}{{m^2_W}({m^2_{\phi^+}} 
-{m^2_W} + {m^2_{\phi^0}})}\right)^2 
\right\}\right]\,\, ,\nonumber\\
\\
\langle{\sigma {\rm{v}}}_{\phi^0\phi^0\rightarrow ZZ}\rangle &=& \frac{1}{4\pi}
\left(1-\frac{{m^2_Z}}{{m^2_{\phi^0}}}\right)^{1/2}\left[{m^2_{\phi^0}}
\left(1-\frac{{m^2_Z}}
{m^2_{\phi^0}}+\frac{3{m^4_Z}}{4{m^4_{\phi^0}}}\right) 
\left\{\left(\frac{2\,\,{g_{\phi^0\phi^0ZZ}}}{m^2_Z}\right)^2 +
\right.\right.\nonumber \\&&\left.\left.
 \frac{\left(\frac{2\,\,g_{\phi^0\phi^0h}}{v}\right)^2}
 {\left[(4{m^2_{\phi^0}}-{m^2_h})^2+
(\Gamma_h m_h)^2\right]} - \frac{\left(\frac{8\,\,g_{\phi^0\phi^0h} \,\,\,
g_{\phi^0\phi^0ZZ}}{{m^2_Z}v}\right)(4{m^2_{\phi^0}}-{m^2_h})}
{\left[(4{m^2_{\phi^0}}-{m^2_h})^2+(\Gamma_h m_h)^2\right]}\right\} 
\right.\nonumber \\&&\left.+\frac{g^2}{4\,Cos^2{\theta_W}}
\left\{
2\left(\frac{({m^4_Z}-3{m^2_Z}{m^2_{\phi^0}} + 2{m^4_{\phi^0}})}{{m^4_Z}({m^2_Z}-{m^2_{A^0}}-
{m^2_{\phi^0}})}\right)
\right.\right.\nonumber \\&&\left.\left.\times
\left(2\,g_{\phi^0\phi^0ZZ}-\frac{2\,\,g_{\phi^0\phi^0h}
(4{m^2_{\phi^0}}-{m^2_h})\,{m^2_Z}}
{v{\left[(4{m^2_{\phi^0}}-{m^2_h})^2+(\Gamma_h m_h)^2\right]}}\right)
\right.\right.\nonumber \\ &&\left.\left.
+\frac{g^2}{Cos^2{\theta_W}}\left(\frac{m_{\phi^0}({m^2_Z}-
{m^2_{\phi^0}})}{{m^2_Z}({m^2_{A^0}} -{m^2_Z} + {m^2_{\phi^0}})}\right)^2
\right\}\right] \,\, ,\\
\langle{\sigma {\rm{v}}}_{\phi^0\phi^0\rightarrow hh}\rangle &=& \frac{1}{4\pi
{m^2_{\phi^0}}}\left(1-\frac{{m^2_h}}{{m^2_{\phi^0}}}\right)^{1/2}
\left[\left(\frac{3\,\,g_{\phi^0\phi^0h}\,\,m^2_h}{2v(4{m^2_{\phi^0}}-{m^2_h})}\right)^2
+ {g^2_{\phi^0\phi^0hh}} \right.\nonumber\\&&\left. 
+ \frac{3\,\,g_{\phi^0\phi^0h}\,\,\,g_{\phi^0\phi^0hh}\,\,\,m^2_h}
{v(4{m^2_{\phi^0}}-{m^2_h})} 
+\left(\frac{2\,\,{g^2_{\phi^0\phi^0h}}}{2{m^2_{\phi^0}}-{m^2_h}}\right)^2 +\frac{4\,\,{g^2_{\phi^0\phi^0h}}
g_{\phi^0\phi^0hh}}{2{m^2_{\phi^0}}-{m^2_h}} 
\right.\nonumber \\ &&\left.
+\frac{6\,\,{g^3_{\phi^0\phi^0h}}\,m^2_h}{v(2{m^2_{\phi^0}}-{m^2_h})
(4{m^2_{\phi^0}}-{m^2_h})}\right].
\end{eqnarray}

\begin{figure}[h]
\centering
\includegraphics[width=6cm,height=3cm]{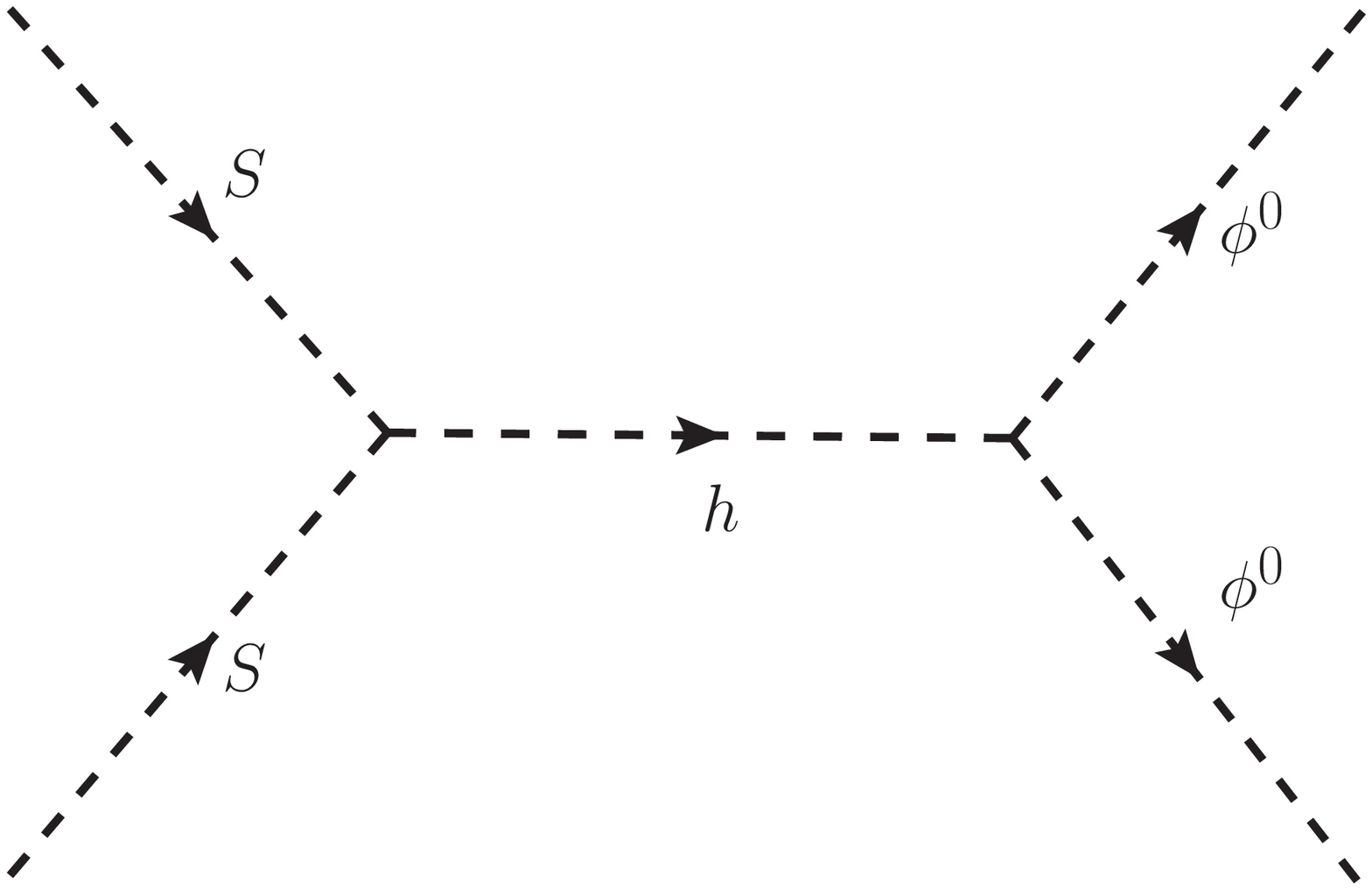}
\hspace{2.0 cm}
\includegraphics[width=4cm,height=3cm]{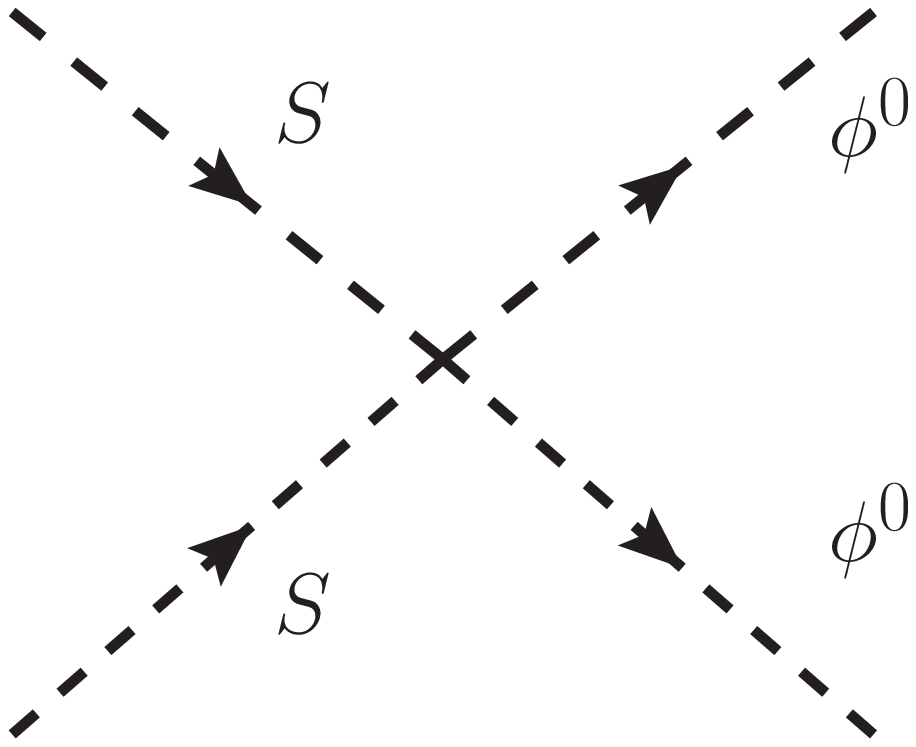}
\caption{Feynman diagrams for the annihilation channel $SS\rightarrow\phi^0\phi^0$}
\label {feyn_dia_phi0s}
\end{figure}

The other cross section involved in the coupled Boltzmann equations is 
$\langle{\sigma {\rm{v}}}_{SS\rightarrow\phi^0\phi^0}\rangle$ 
(due to the presence of the interaction between the two dark matter components 
$S$ and $\phi^0$). The expression for this cross section is given by 
\begin{eqnarray}
\langle{\sigma {\rm{v}}}_{SS\rightarrow\phi^0\phi^0}\rangle &=& \frac{1}{4\pi {m^2_S}}\left(1-
\frac{{m^2_{\phi^0}}}{{m^2_S}}\right)^{1/2}
\left[\frac{\left(g_{\phi^0\phi^0h}\,\,g_{SSh}\right)^2}
{\left[(4{m^2_{S}}-{m^2_h})^2+(\Gamma_h m_h)^2\right]} + {g^2_{\phi^0\phi^0SS}}\right.
\nonumber \\ &&\left. -\frac{\left(2\,\,g_{\phi^0\phi^0h}\,\,g_{SSh}\,\,g_{\phi^0\phi^0SS}\right)
(4{m^2_S}-{m^2_h})}{\left[(4{m^2_{S}}-{m^2_h})^2+(\Gamma_h m_h)^2\right]}\right]
\label{sigmaV2211}\,\, .
\end{eqnarray}
The heavier component of DM is chosen to be $S$ in our case\footnote{If $m_S < m_{\phi^0}$, $\phi^0$
component would be annihilated into $S$ (via the process $\phi^0\phi^0 \rightarrow SS$).
This would reduce the contribution of $\phi^0$ to the combined relic density and hence the
$\gamma-$ray flux originated from $\phi^0 \phi^0$ annihilation would be suppressed.}.
Since only $\phi^0$ will contribute to the 130 GeV gamma-ray production (the annihilation cross section  
$\langle{\sigma {\rm{v}}}_{SS\rightarrow\gamma\gamma}\rangle$ for the singlet $S$ is not enough to
produce Fermi-LAT observed 130 GeV $\gamma-$ray flux as we argued before), 
its mass should be $m_{\phi^0} \sim$ 130 GeV. We will show later that 
$\phi^0$ with its mass $\sim$ 130 GeV can not accommodate the total relic 
density consistent with the WMAP data. So the rest of the relic density 
should be provided by the other component $S$. And we find that it would 
be a good choice to make $m_S > m_{\phi^0}$. The involvement of this interaction
between two dark matter components is a salient feature of our analysis.
It would enhance the number density of $\phi^0$ 
through the annihilation of $SS$ during evolution. The Feynman diagrams for this
process ($SS\rightarrow\phi^0\phi^0$) are given in Fig. {\ref{feyn_dia_phi0s}}.  

Note that the relic density of the component $\phi^0$ also depends on
its co-annihilation with the charged scalars $\phi^\pm$ and CP odd neutral scalar
$A^0$. As the present scenario demands $m_{\phi^\pm}\ga m_{\phi^0}$ and $m_{A^0}>> m_{\phi^0}$
for the explanation of 130 GeV gamma-line and the exclusion of charged relic in the model simultaneously
(see Section \ref{calc} and \ref{gamma} for more discussion on this topic),
the effect of co-annihilation term between
$\phi^0$ and $\phi^\pm$ could have a significant contribution compared to the other
term between $\phi^0$ and $A^0$ \cite{gondolo, griest}. Therefore the Boltzmann's Equation (Eq. (\ref{eq1}))
for $\phi^0$
should have another term describing the co-annihilation of $\phi^0$ and $\phi^\pm$. This term can be
written as $\langle {\sigma {\rm v}}_{\phi^0\phi^\pm\rightarrow \chi \chi^\prime}\rangle
\left(n_{\phi^0}n_{\phi^\pm}-n^{eq}_{\phi^0}n^{eq}_{\phi^\pm}\right)$. Since both $\phi^\pm$
have electromagnetic charges, they exhibit electromagnetic interaction
beside weak interaction. Thus they are expected to be in thermal equilibrium with the thermal
plasma by interacting electromagnetically. Therefore the number density $n_{\phi^\pm}
= n^{eq}_{\phi^\pm}$. With this the above co-annihilation term is expressed as
\begin{eqnarray}
\langle {\sigma {\rm v}}_{\phi^0\phi^\pm\rightarrow \chi \chi^\prime}\rangle
\left(n_{\phi^0}n_{\phi^\pm}-n^{eq}_{\phi^0}n^{eq}_{\phi^\pm}\right) = 
\left[\langle {\sigma {\rm v}}_{\phi^0\phi^\pm\rightarrow \chi \chi^\prime}\rangle
\right]\,\,n^{eq}_{\phi^\pm}\left(n_{\phi^0}-n^{eq}_{\phi^0}\right) \,\, .
\label{co-anni} 
\end{eqnarray}
Since the number density of a non relativistic particle ($m_{\phi^{\pm}}>>T$) in thermal equilibrium is $\sim
T^{3/2}\exp \left(-\frac{m_{\phi^\pm}}{T}\right)$, the contribution of the $\phi^0 \phi^\pm$ co-annihilation term
(R.H.S. of Eq. \ref{co-anni}) is exponentially suppressed. Hence this term contributes very little
to the relic density of $\phi^0$ even if $\langle {\sigma {\rm v}}_{\phi^0\phi^\pm\rightarrow \chi
\chi^\prime}\rangle$ is considerably large ($\sim 10^{-8} {\rm GeV^{-2}}$).

\begin{figure}[h]
\centering
\includegraphics[width=8.3cm,height=8.5cm,angle=-90.0]{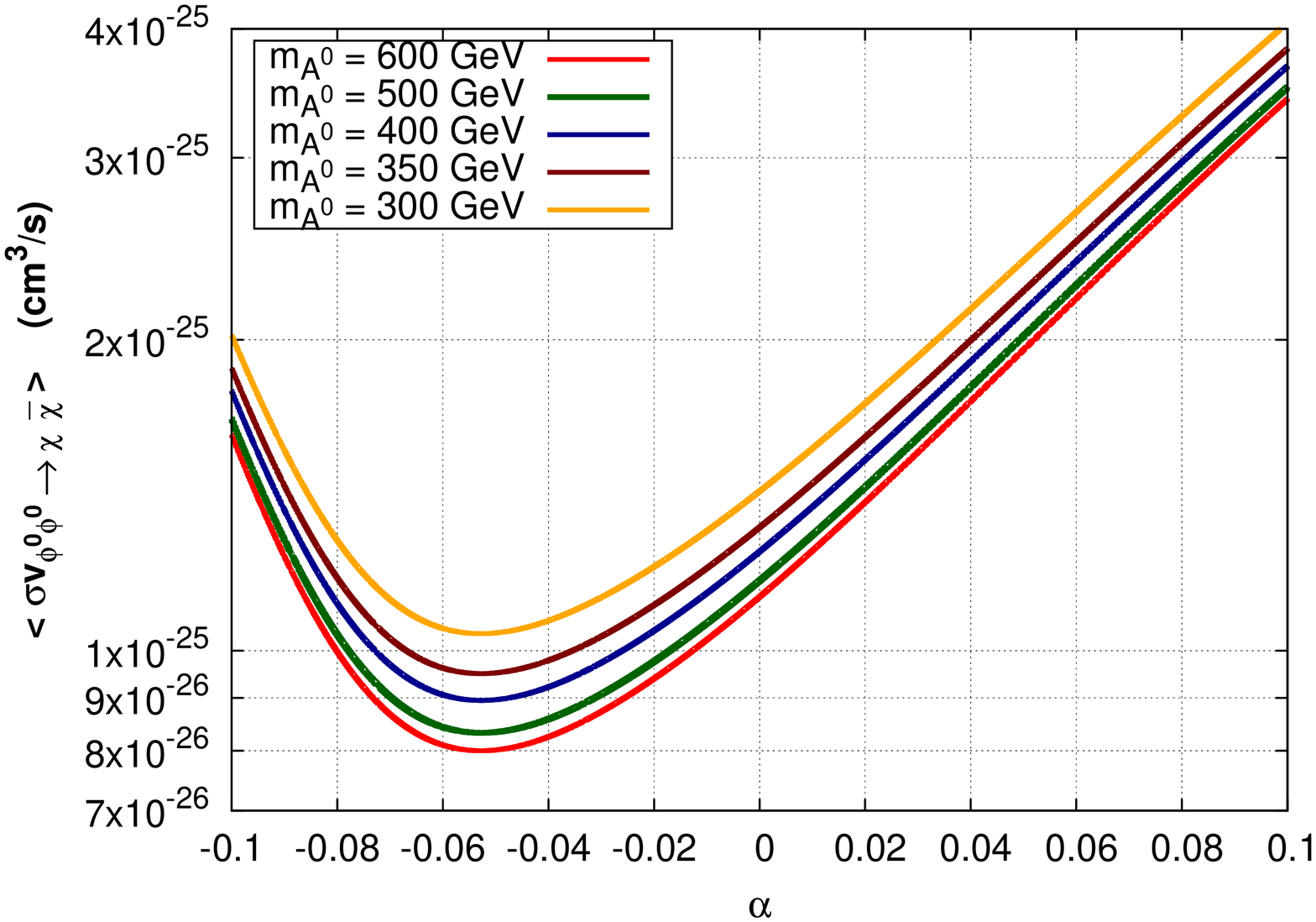}
\includegraphics[width=8.5cm,height=8.5cm,angle=-90.0]{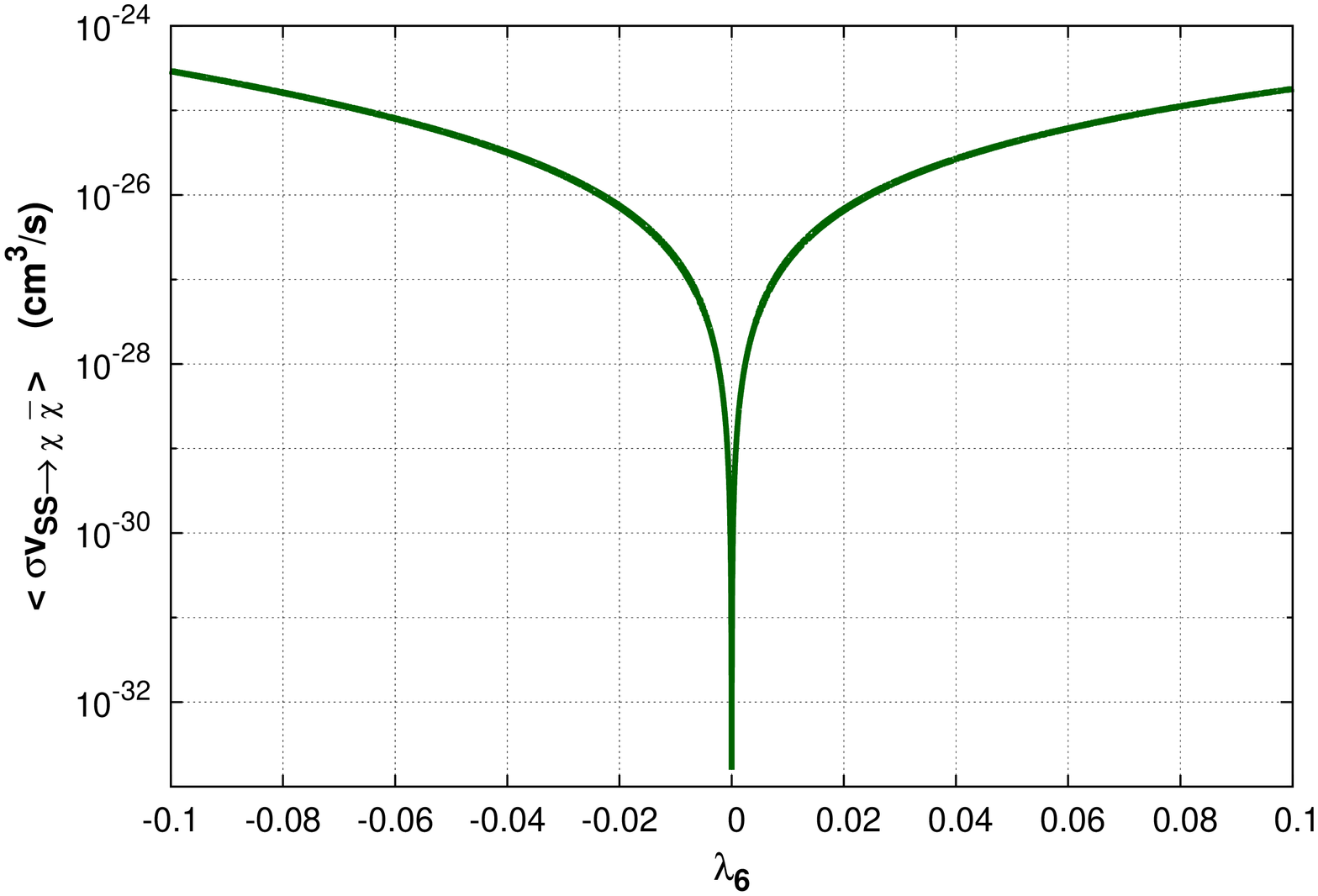}
\caption{Left panel - Variation of 
$\langle {\sigma {\rm{v}}}_{\phi^0\phi^0\rightarrow {\chi \bar{\chi}}}\rangle$
with $\alpha$ for different values of $m_{A^0}$, 
Right panel -  Variation of $ \langle {\sigma {\rm{v}}}_{SS\rightarrow {\chi \bar{\chi}}}
\rangle$ with $\lambda_6$ for $m_S = 130.5$ GeV.}
\label {sigmaV}
\end{figure}

In the left panel of Fig. \ref{sigmaV}, we have shown the variation of 
$\langle {\sigma {\rm{v}}}_{\phi^0\phi^0\rightarrow {\chi \bar{\chi}}} \rangle $ with
parameter $\alpha$ (Eq. (\ref{5})) for different values of $m_{A^0}$, namely $m_{A^0} = $ 600 GeV,
500 GeV, 400 GeV, 350 GeV, 300 GeV. In the right panel, the variation of 
$\langle{\sigma {\rm{v}}}_{SS\rightarrow {\chi \bar{\chi}}}\rangle$ with 
parameter $\lambda_6$ (Eq. (\ref{RSDM})) is shown. We have chosen 
$m_{\phi^0} = 130$ GeV, $m_{\phi^+}$ = 130.2 GeV and $m_S$ = 130.5 GeV 
for drawing these plots.
The reason behind this choice of parameters should be cleared in the next section.

\section{Calculation of Model Parameters}
\label{calc}
In this section we describe the  procedure adopted in this work in order to estimate
the values of the parameters involved in our proposed model. At the very outset, we
summarise few basic requirements. In the present two component dark matter model, only the 
component $\phi^0$ can account for the observed 130 GeV $\gamma$-line through the annihilation 
channel $\phi^0\phi^0\rightarrow\gamma\gamma$ details of which is discussed in the next 
section. This is due to the fact that the other component $S$ being a scalar singlet 
cannot produce sufficient annihilation cross section for the channel $SS\rightarrow 
\gamma\gamma$. So we conclude that the mass of the $\phi^0$ component needs to be $\sim$ 130 
GeV. We also find that a choice for the mass of the charged scalars, $\phi^{\pm}$ 
(involved in the loop of the $\phi^0 \phi^0 \rightarrow \gamma\gamma$ process,
see Fig. \ref{feyn_dia_gamma}) close to $m_{\phi^0}$
enhances the annihilation cross section $\langle \sigma {\rm{v}}_{\phi^0 \phi^0 \rightarrow 
\gamma \gamma} \rangle$ (see Eq.(\ref{sigmav_gamma})). Therefore, in order to maximise this contribution 
(required to achieve the cross section in the right ball park without taking a very high value 
for the relevant parameter in the model), 
we consider \footnote{The reason behind this consideration is explained
in Section \ref{gamma} where a complete study between the gamma-ray flux,
$\langle {\sigma \rm v}_{\phi^0 \phi^0 \rightarrow \gamma \gamma}\rangle$
and $m_{\phi^+}$ (see Fig. \ref{fluxplot}) is discussed in view of
Eq. (\ref{gammaflux}, \ref{sigmav_gamma}, \ref{sigmav_gamma1}).} $m_{\phi^{\pm}} =$ 130.2 GeV 
throughout the present discussion. Note that such a choice is consistent with the LEP 
bound \cite{LEP-charged}. 
With this consideration, the parameter $\alpha$ is reduced to $\sim \lambda_1/2$. 

Of course $\phi^0$ with $m_{\phi^0}$ = 130 GeV cannot individually account for 
the WMAP results on relic density. In our scenario, this deficit will be compensated by 
the contribution of the other component $S$. However we need to maximise the contribution 
of $\phi^0$ to the combined relic density so as to keep the flux of the 
observed gamma-ray (130 GeV) from the Galactic centre at an adequate level 
(see Eqs.(\ref{jint}, \ref{rho-prime})). For example if $\phi^0$ contributes to 60$\%$ 
compared to a case where it contributes only 30$\%$ to the total relic density then 
the flux for the gamma-ray originated from DM ($\phi^0$) annihilation will also be proportionately 
higher compared to the latter case. Apart from the WMAP data, we also use the 
limits obtained from the dark matter direct detection experiments. Needless to mention that 
in doing so the conditions obtained in Eqs. (\ref{cons1}-\ref{cons4}, \ref{cons5}, 
\ref{cons6}) are always satisfied.

\begin{figure}[h]
\centering
\includegraphics[width=8.5cm,height=8.5cm,angle=-90]{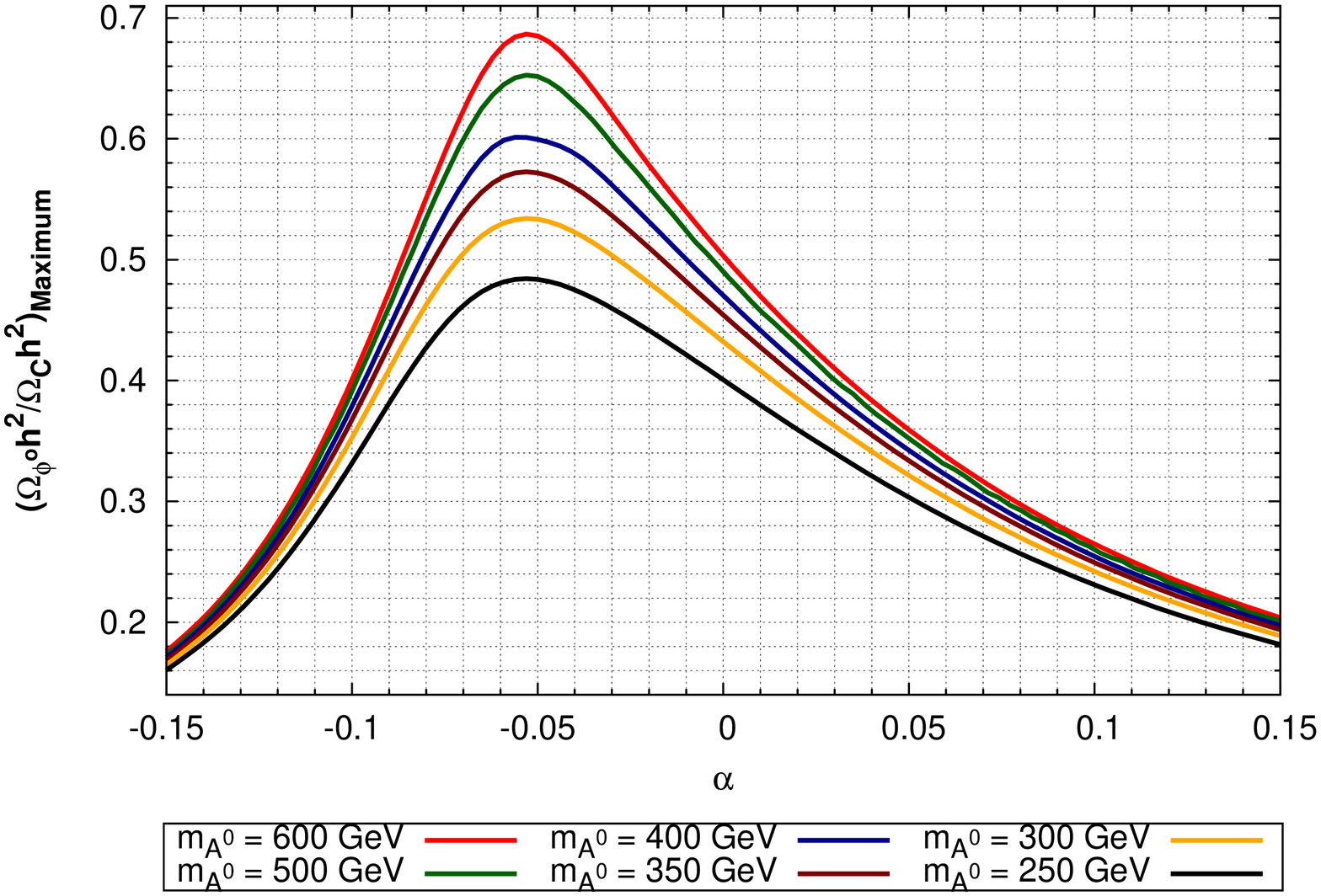}
\includegraphics[width=8.0cm,height=8.5cm,angle=-90]{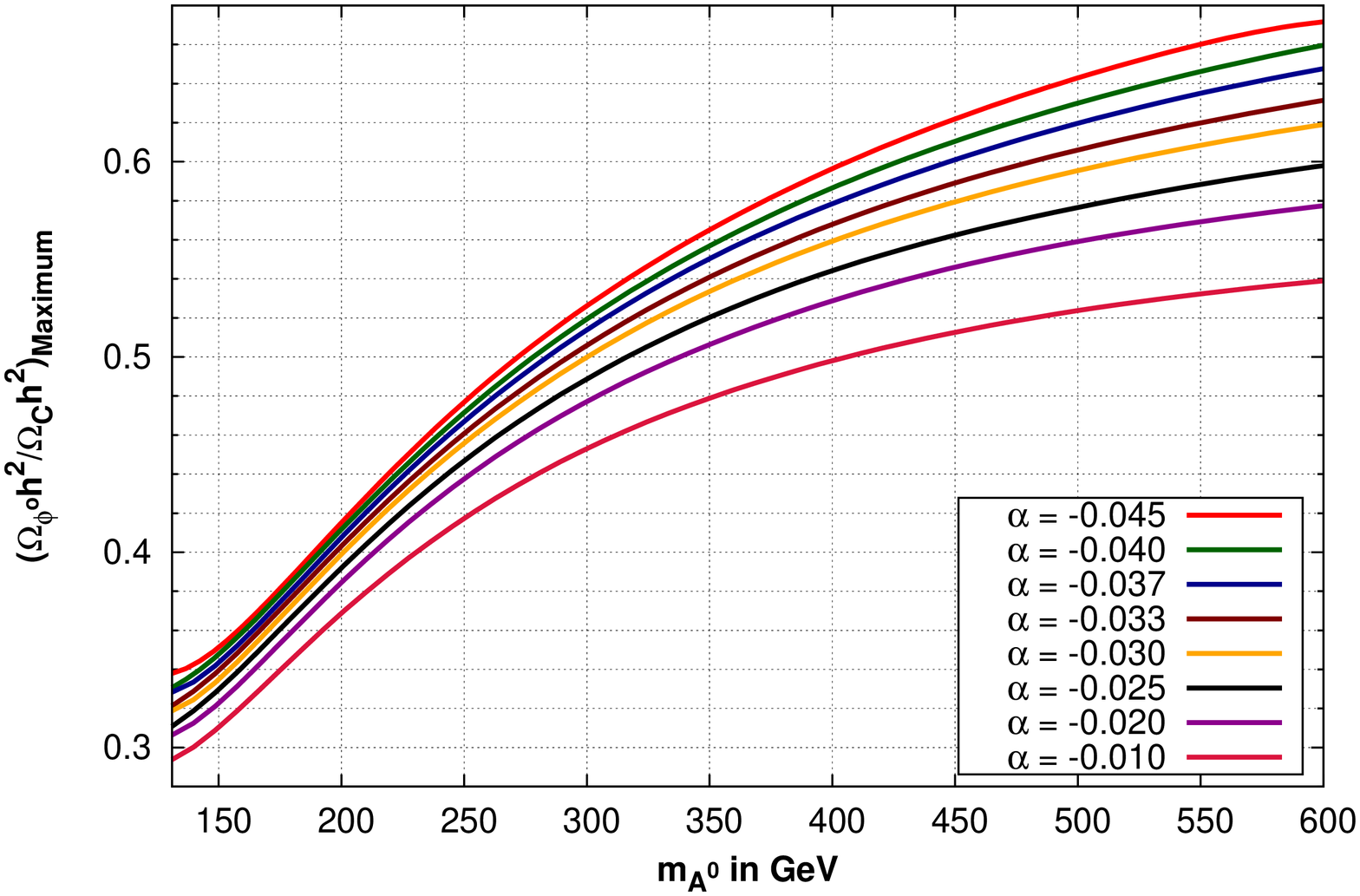}
\caption{Left panel - Variations of $\left(\frac{\Omega_{\phi^0}h^2}{\Omega_c h^2}\right)_{\rm Maximum}$
with parameter $\alpha$ for different values of $m_{A^0}$,
Right panel - Variations of $\left(\frac{\Omega_{\phi^0}h^2}{\Omega_c h^2}\right)_{\rm Maximum}$ with $m_{A^0}$ for different values of parameter $\alpha$.}
\label{result1}
\end{figure}

\begin{figure}[h]
\centering
\includegraphics[width=9cm,height=9cm,angle=-90]{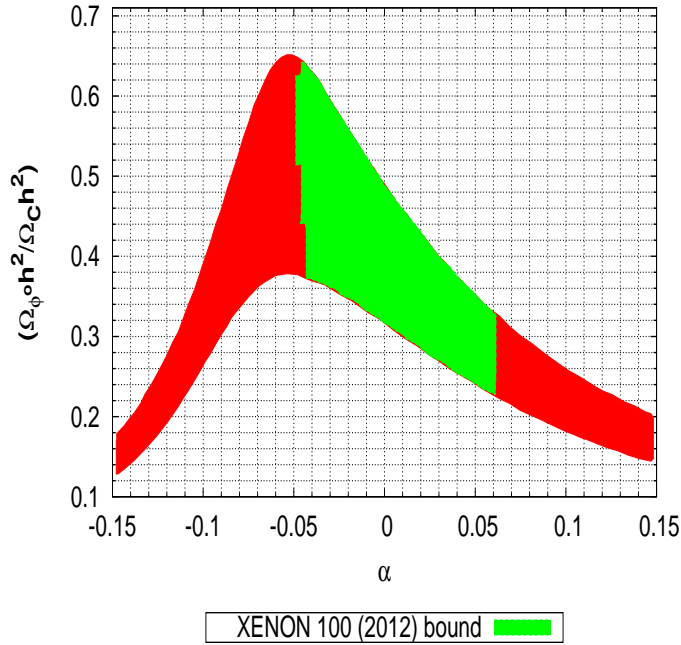}
\caption{Variations of $\left(\frac{\Omega_{\phi^0}h^2}{\Omega_c h^2}\right)$ with parameter $\alpha$ 
for a particular value of $m_{A^0} = 500$ GeV. The green patch corresponds to the allowed region when
XENON 100 (2012) data are considered; see text for details.}
\label{result1.1}
\end{figure}
In order to find a choice of parameter space for which the contribution
from $\phi^0$ towards relic density can be maximised, we calculate
the ratio $\left(\frac{\Omega_{\phi^0}h^2}{\Omega_c h^2}\right)$. The relic
densities $\Omega_{\phi^0}h^2$ and $\Omega_c h^2$ are computed using
Eqs. (\ref{eq3}, \ref{eq4}, \ref{omega_individual}, \ref{totalDM}).
In these computations we note that the inclusion of co-annihilation term contributes
only $\sim 0.2\%$ to the relic density of the dark matter component $\phi^0$.
Therefore we do not consider this co-annihilation term in Boltzmann's Equation
for the rest of our analysis. 
In the left panel
of Fig. \ref{result1} we plot $\left(\frac{\Omega_{\phi^0}h^2}{\Omega_c h^2}\right)_{\rm Maximum}$,
as a function of the parameter $\alpha$ for different values of $m_{A^0}$.
Here by $\left(\frac{\Omega_{\phi^0}h^2}{\Omega_c h^2}\right)_{\rm Maximum}$,
we mean that we have chosen only the largest possible value of the ratio
corresponding to a particular $\alpha$ while other parameters are scanned over their
entire range ($-0.1 \le\lambda_6\le 0.1$, $0.01\le\lambda_5\le 1.0$).
From the left panel of Fig.\ref{result1}, our aim is to find a suitable
value of $\alpha$ for which we can have a maximum contribution to the
ratio $\left(\frac{\Omega_{\phi^0}h^2}{\Omega_c h^2}\right)$. Note
that this ratio depends on the choice of $m_{A^0}$.
For more insight into this dependence, we show the
variations of $\left(\frac{\Omega_{\phi^0}h^2}{\Omega_c h^2}\right)_{\rm Maximum}$ with $m_{A^0}$
for different values of $\alpha$ in the right panel of Fig. {\ref{result1}}.
It is evident that the contribution of $\phi^0$ increases as $m_{A^0}$ increases.
This can also be noted from the same figure that for a fixed value of $\alpha$,
the increase of $\Omega_{\phi^0}h^2$ with $m_{A^0}$ is
steeper for lower values of $m_{A^0}$. Note that we cannot choose a value
of $m_{A^0}$ which is arbitrarily large
since our choice $m_{\phi^0} \sim m_{\phi^{\pm}} \sim $ 130 GeV imposes a relation between parameters
$\lambda_2$ and $\lambda_3$, thereby one parameter can be very large.
For example, to achieve $m_{A^0} = 500$ GeV,
we find $\lambda_2$ = 3.850, $\lambda_3 = -1.926$ when $\alpha =$ -0.037 is considered.
The value of $\alpha$ is so chosen that it satisfies the condition obtained from
XENON 100 (2012) \cite {xenon2012} direct detection experiment bound
(Eq. (\ref{xenon-number})). The allowed region for $\alpha$
satisfying this bound is shown in Fig. \ref{result1.1} (see below for discussions on Fig. \ref{result1.1}). 
A choice of $m_{A^0} =$ 900 GeV would require a value of  $\lambda_2 \sim 4\pi$, which
poses a threat to the perturbativity. So for the rest of our analysis, we
mostly consider $m_{A^0} = 500$ GeV unless otherwise mentioned.
With this particular choice of $\alpha = -0.037$, the contribution of $\phi^0$
to the total relic density to be at most 62 \%.

In Fig. {\ref{result1.1}}, the same ratio (as in Fig. {\ref{result1}}) is plotted against
$\alpha$ for a fixed value of $m_{A^0} = 500$ GeV. Note that here the
band corresponds to the variations of the parameters $\lambda_5$ and $\lambda_6$ respectively.
In the previous Figure (left panel of Fig. \ref{result1}),
only maximum value of the ratio $\left(\frac{\Omega_{\phi^0}h^2}{\Omega_c h^2}\right)$ was considered
and so it was a line. Now in this Figure once we consider the XENON 100 (2012) limit on parameters
$\alpha$ and $\lambda_6$ through Eq. ({\ref{xenon-number}}), the initial band
is restricted to only the intermediate green colour patch. We find that the
ratio would be maximised to $\sim$ 0.64 for the value of $\alpha \simeq -0.045$,
once we impose the XENON 100 (2012) data.

\begin{figure}[h]
\centering
\includegraphics[width=8cm,height=8.5cm,angle=-90]{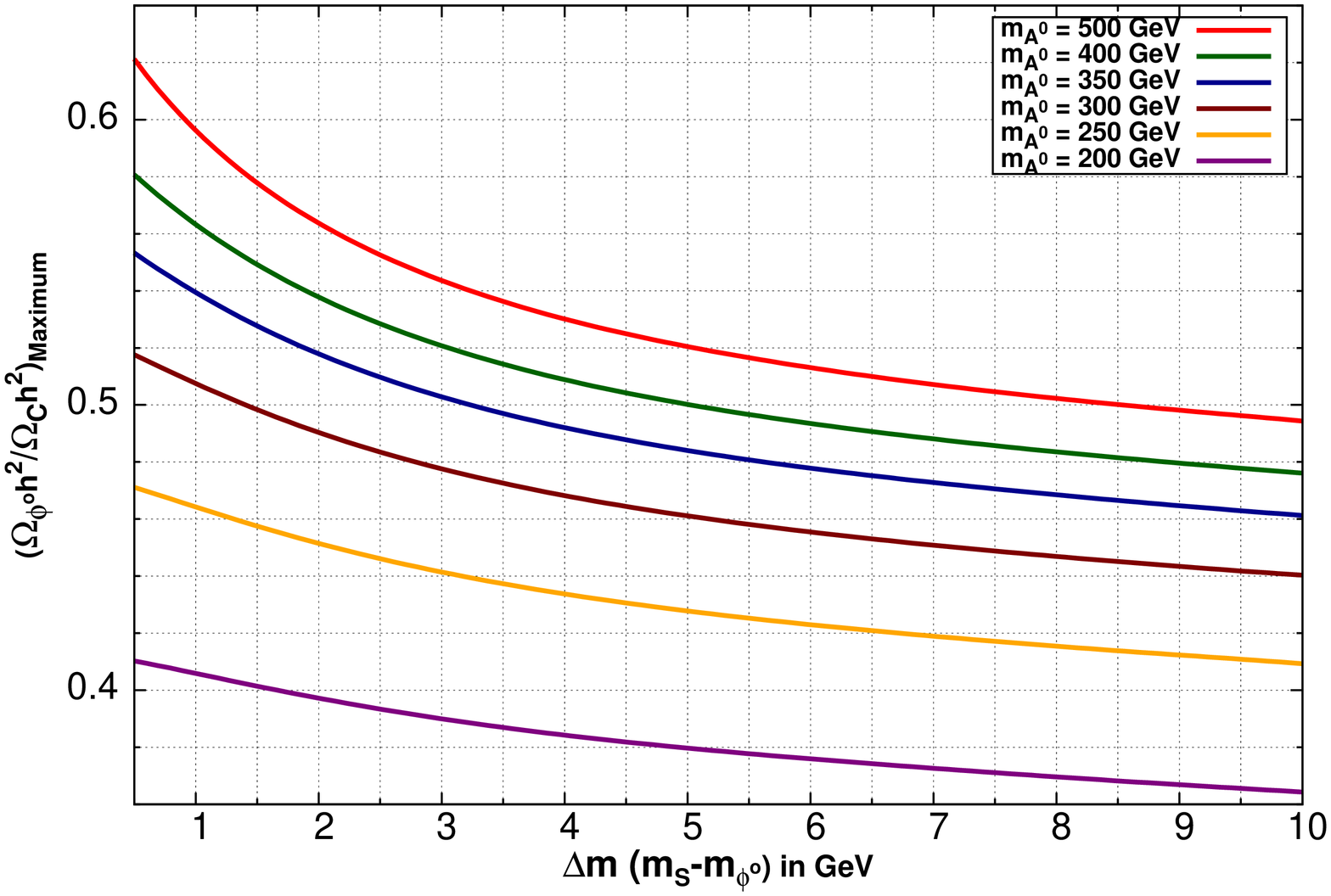}
\includegraphics[width=8cm,height=8.5cm,angle=-90]{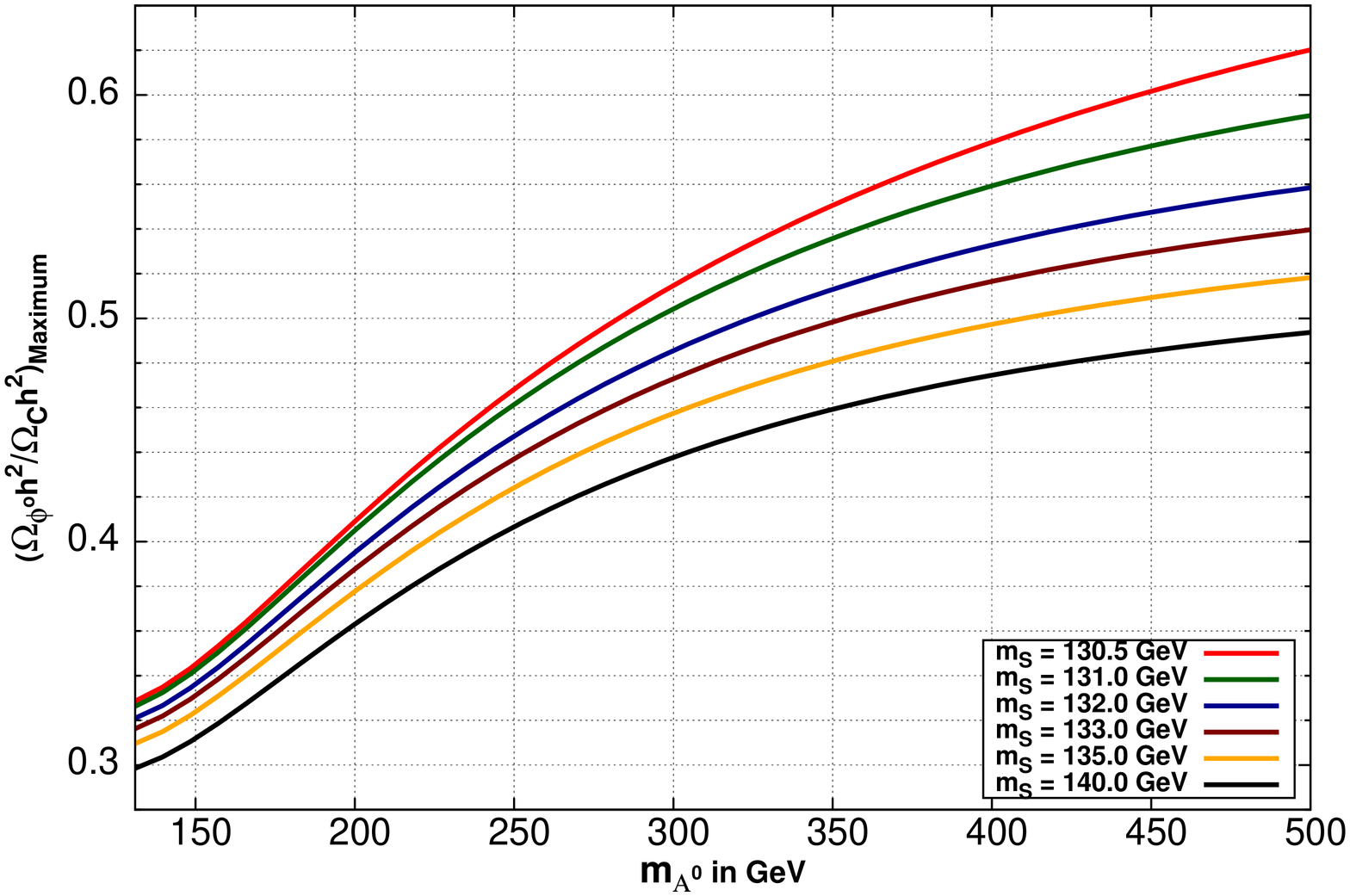}
\caption{Left panel - Variations of $\left(\frac{\Omega_{\phi^0}h^2}{\Omega_c h^2}\right)_{\rm Maximum}$
with $\Delta m (m_S-m_{\phi^0})$ for different values of $m_{A^0}$,
Right panel - Variations of $\left(\frac{\Omega_{\phi^0}h^2}{\Omega_c h^2}\right)_{\rm Maximum}$ 
with $m_{A^0}$ for different values of $\Delta m (m_S-m_{\phi^0})$.}
\label{result2}
\end{figure}

As it is now evident that the inert doublet component alone
with $m_{\phi^0} = 130$ GeV cannot account for the total dark matter
content of the universe and one needs to add the relic density of the singlet component $S$ for
producing the WMAP satisfied total relic density. In fact this is one of the motivations
for choosing this two component (inert doublet + scalar singlet) dark matter model. 
In order to choose a suitable mass for $S$ in the present scenario,
we define a quantity $\Delta m \,\,( = m_S-m_{\phi^0})$ remembering that 
$m_S$ should be heavier than $m_{\phi^0}$ as discussed before. We then study the 
variations of $\left(\frac{\Omega_{\phi^0}h^2}{\Omega_c h^2}\right)_{\rm Maximum}$
with $\Delta m$ for different values of $m_{A^0}$ which is shown in
the left panel of Fig. \ref{result2}. In the right panel of Fig. \ref{result2}
we show the variations of the fraction $\left(\frac{\Omega_{\phi^0}h^2}{\Omega_c h^2}\right)_{\rm Maximum}$ 
with $m_A^0$ ($\leq 500$ GeV) for different values of $m_S$. The plots in 
both the panels are obtained with $\alpha = -0.037$ (which satisfy latest XENON 100 data).
Comparing both the panels of Fig. \ref{result2} one
concludes that the contribution of $\phi^0$ to the combined relic density
increases as the mass splitting between the two components of the dark matter
decreases and for this particular value of $\alpha$ = -0.037,  
the ratio $\left(\frac{\Omega_{\phi^0}h^2}{\Omega_c h^2}\right)_{\rm Maximum}$  is 
$\sim 62\%$ when $\Delta m = 0.5$ GeV and $m_{A^0} = 500$ GeV respectively.  

\begin{figure}[h]
\centering 
\includegraphics[width=8cm,height=8.5cm,angle=-90]{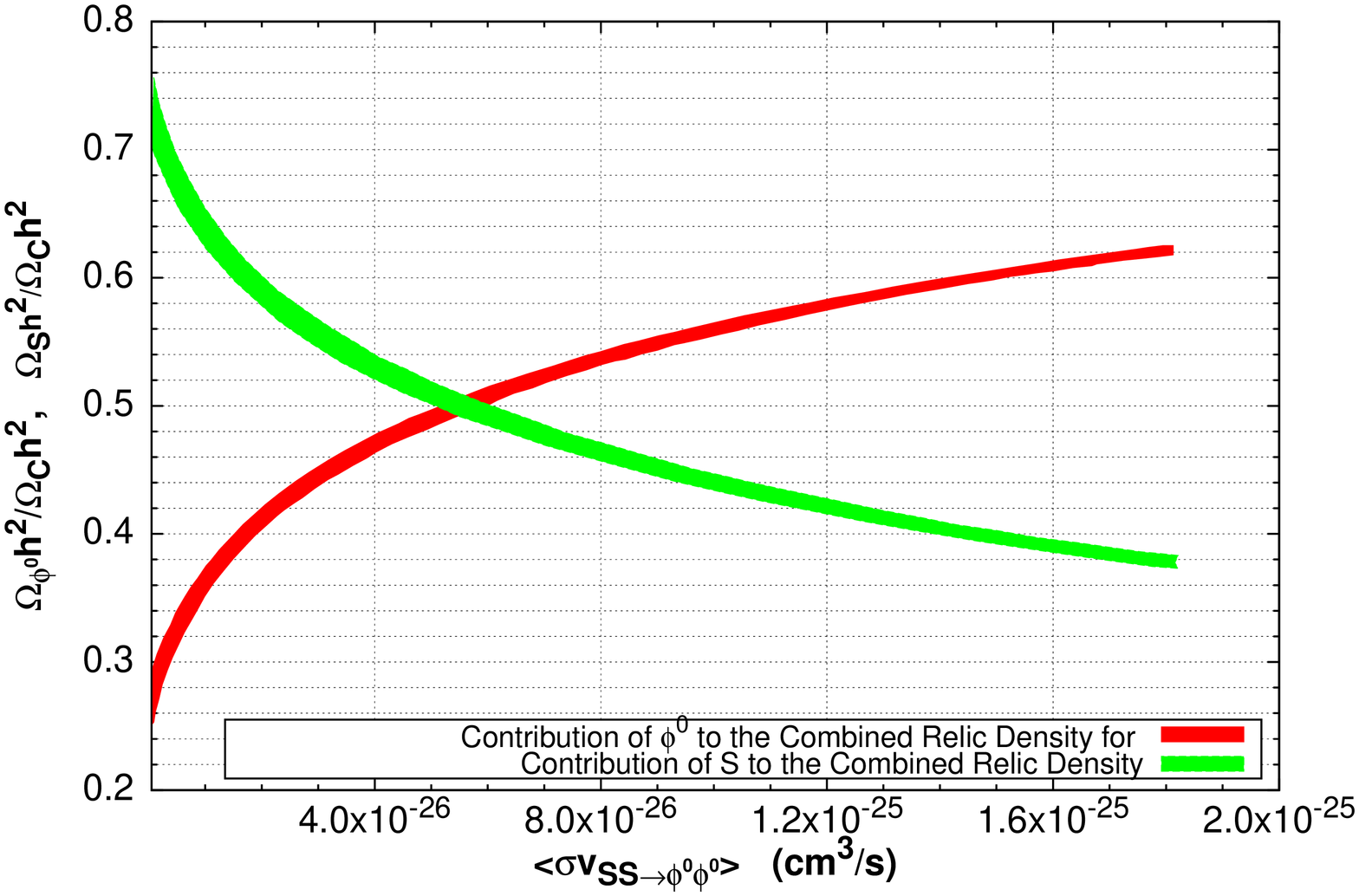}
\includegraphics[width=8cm,height=8.5cm,angle=-90]{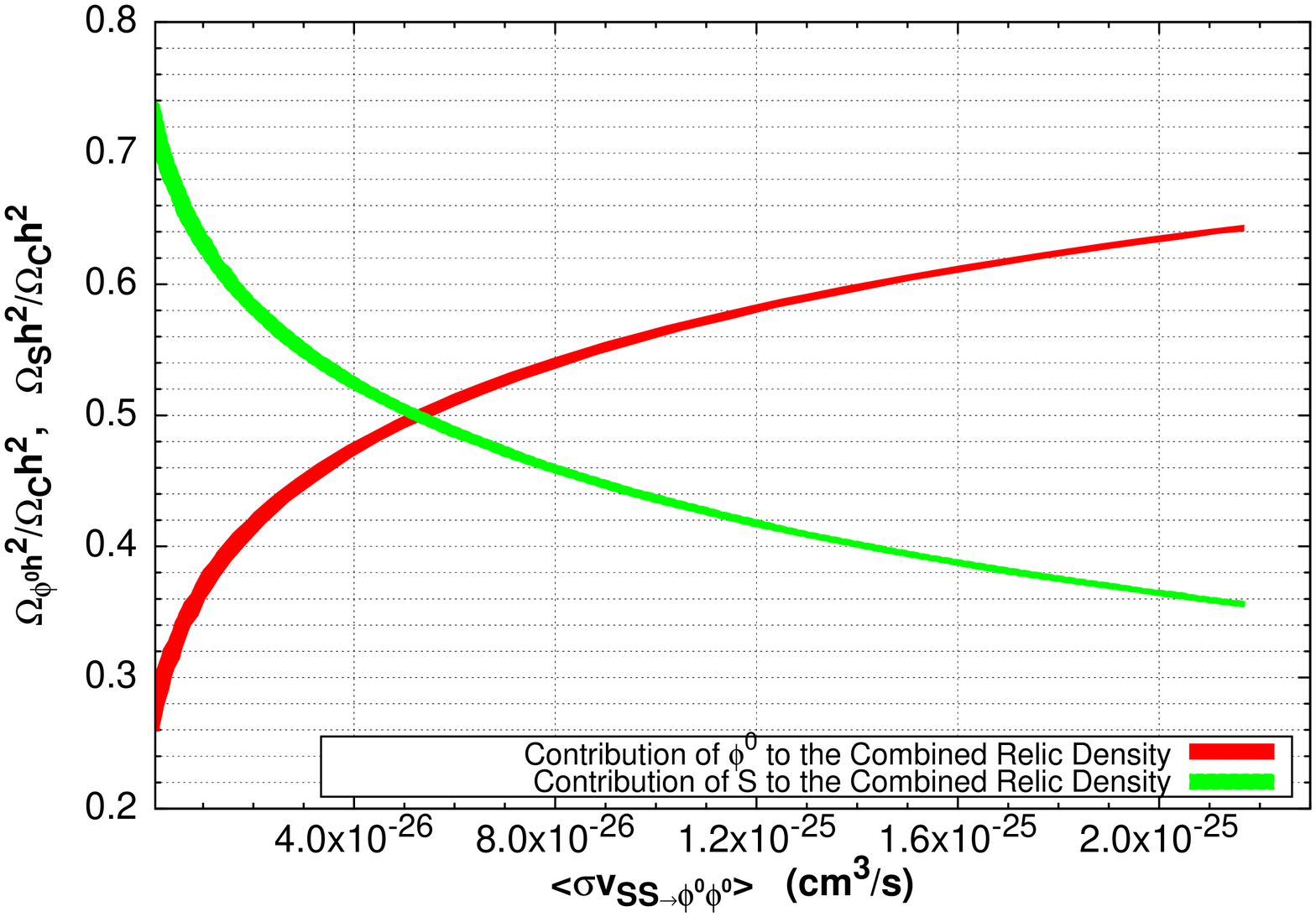}
\caption{Variations of the contributions of $S$ ($\Omega_{S}h^2$)
and $\phi^0$ ($\Omega_{\phi^0}h^2$) to the combined relic density ($\Omega_{c}h^2$)
with the annihilation cross section $\langle {\sigma {\rm{v}}}_{SS\rightarrow\phi^0\phi^0} \rangle$ for
two different values of $\alpha = -0.037$ (left panel) and -0.045 (right panel)}
\label{result3}
\end{figure}

\begin{figure}[h]
\centering 
\includegraphics[width=7.0cm,height=8.5cm,angle=-90]{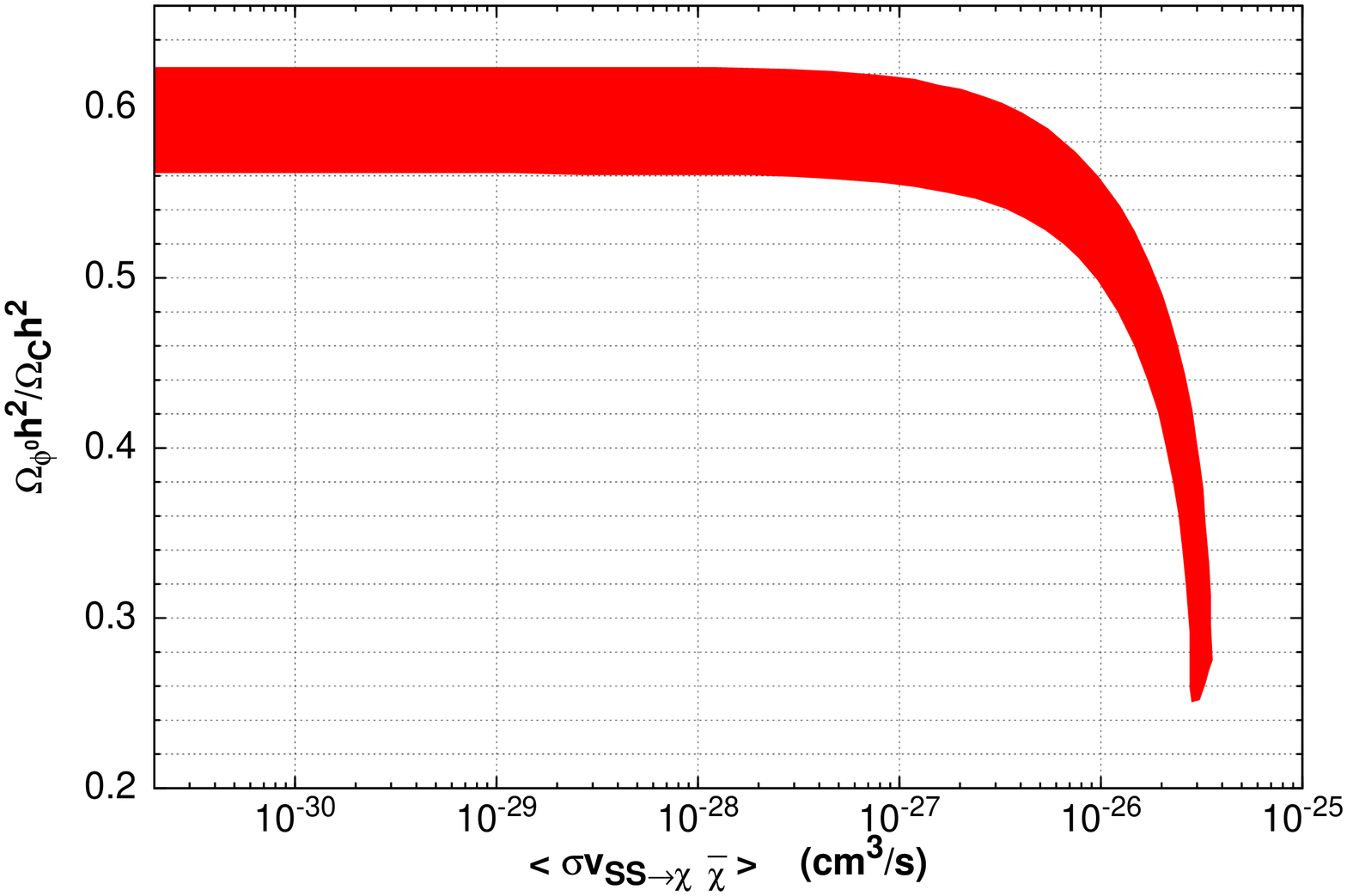}
\includegraphics[width=7.0cm,height=8.5cm,angle=-90]{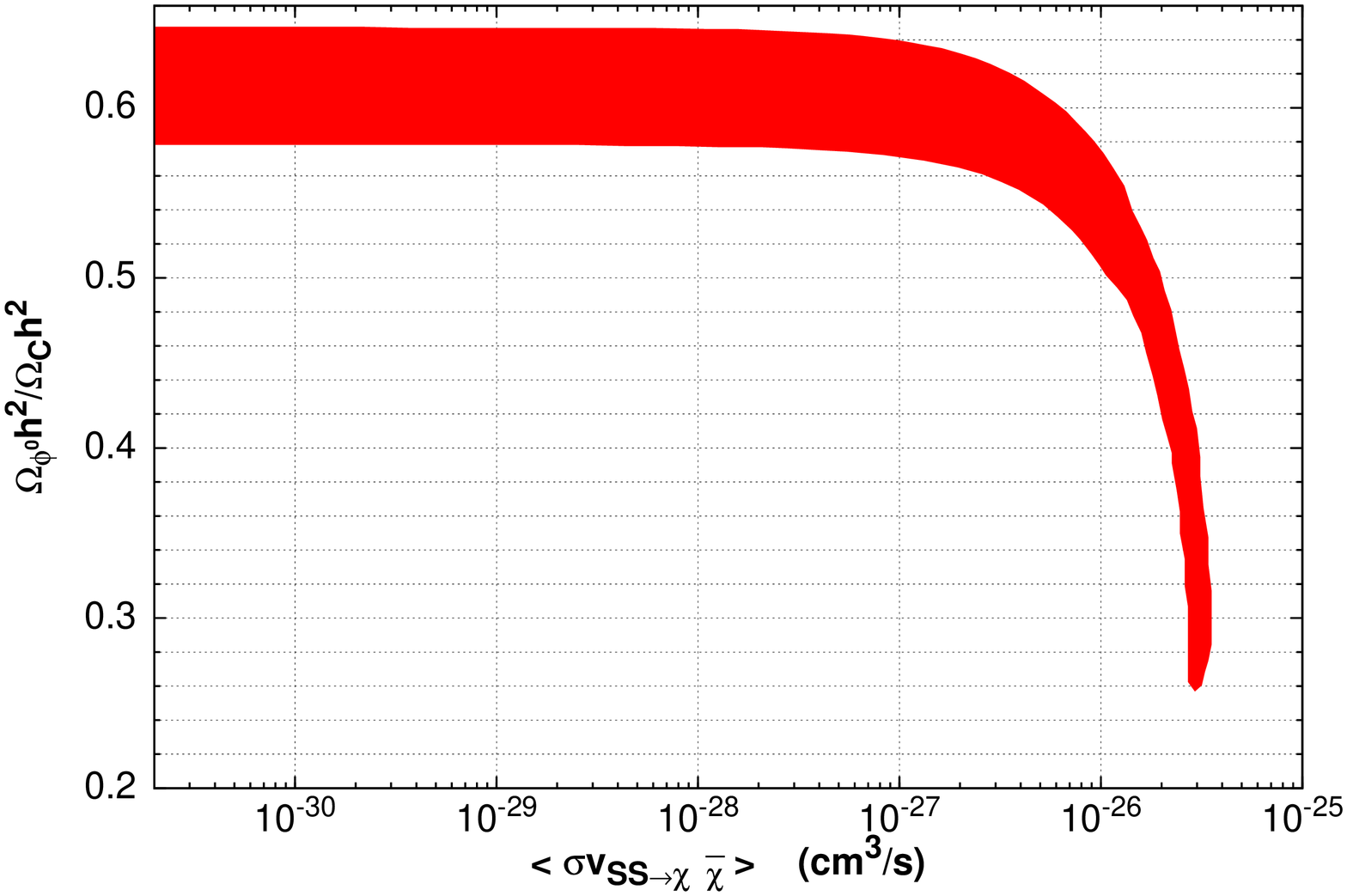}
\caption{Variations of the contributions of $\phi^0$ ($\Omega_{\phi^0}h^2$)
to the combined relic density ($\Omega_{c}h^2$) with the annihilation cross section 
$\langle {\sigma {\rm{v}}}_{SS\rightarrow\chi\bar{\chi}} \rangle$ for two different values of
$\alpha = -0.037$ (left panel) and -0.045 (right panel).}
\label{result3.1}
\end{figure}

We also calculate the variations of $\left(\frac{\Omega_{\phi^0}h^2}{\Omega_{c}h^2}\right)$ and
$\left(\frac{\Omega_{S}h^2}{\Omega_{c}h^2}\right)$ with the cross section 
$\langle{\sigma {\rm{v}}}_{SS \rightarrow \phi^0 \phi^0}\rangle$ and the results are plotted in the 
Fig. \ref{result3}. The plots in Fig. \ref{result3} are generated with
our standard set of parameter values such as  $m_{\phi^0} = 130$ GeV, $m_{\phi^+} = 130.2$
GeV, $m_S = 130.5$ GeV and $m_{A^0} = 500$ GeV. The left panel of Fig. \ref{result3}
is for the value of the parameter $\alpha = -0.037$ which lies within the
allowed range for $\alpha$ as shown in Fig. \ref{result1.1} earlier.
Similar plots in the right panel of Fig. \ref{result3} are given for comparison
for the chosen value of $\alpha = -0.045$ that corresponds to the largest allowed value
of $\left(\frac{\Omega_{\phi^0}h^2}{\Omega_{c}h^2}\right)$ in the present framework for
XENON 100 (2012) bound (Fig. \ref{result1.1}).     
From these plots, it is clear that initially when $\langle{\sigma {\rm{v}}}_{SS \rightarrow \phi^0\phi^0}\rangle$ 
is very small (or nearly zero) the contribution of $\phi^0$ is only $\sim 25 \%$ of the combined relic density. 
This is because for small values of $\langle{\sigma {\rm{v}}}_{SS \rightarrow \phi^0\phi^0}\rangle$,
Eqs. (\ref{eq3}, \ref{eq4}) effectively become two decoupled equations that represent the Boltzmann's
equations for RSDM and IDM respectively. Under such circumstance, the calculations of the individual
contributions for $S$ and $\phi^0$ are pursued .  

Note that in this decoupled scenario, individual relic density contribution would be 
inversely proportional to the corresponding annihilation cross section. Therefore we 
need to have an estimate for $\langle \sigma {\rm{v}}_{\phi^0 \phi^0 \rightarrow \chi {\bar{\chi}}} \rangle $ 
and $\langle \sigma {\rm{v}}_{S S \rightarrow \chi {\bar{\chi}}} \rangle$. 
In order to understand this in more detail, we refer to Fig. \ref{sigmaV}.
From the left panel of Fig. \ref{sigmaV}, 
we see that the value of $\langle{\sigma {\rm{v}}}_{\phi^0\phi^0 \rightarrow 
{\chi \bar{\chi}}}\rangle$ is $\sim 8.5\times 10^{-26}$ cm$^{3}$/s for $m_{A^0}
= 500$ GeV, $\alpha = -0.037$, $m_{\phi^0} = 130$ GeV, $m_{\phi^+} = 130.2$ GeV.  
This value of  $\langle{\sigma {\rm{v}}}_{\phi^0\phi^0 \rightarrow {\chi \bar{\chi}}}\rangle$
is nearly 4 times larger than what is required to get WMAP satisfied relic density for dark matter mass of
130 GeV, thereby contributing only $\sim$ 25$\%$ to the total relic density (Eq. (\ref{wmap})) as seen from
left panel of Fig. \ref{result3}. So we would expect that the rest $\sim$ 75$\%$ contribution should come from 
the singlet scalar component $S$ in the present model. This is indeed possible if we consider the right
panel of Fig. \ref{sigmaV}. This corresponds to a case where practically there is no interactions
between $S$ and $\phi^0$, i.e. 
$\lambda_5 \simeq 0$. As the interaction between $S$ and $\phi^0$ becomes increasingly stronger (i.e. 
$|\lambda_5|$ starts to have nonzero value), more and more $S$ particles annihilate to produce 
$\phi^0$ particles and contribution of $\phi^0$ to the combined relic density will 
be boosted. It reaches a maximum which comes out to be $\sim 62\%$ ($\sim 64\%$) in our present 
analysis for $\alpha = -0.037$ ($\alpha = -0.045$)
for $\langle{\sigma {\rm{v}}}_{SS \rightarrow \phi^0\phi^0}\rangle$ 
$\sim 1.8\times 10^{-25} {{\rm cm}^{3}}/{\rm s}$ ($\sim 2.2\times10^{-25}{{\rm cm}^{3}}/{\rm s}$)
seen from the left panel (right panel) of Fig. \ref{result3}.

\begin{figure}[h]
\centering
\includegraphics[width=5cm,height=8cm,angle=-90]{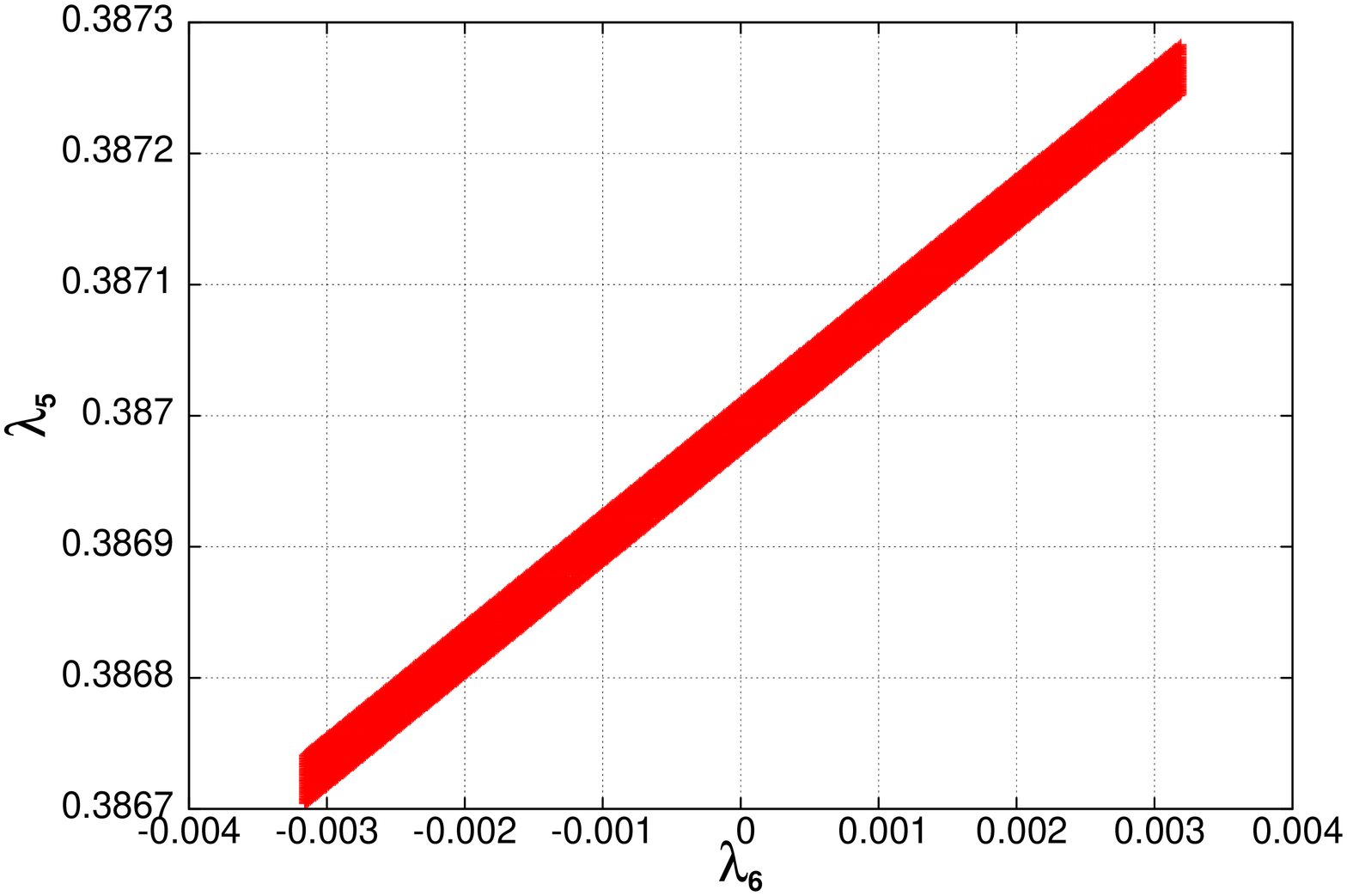}
\includegraphics[width=5cm,height=8cm,angle=-90]{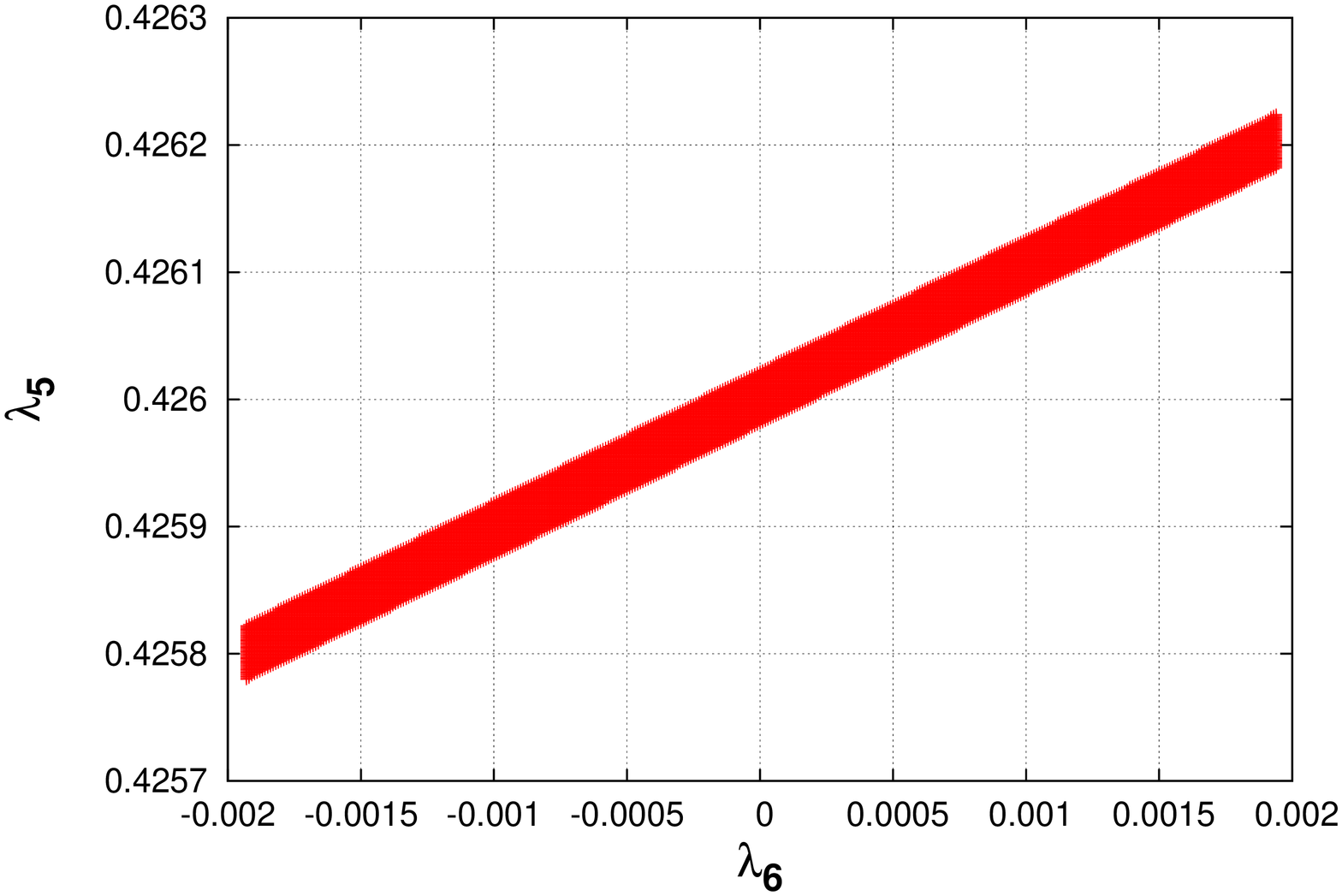}
\caption{Left panel - Allowed range of the parameters $\lambda_5, \lambda_6$ for the parameter $\alpha = -0.037$.
Right panel - Allowed range of the parameters $\lambda_5, \lambda_6$ for the parameter $\alpha = -0.045$.
Both plots are drawn for $m_{\phi^0} = 130$ GeV, $m_{\phi^+} = 130.2$ GeV, $\Delta m = 0.5$ GeV,
$m_{A^0} = 500$ GeV.}
\label{lam5-lam6}
\end{figure}

Fig. \ref{result3.1} shows the variation of  
$\left(\frac{\Omega_{\phi^0}h^2}{\Omega_{c}h^2}\right)$ with
$\langle{\sigma {\rm{v}}}_{SS \rightarrow {\chi \bar{\chi}}}\rangle$.
Similar set of parameters as in Fig. \ref{result3} are adopted in generating
the two plots of Fig. \ref{result3.1}. Likewise in Fig. \ref{result3} the right panel
of Fig. \ref{result3.1} corresponds to $\alpha = -0.045$ and is given for comparison
with the left panel ($\alpha = -0.037$, adopted value in this calculation) of this figure. 
The plots indicate that the contribution of $\phi^0$ to the combined relic density decreases as the value 
of $\langle{\sigma {\rm{v}}}_{SS \rightarrow {\chi \bar{\chi}}}\rangle$ 
increases. This is because of the increment of $\langle{\sigma 
{\rm{v}}}_{SS \rightarrow {\chi \bar{\chi}}}\rangle$ 
signifies large number $S$ particles annihilate into SM particles and 
consequently less number of particles are available for annihilation of $S$ to  
produce $\phi^0$ in the final state. As a result the relic density of 
$\phi^0$ decreases. From the left panel of Fig. \ref{result3.1}, we conclude that in order to 
obtain the relic density contribution of $\phi^0$ close to 62$\%$ or 
above the value of $\langle{\sigma {\rm{v}}}_{SS \rightarrow {\chi \bar{\chi}}}\rangle$ 
should be $\la 1.74\times 10^{-28} {{\rm cm}^{3}}/{\rm s}$. Note that the parameter
$\lambda_6$ is involved both in $\langle{\sigma {\rm{v}}}_{SS \rightarrow {\chi \bar{\chi}}}\rangle$
and $\langle{\sigma {\rm{v}}}_{SS \rightarrow \phi^0\phi^0}\rangle$, whereas 
the parameter $\lambda_5$ appeared only in  
$\langle{\sigma {\rm{v}}}_{SS \rightarrow \phi^0\phi^0}\rangle$. Therefore 
this limit (on $\langle{\sigma {\rm{v}}}_{SS \rightarrow {\chi \bar{\chi}}}\rangle$) 
along with the other one we just discussed above, $\langle{\sigma {\rm{v}}}_{SS \rightarrow \phi^0\phi^0}\rangle$
$\sim 1.8\times 10^{-25} {{\rm cm}^{3}}/{\rm s}$, set up a range 
of allowed region of $\lambda_5$ and $\lambda_6$ if we restrict ourselves
with $\left(\frac{\Omega_{\phi^0}h^2}{\Omega_{c}h^2}\right) \sim$ 62$\%$. 
The allowed range of values of the parameters
$\lambda_5,\,\lambda_6$ for the present case ($m_{A^0} = 500$ GeV)
is shown in the left panel of Fig. \ref{lam5-lam6}.
In Table \ref{tab3} we furnish the values of the parameters
$\lambda_5$ and $\lambda_6$ along with other model parameters.

Similarly, from the right panel of Fig. \ref{result3.1}
we conclude that in order to obtain
$\left(\frac{\Omega_{\phi^0}h^2}{\Omega_{c}h^2}\right) \sim 64\%$, one should have 
$\langle{\sigma {\rm{v}}}_{SS \rightarrow {\chi \bar{\chi}}}\rangle \la 0.65\times10^{-28} {\rm cm}^3/{\rm s}$.
The corresponding allowed region in $\lambda_5$-$\lambda_6$ parameter space is shown in the right panel of
Fig. \ref{lam5-lam6}. The values of other parameters for this present scenario ($\alpha = -0.045$)
which produces the largest possible contribution of $\phi^0$ towards the total relic density are also
tabulated in Table \ref{tab3.1}.

\begin{table}[h]
\begin{center}
\vskip 0.5 cm
\begin{tabular} {|c|c|c|c|c|c|c|c|}
\hline
{Mass of} & {Contribution of}&$\mu_2$ &$\lambda_1$&$\lambda_5$&$
|\lambda_6|$&$\lambda_2$&$\lambda_3$\\
{$A^0$ $(m_A^0)$}& $\phi^0$ in the combined &(GeV)&&&&&\\
(GeV) & Relic Density &&&&&&  \\  
\hline
300 &51.7$\%$ &138.344&-0.072&0.2698\,-\,0.2703&$\leq 2.6\times{10^{-3}}$&1.206&-0.604\\
\hline
350 &55.3$\%$ &138.344&-0.072&0.3014\,-\,0.3020&$\leq 2.8\times{10^{-3}}$&1.743&-0.872\\
\hline
400 & $58.1\%$ &138.344&-0.072& 0.3322\,-\,0.3327&$\leq 2.8\times{10^{-3}}$&2.363&-1.182\\ 
\hline
500 & $62.1\%$ &138.344&-0.072& 0.3867\,-\,0.3873&$\leq3.2\times{10^{-3}}$&3.850&-1.926\\
\hline
\end{tabular}
\end{center}
\caption{Relic density contribution of $\phi^0$ for different values of $m_A^0$ with the corresponding
values of other model parameters are $\alpha = -0.037$, $\Delta m$ = 0.5 GeV ($m_S$ = 130.5
GeV, $m_{\phi^0}$ = 130.0 GeV).}
\label{tab3}
\end{table}

\begin{table}[h]
\begin{center}
\vskip 0.5 cm
\begin{tabular} {|c|c|c|c|c|c|c|c|}
\hline
{Mass of} & {Contribution of}&$\mu_2$ &$\lambda_1$&$\lambda_5$&$
|\lambda_6|$&$\lambda_2$&$\lambda_3$\\
{$A^0$ $(m_A^0)$}& $\phi^0$ in the combined &(GeV)&&&&&\\
(GeV) & Relic Density &&&&&&  \\  
\hline
500 & $64.4\%$ &140.083&-0.088& 0.4258\,-\,0.4262&$\leq1.9\times{10^{-3}}$&3.850&-1.926\\
\hline
\end{tabular}
\end{center}
\caption{Maximum Relic density contribution of $\phi^0$ for $m_A^0 = 500$ GeV, $\alpha = -0.045$,
$\Delta m$ = 0.5 GeV ($m_S$ = 130.5 GeV, $m_{\phi^0}$ = 130.0 GeV).}
\label{tab3.1}
\end{table}
\section{130 GeV $\gamma$-ray line from Dark Matter annihilation}
\label{gamma}
In this section our endeavour will be to explain the recently observed
130 GeV $\gamma$-line from the Galactic centre originated from
dark matter annihilation in the framework of the present two component 
dark matter model. In order to produce such a 130 GeV $\gamma$-line, 
the required DM annihilation cross section into two photons should be 
$\sim 10^{-27}$ cm$^3/$s as predicted from the analysis \cite{data-analysis, Tempel:2012ey} of 
Fermi-LAT data \cite{fermi-data}. In the context of the present two component 
dark matter model, only the $\phi^0$ component having mass 130 GeV 
can contribute to the production of 130 GeV $\gamma-$ray line. The annihilation 
of $\phi^0\phi^0$ into $\gamma\gamma$ can take place only via charged scalar
$\phi^\pm$ and $W^{\pm}$ loops. It can indeed produce the required cross section
$\sim 10^{-27}$ cm$^3/$s. The cross sections for other annihilation channels that
can produce $\gamma\gamma$ (e.g. via Higgs) for both the components $\phi^0$ and $S$
are orders of magnitude less than this value \cite{Biswas}.

\begin{figure}[h]
\centering
\includegraphics[width=5cm,height=3cm]{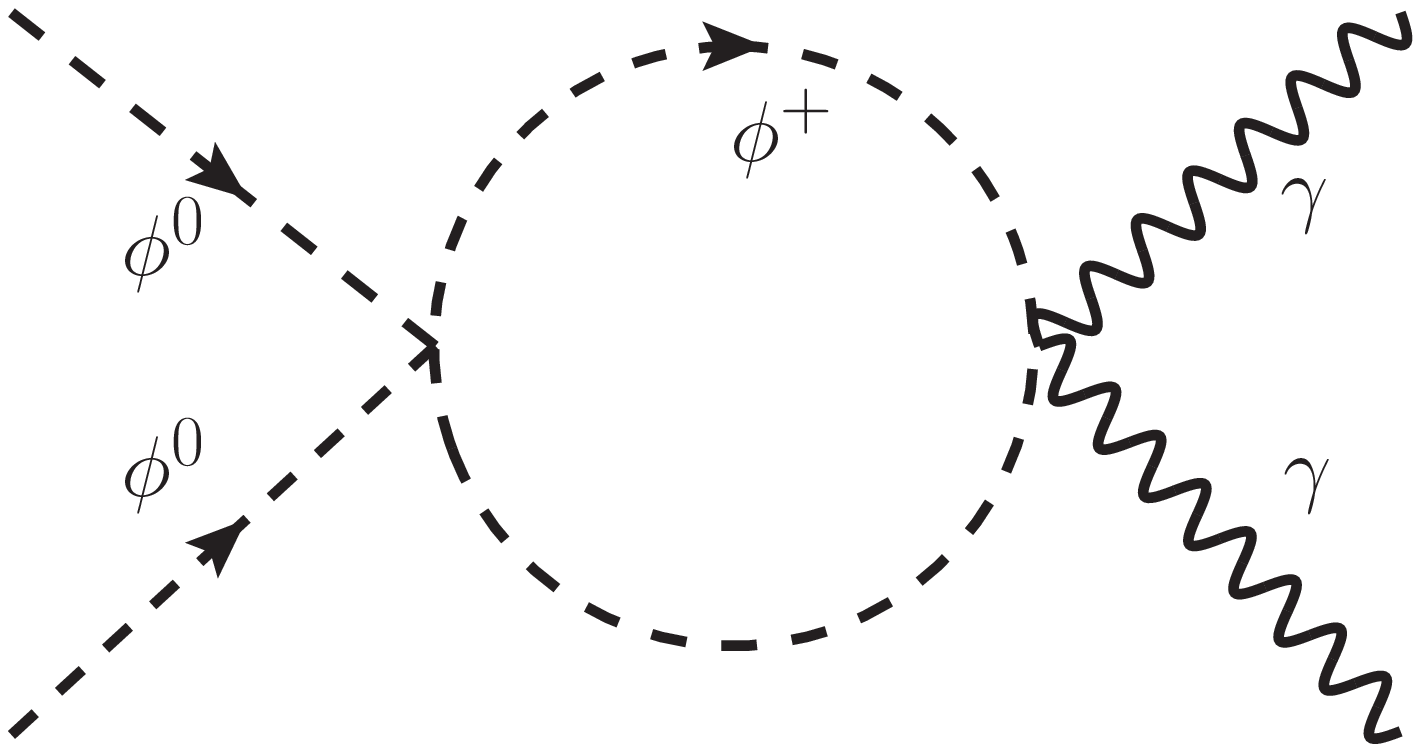}
\hspace{0.75 cm}
\includegraphics[width=5cm,height=3cm]{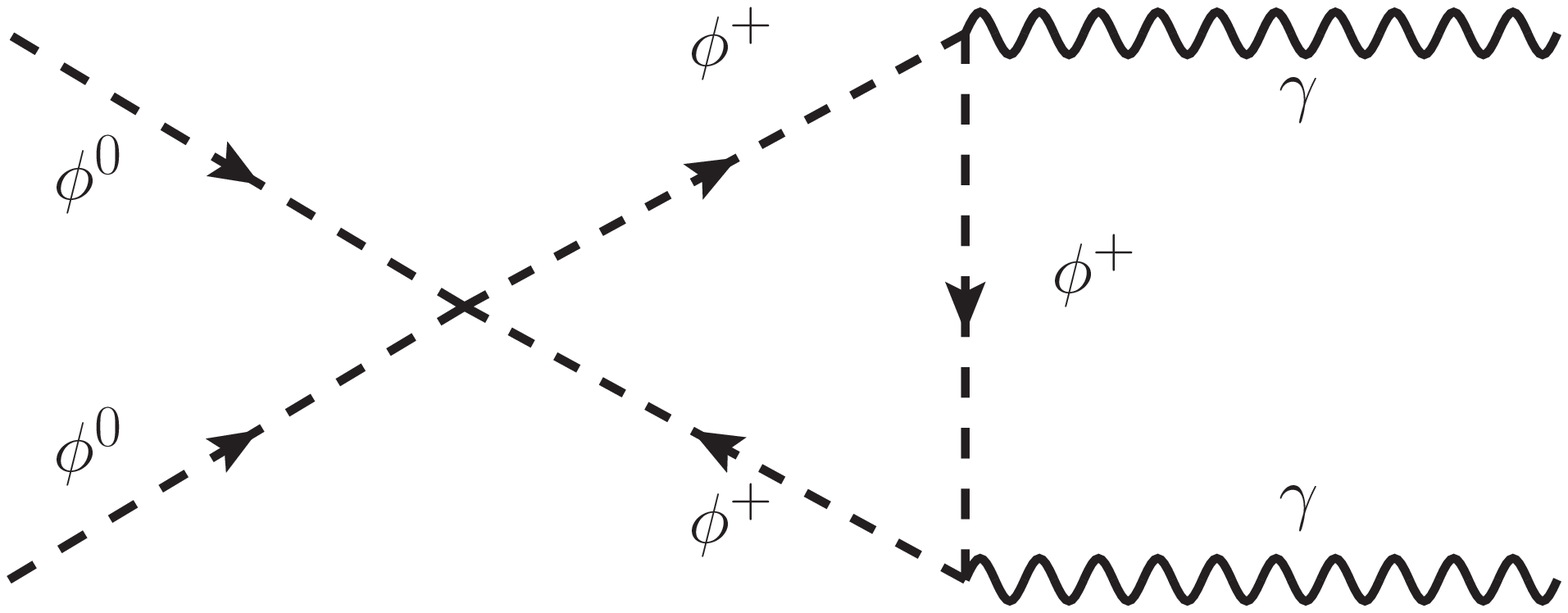}
\hspace{0.75 cm}
\includegraphics[width=5cm,height=3cm]{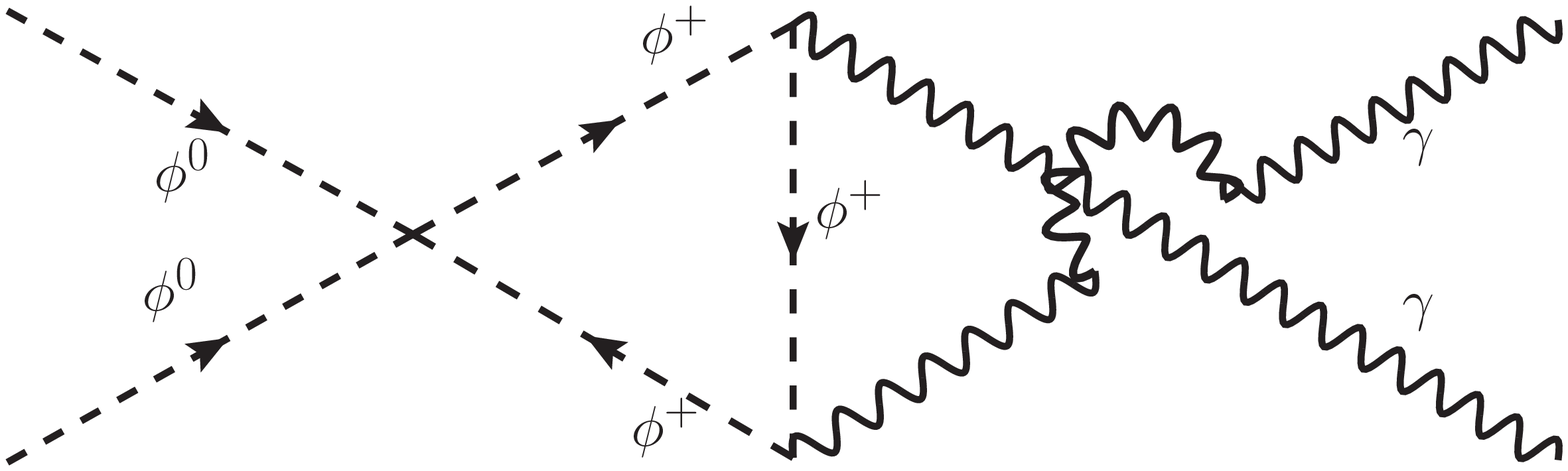} \\
\,\,\ \\
\,\,\ \\
\includegraphics[width=5cm,height=3cm]{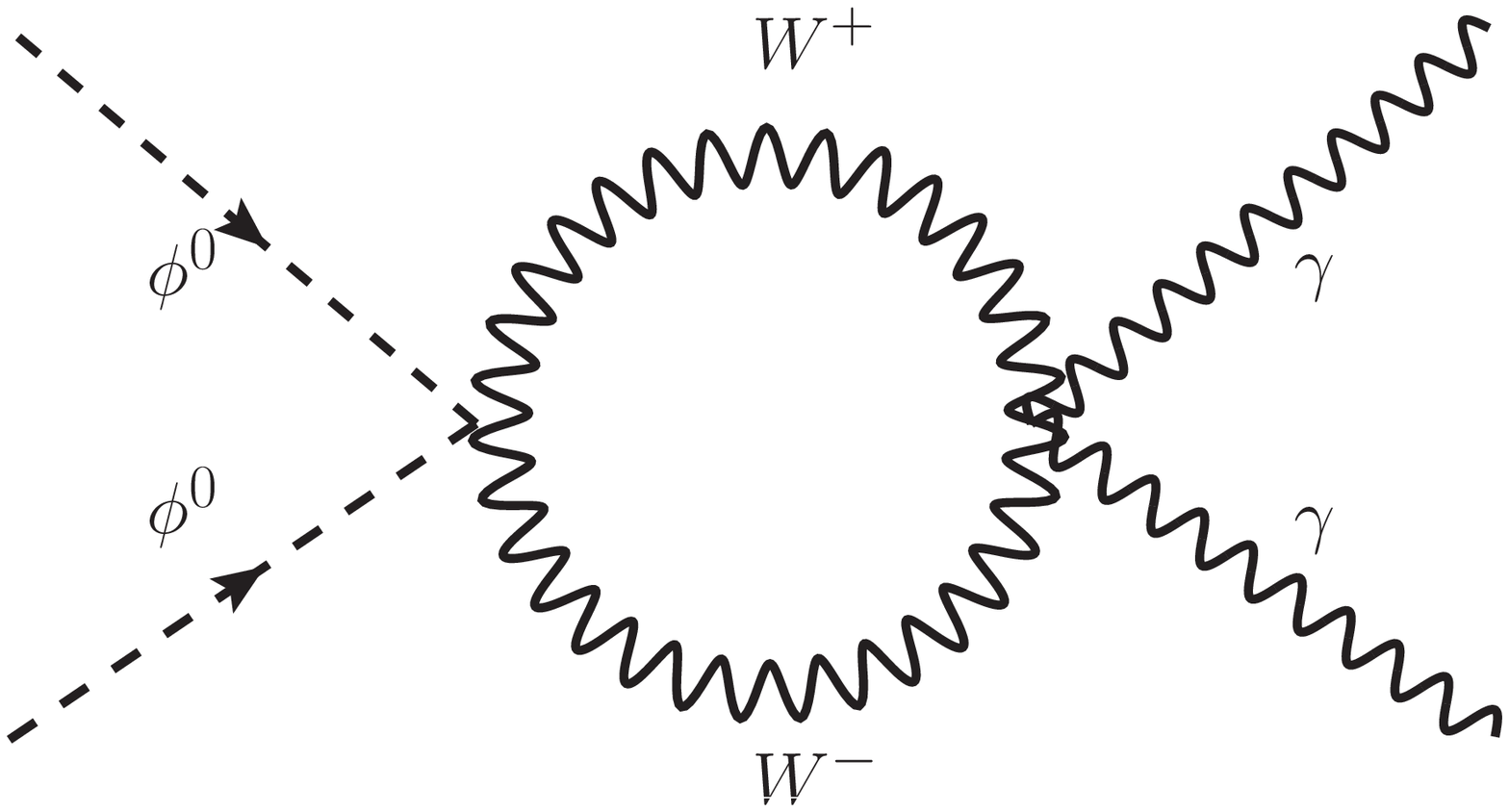}
\hspace{0.75 cm}
\includegraphics[width=5cm,height=3cm]{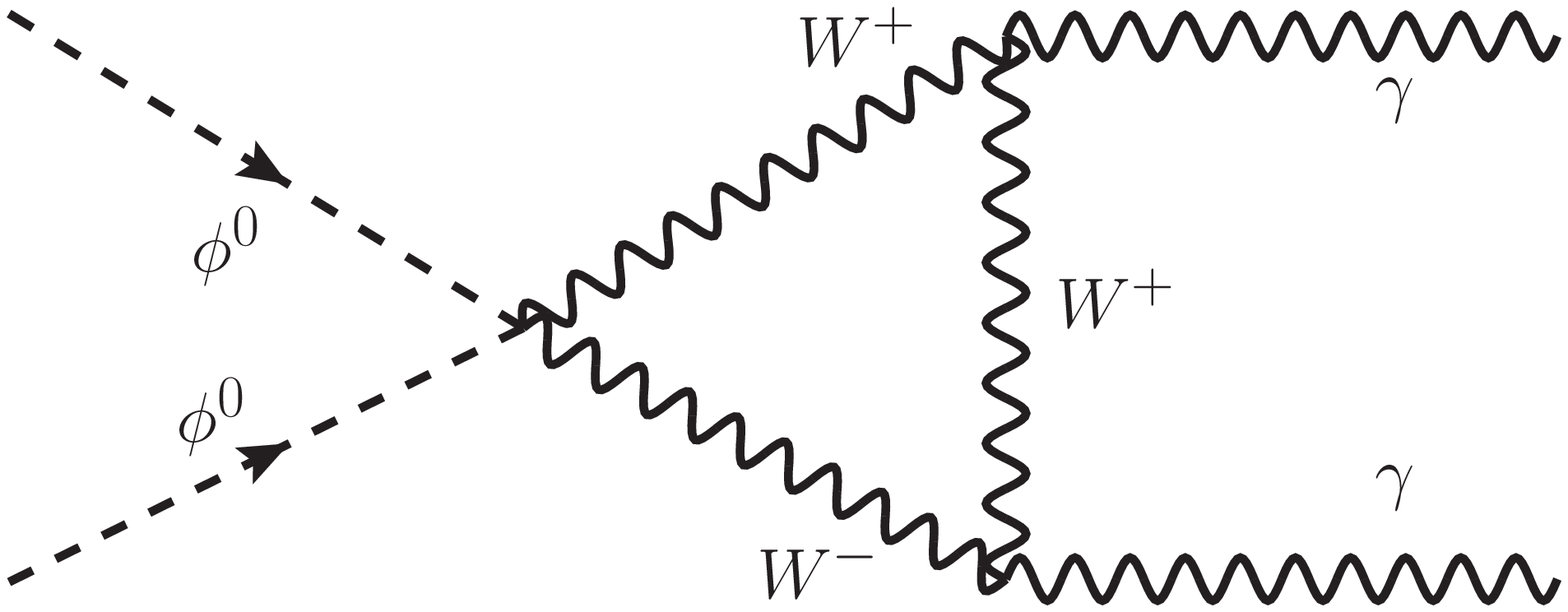}
\hspace{0.75 cm}
\includegraphics[width=5cm,height=3cm]{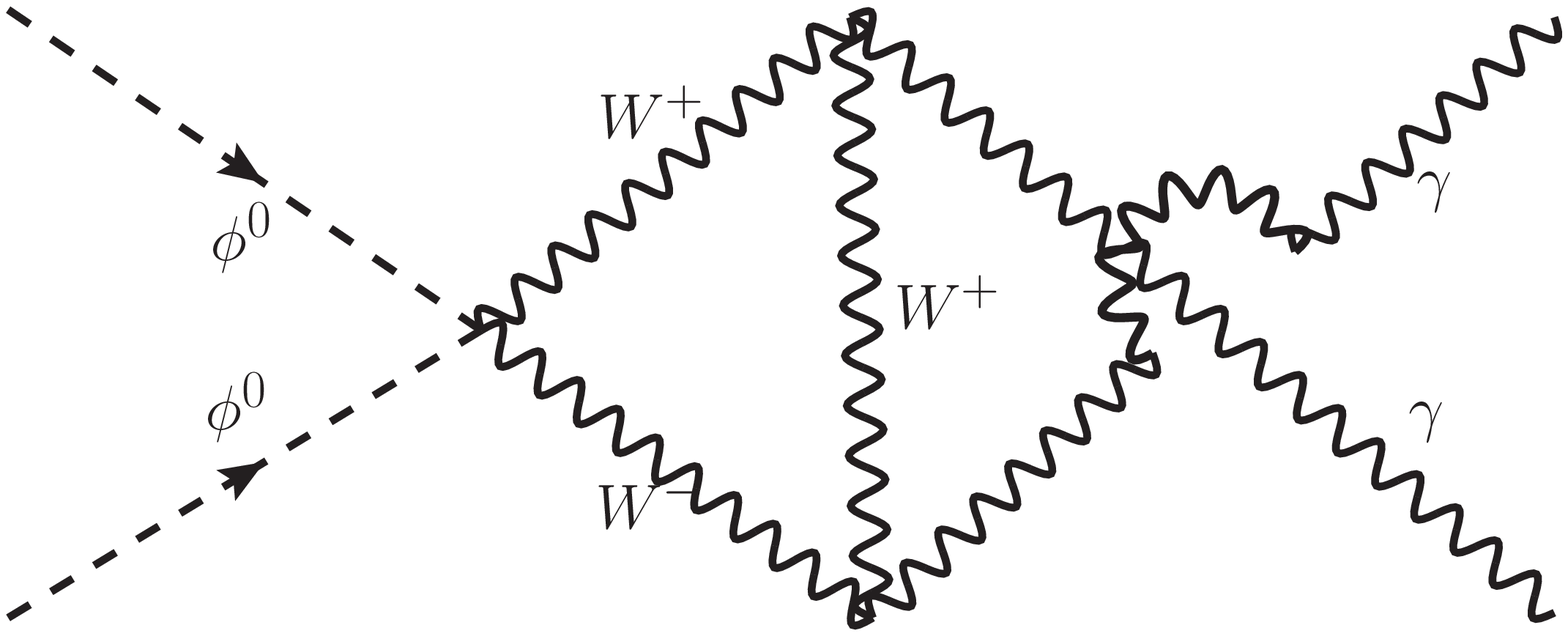}
\caption{Lowest order Feynman diagrams for the process $\phi^0 \phi^0 \rightarrow
\gamma \gamma$}
\label{feyn_dia_gamma}
\end{figure}

We calculate the $\gamma$-ray flux due to annihilation of $\phi^0$
in the ``central region" of our Milky way galaxy. The lowest order Feynman diagrams for the process
$\phi^0 \phi^0 \rightarrow \gamma \gamma$ via $\phi^\pm$ and $W^{\pm}$ loops
are shown in Fig. \ref{feyn_dia_gamma}.
The expression of differential $\gamma-$ray flux due to dark matter annihilation in 
galactic halo is given by \cite{Cirelli}, 
\begin{eqnarray}
\frac{d\Phi_{\gamma}}{dE_{\gamma}} = \frac{1}{8\pi}
\frac{\langle{\sigma {\rm{v}}}_{\phi^0 \phi^0 \rightarrow \gamma \gamma}\rangle}{m^2_{\phi^0}}
\frac{dN_{\gamma}}{dE_{\gamma}} r_{\odot} \rho^2_{\odot} \bar{J} \,\, ,
\label{gammaflux}
\end{eqnarray}
where $r_{\odot} = 8.5$ kpc is the distance of the sun from the Galactic centre and
$\rho_{\odot} = 0.4$ GeV/cm$^3$ is the the local dark matter density at the solar neighbourhood.
The quantity $\bar{J}$ in the above is given by
\begin{eqnarray}
\bar{J} = \frac{4}{\Delta\Omega}\int dl \int db \,\,\cos b \,\, J(l,b) \,\, ,
\label{jbar}
\end{eqnarray}
with
\begin{eqnarray}
J(l,b) = \int_{l.o.s} \frac{ds}{r_\odot}
\left(\frac{\rho(r)}{\rho_\odot}\right)^2 \,\, ,
\label {jint} 
\end{eqnarray}
and
\begin{eqnarray}
\Delta \Omega = 4\int dl \int db\,\, \cos b \,\, .
\label{solidangle}
\end{eqnarray}  
Here $l$ and $b$ in Eqs. (\ref{jbar}, \ref{solidangle}) are galactic longitude and latitude respectively.
In the above, the integration is performed around a radius of 3$^{o}$ (galactic central region)
around a centre with coordinates ($l$, $b$) $=$ ($-1^{o}$ , $-0.7^{o}$) \cite{Tempel:2012ey}.
In Eq. (\ref{jint}) $r$ and $s$ are related by,
\begin{eqnarray}
r = \left( s^2 + r^2_{\odot} - 2sr_{\odot} {\rm cos}\,{l} \,{\rm cos}\,{b}\right)^{1/2}\, .
\end{eqnarray}
The expression of  the energy spectrum of $\gamma$, denoted by $\frac{dN_{\gamma}}{dE_{\gamma}}$ is given by,
\begin{eqnarray}
\frac{dN_{\gamma}}{dE_{\gamma}} = 2\delta(E-E_{\gamma})\,\, . 
\end{eqnarray}
We have performed $l$, $b$ integration (in Eqs. (\ref{jbar}, \ref{solidangle})) over the ``central region" of
our galaxy and the $s$ integration (in Eq. \ref{jint}) is along the line of sight (l.o.s). 

We explicitly calculate all the Feynman diagrams given in Fig. \ref{feyn_dia_gamma}    
and obtain the expression for  $\langle{\sigma  {\rm{v}}}_{\phi^0\phi^0\rightarrow\gamma\gamma}\rangle$ as
\begin{eqnarray}
\langle{\sigma {\rm{v}}}_{\phi^0\phi^0\rightarrow\gamma\gamma}\rangle = \frac{\alpha^2\,\,m^2_{\phi^0}}{32\pi^3}
\left|\frac{g_{\phi^0\phi^0\phi^+\phi^-}F_{\phi^+}}{m^2_{\phi^+}} -  
\frac{g_{\phi^0\phi^0W^+W^-}F_{W}}{m^2_W}\right|^2 \,\, .
\label{sigmav_gamma}
\end{eqnarray} 
Where 
\begin{eqnarray}
F_{\phi^+} &=& \tau[1-\tau f(\tau)] \,\, , \nonumber \\ 
F_{W} &=& 2 + 3\tau + 3\tau(2-\tau) f(\tau)\,\, ,
\end{eqnarray}
with $\tau = \frac{m^2_i}{m^2_{\phi^0}}, \;\;\;\;\;\;\;i = \phi^{\pm}, W^{\pm}$ and 
\begin{equation}
f(\tau)=\left\{\begin{array}{lcc}
\left[\sin^{-1}(\sqrt{1/\tau})\right]^2\,,&\mbox{for}&\tau\geq 1\,,\\[5mm]
-{\displaystyle\frac{1}{4}\,\left[\ln\frac{1+\sqrt{1-\tau}}{1-\sqrt{1-\tau}}-i\pi\right]^2}\,,
&\mbox{for}&\tau<1\,.\end{array}\right.
\label{ftau}
\end{equation}
Since the couplings $g_{\phi^0\phi^0\phi^+\phi^-} = -\rho_2$ and $g_{\phi^0\phi^0 W^+ W^-} = \frac{m^2_W}{v^2}$
(see Table \ref{tab2}), Eq. (\ref{sigmav_gamma}) can be written as

\begin{eqnarray}
\langle{\sigma {\rm{v}}}_{\phi^0\phi^0\rightarrow\gamma\gamma}\rangle = \frac{\alpha^2\,\,m^2_{\phi^0}}{32\pi^3}
\left|\frac{\rho_2 F_{\phi^+}}{m^2_{\phi^+}} +  
\frac{F_{W}}{v^2}\right|^2 \,\, .
\label{sigmav_gamma1}
\end{eqnarray}   

In the present work, $\gamma$-ray flux is calculated for two different dark matter halo profiles namely the 
NFW profile \cite{nfw} and the Einasto profile \cite{einasto}. These halo profiles give the functional
dependence of dark matter density
$\rho(r)$ with $r$. The expression of $\rho(r)$ for the NFW profile is given  by,
\begin{eqnarray}
\rho_{\rm NFW}(r) = \frac{\rho_s}{\left(r/r_s\right)\left(1+r/r_s\right)^2} \,\, ,
\label{nfw}
\end{eqnarray}
and for the Einasto profile
\begin{eqnarray}
\rho_{\rm Einasto} (r) = \rho_s {\rm
exp}\left\{-\frac{2}{\gamma}\left[\left(\frac{r}{r_s}\right)^{\gamma}-1\right]\right\}\,\, ,
\label{einasto}
\end{eqnarray}
where $r_s$ in Eqs. (\ref{nfw}-\ref{einasto}) is taken to be 20 kpc and $\gamma = 0.17$ in
Eq. (\ref{einasto}). In the above the value of the normalisation
constant $\rho_s$ is determined by demanding that at $r = r_{\odot}$,
the density $\rho(r) = \rho_{\odot}$.
\begin{table}[h]
\begin{center}
\vskip 0.5 cm
\begin{tabular} {|c|c|c|c|c|c|}
\hline
{Mass of}& ${\left(\frac{\Omega_{\phi^0}h^2}{\Omega_{c}h^2}\right)}$&${ \rho^{\prime}_{\odot}}$&${\rho_s}$& 
${\langle{\sigma {\rm{v}}}_{\phi^0\phi^0\rightarrow\gamma \gamma}\rangle}$ &${\rho_2}$ \\
{$A^0$}& & (GeV/cm$^3$) &(GeV/cm$^3$)& (cm$^3$/s)& \\
(GeV) & &&&&  \\
\hline
&&&&&\\
500 & 0.621& 0.248 & 0.214&${{1.187}^{+0.405}
_{-0.344}}\times10^{-27}$ & $5.34^{+0.72}_{-0.72}$\\
&&&&&\\
\hline
&&&&&\\
400 & 0.581& 0.232 & 0.201&${{1.347}^{+0.460}
_{-0.392}}\times10^{-27}$ & $5.64^{+0.76}_{-0.77}$\\
&&&&&\\
\hline
&&&&&\\
350 & 0.553& 0.221 & 0.191&${{1.490}^{+0.505}
_{-0.434}}\times10^{-27}$ & $5.89^{+0.79}_{-0.81}$\\
&&&&&\\  
\hline
&&&&&\\
300 & 0.517& 0.207 & 0.178&${{1.717}^{+0.586}
_{-0.500}\times10^{-27}}$ & $6.26^{+0.85}_{-0.86}$\\
&&&&&\\
\hline
\end{tabular}
\end{center}
\caption{Results for the NFW Profile for $\alpha = -0.037$.}
\label{tab4}
\end{table}
We have seen earlier that in the present model of two component dark matter, the inert doublet
component $\phi^0$ contributes to $\sim$ 62\% $\left(\frac{\Omega_{\phi^0}h^2}{\Omega_{c}h^2}\sim 0.62\right)$
of the total dark matter relic density. Therefore in calculating the $\gamma-$ray flux from the process
$\phi^0\phi^0\rightarrow\gamma\gamma$ we compute $\rho_s$ by taking
\begin{equation}
\rho^{\prime}_{\odot} = \rho_{\odot}\times \frac{\Omega_{\phi^0}h^2}{\Omega_{c}h^2},
\label{rho-prime}
\end{equation}
and demanding that for the dark matter component $\phi^0$, $\rho(r) = \rho^{\prime}_{\odot}$ at
$r = r_{\odot}$ with $\rho_{\odot} = 0.4$ GeV/cm$^3$ \cite{rho0.4}.  

\begin{table}[h]
\begin{center}
\vskip 0.5 cm
\begin{tabular} {|c|c|c|c|c|c|}
\hline
{Mass of}& ${\left(\frac{\Omega_{\phi^0}h^2}{\Omega_{c}h^2}\right)}$&${ \rho^{\prime}_{\odot}}$&${\rho_s}$& 
${\langle{\sigma {\rm{v}}}_{\phi^0\phi^0\rightarrow\gamma \gamma}\rangle}$ &${\rho_2}$ \\
{$A^0$}& & (GeV/cm$^3$) &(GeV/cm$^3$)& (cm$^3$/s)& \\
(GeV) & &&&&  \\
\hline
&&&&&\\
500 & 0.621& 0.248 & 0.051&${{0.606}^{+0.207}
_{-0.176}}\times10^{-27}$ & $4.01^{+0.54}_{-0.55}$\\
&&&&&\\
\hline
&&&&&\\
400 & 0.581& 0.232 & 0.047&${{0.714}^{+0.243}
_{-0.208}}\times10^{-27}$ & $4.30^{+0.58}_{-0.58}$\\
&&&&&\\
\hline
&&&&&\\
350 & 0.553& 0.221 & 0.045&${{0.789}^{+0.255}
_{-0.227}}\times10^{-27}$ & $4.47^{+0.59}_{-0.61}$\\
&&&&&\\  
\hline
&&&&&\\
300 & 0.517& 0.207 & 0.042&${{0.894}^{+0.305}
_{-0.260}}\times10^{-27}$ & $4.74^{+0.63}_{-0.65}$\\
&&&&&\\
\hline
\end{tabular}
\end{center}
\caption{Results for the Einasto Profile for $\alpha = -0.037$.}
\label{tab5}
\end{table}

From the left panel of Fig. 3 of Ref. \cite{Tempel:2012ey} the best fit value of
the $\gamma-$ray flux (in terms of $E^2\Phi$ (GeV cm$^{-2}$ s$^{-1}$ sr$^{-1}$))
observed by Fermi-LAT from the central signal region of the Galaxy can be read as $E^2\Phi =  5.6\times 10^{-5}$
GeV cm$^{-2}$ s$^{-1}$ sr$^{-1}$ with 95\% C.L. error band that lies in the range
$3.97 \times 10^{-5} \leq E^2\Phi \leq 7.51 \times 10^{-5}$ (GeV cm$^{-2}$ s$^{-1}$ sr$^{-1}$).
We use these values of the flux in Eq. (\ref{gammaflux}) and compute $\langle{\sigma
{\rm{v}}}_{\phi^0\phi^0\rightarrow\gamma\gamma}\rangle$ for the best fit value of the flux as also
the two extremities of its error band at 130 GeV. Note that the parameter $\rho_2$ is yet to be determined.
This can be estimated by calculating the cross section $\langle{\sigma
{\rm{v}}}_{\phi^0\phi^0\rightarrow\gamma\gamma}\rangle$
(given by Eq. (\ref{sigmav_gamma1})) and hence the flux ($E^2\Phi$) and then 
comparing this flux with that given by the Fermi-LAT data.

The results are furnished in Table \ref{tab4} and Table \ref{tab5} for
the NFW profile and the Einasto profile respectively. They are given for the chosen mass $m_{A^0}$ of
500 GeV as also for three other values
of $m_{A^0}$ namely 400, 350, 300 GeV for the purpose of demonstration where $\alpha = -0.037$. In
both the Tables \ref{tab4}, \ref{tab5} the values of the cross sections obtained
for the best fit value of the flux are given. The computed 
cross sections for the two extremities of the error band of the flux are
shown by the subscripts and superscripts of the central values.
The corresponding values of $\rho_2$ that are calculated using
Eq. (\ref{sigmav_gamma1}) are also shown in similar fashion.
It is seen from both the Tables that although the calculated
values for $\rho_2$ depend on the dark matter density profile that
one chooses, they are within the perturbative limit and the corresponding
cross sections ${\langle{\sigma {\rm{v}}}_{\phi^0\phi^0\rightarrow\gamma \gamma}\rangle}$
are also within the desired limits of $\sim 10^{-27}$ cm$^3$/s. In Table \ref{tab6} we
show the similar set of values (like Tables \ref{tab4} and \ref{tab5}) for another value of $\alpha = -0.045$.

\begin{table}[h]
\begin{center}
\vskip 0.5 cm
\begin{tabular} {|c|c|c|c|c|c|c|}
\hline
{Name of}&{Mass of}& ${\left(\frac{\Omega_{\phi^0}h^2}{\Omega_{c}h^2}\right)}$&
${\rho^{\prime}_{\odot}}$&${\rho_s}$& 
${\langle{\sigma {\rm{v}}}_{\phi^0\phi^0\rightarrow\gamma \gamma}\rangle}$ &${\rho_2}$ \\
{the halo profile}&{$A^0$}& & (GeV/cm$^3$) &(GeV/cm$^3$)& (cm$^3$/s)& \\
&(GeV) & &&&&  \\
\hline
&&&&&&\\
NFW&500 & 0.644& 0.258 & 0.222&${{1.104}^{+0.375}
_{-0.324}}\times10^{-27}$ & $5.18^{+0.69}_{-0.71}$\\
&&&&&&\\
\hline
&&&&&&\\
Einasto&500 & 0.644& 0.258 & 0.052&${{0.584}^{+0.196}
_{-0.171}}\times10^{-27}$ & $3.95^{+0.52}_{-0.55}$\\
&&&&&&\\
\hline
\end{tabular}
\end{center}
\caption{Results for NFW and Einasto profile for the value of $\alpha = -0.045$, $m_{A^0} = 500$ GeV,
$m_{\phi^+} = 130.2$ GeV.}
\label{tab6}
\end{table}
\begin{figure}[h]
\centering 
\includegraphics[width=5.5cm,height=8.0cm,angle=-90]{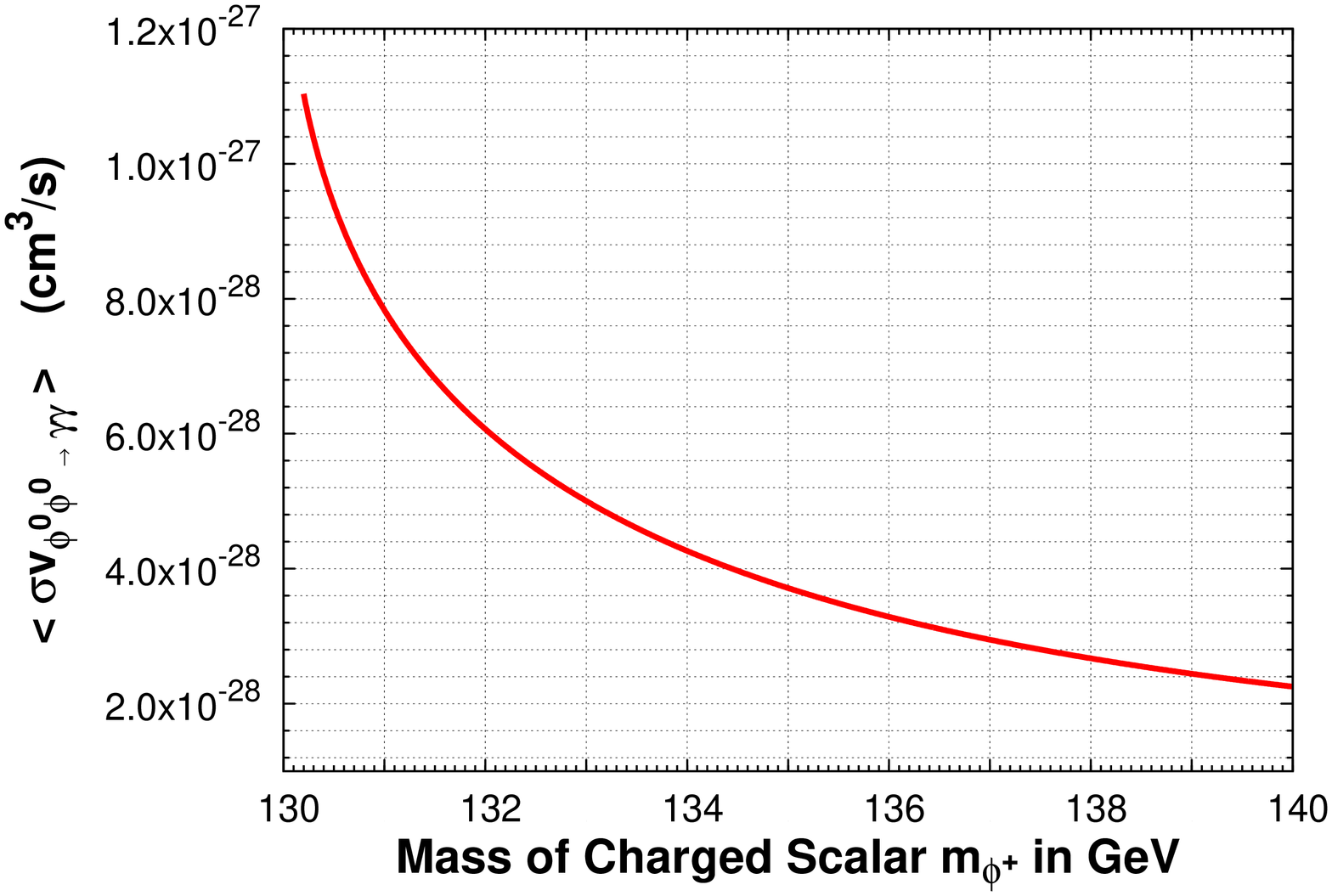}
\hspace{1.0cm}
\includegraphics[width=5.5cm,height=8.0cm,angle=-90]{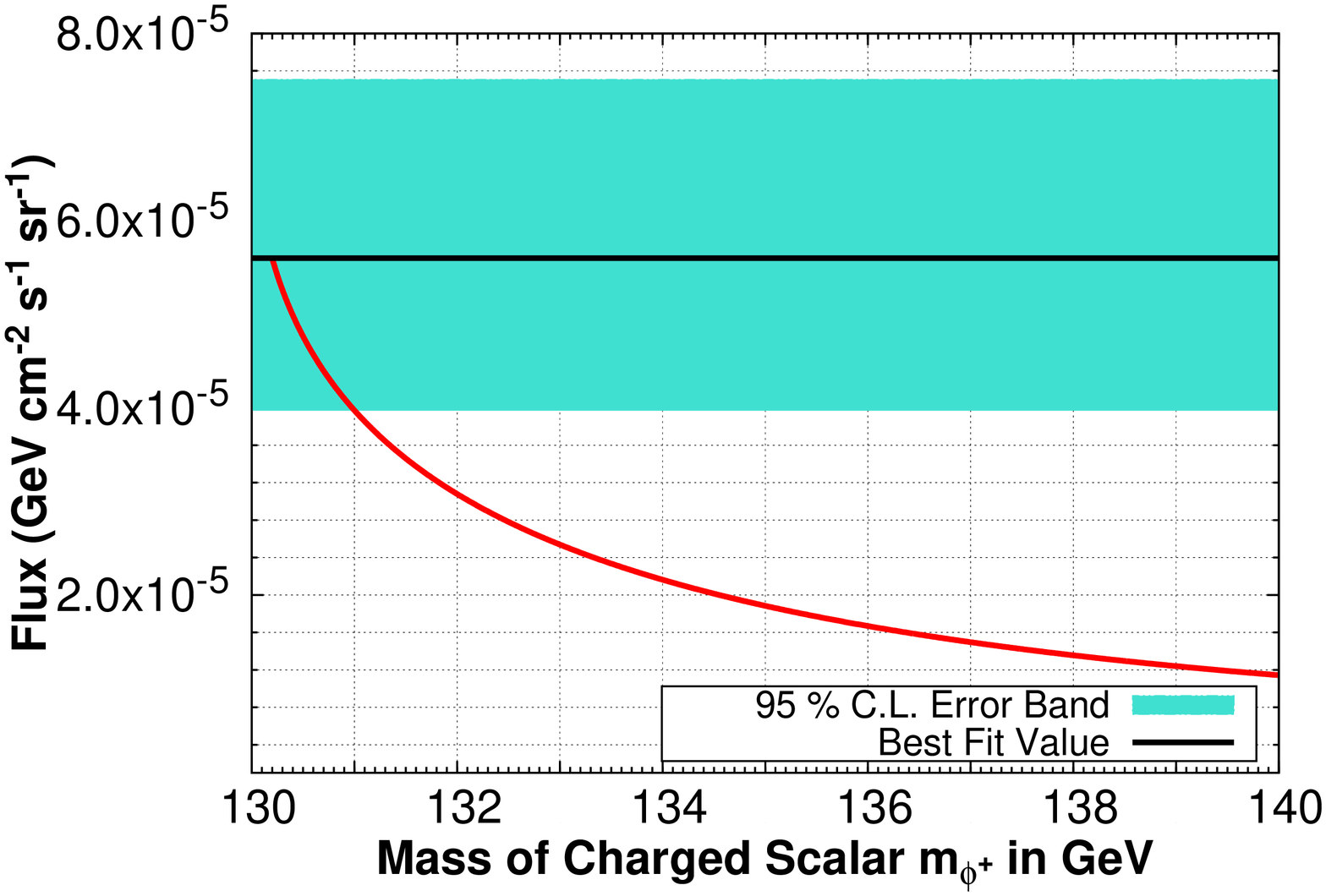}
\caption{Left panel - Variations of
$\langle {\sigma {\rm v}}_{\phi^0 \phi^0 \rightarrow \gamma\gamma}\rangle$ with
the mass of charged scalar ($m_{\phi^+}$).
Right panel - Variations of gamma-ray flux with $m_{\phi^+}$, with 95\%
C.L. error band and the best fit value of the gamma-ray flux from
Fermi-LAT data are represented by turquoise colour band and black solid line respectively.}
\label{fluxplot}
\end{figure}
\begin{figure}[h]
\centering 
\includegraphics[width=5.0cm,height=8.0cm,angle=-90]{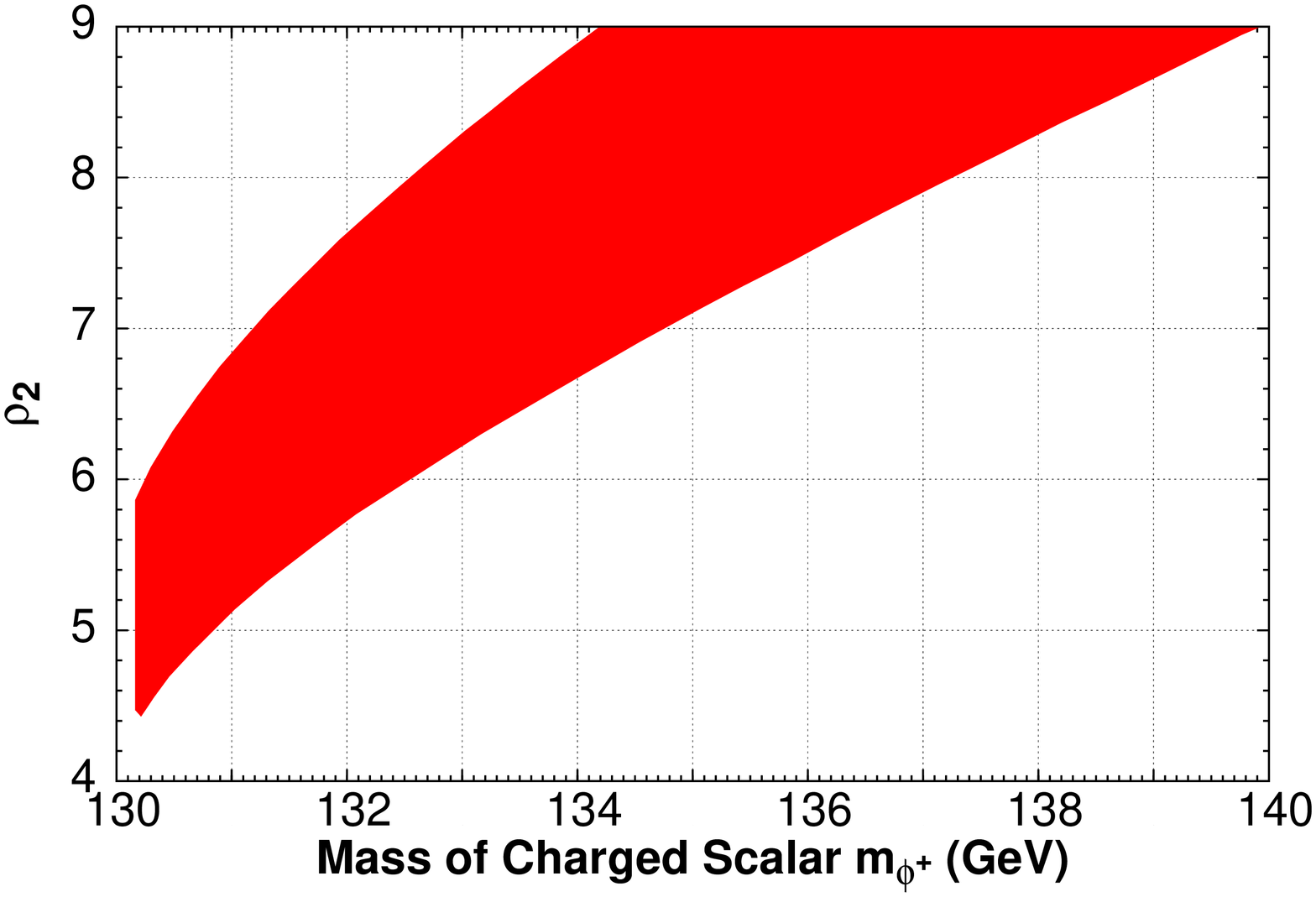}
\hspace{1.0cm}
\includegraphics[width=5.0cm,height=8.0cm,angle=-90]{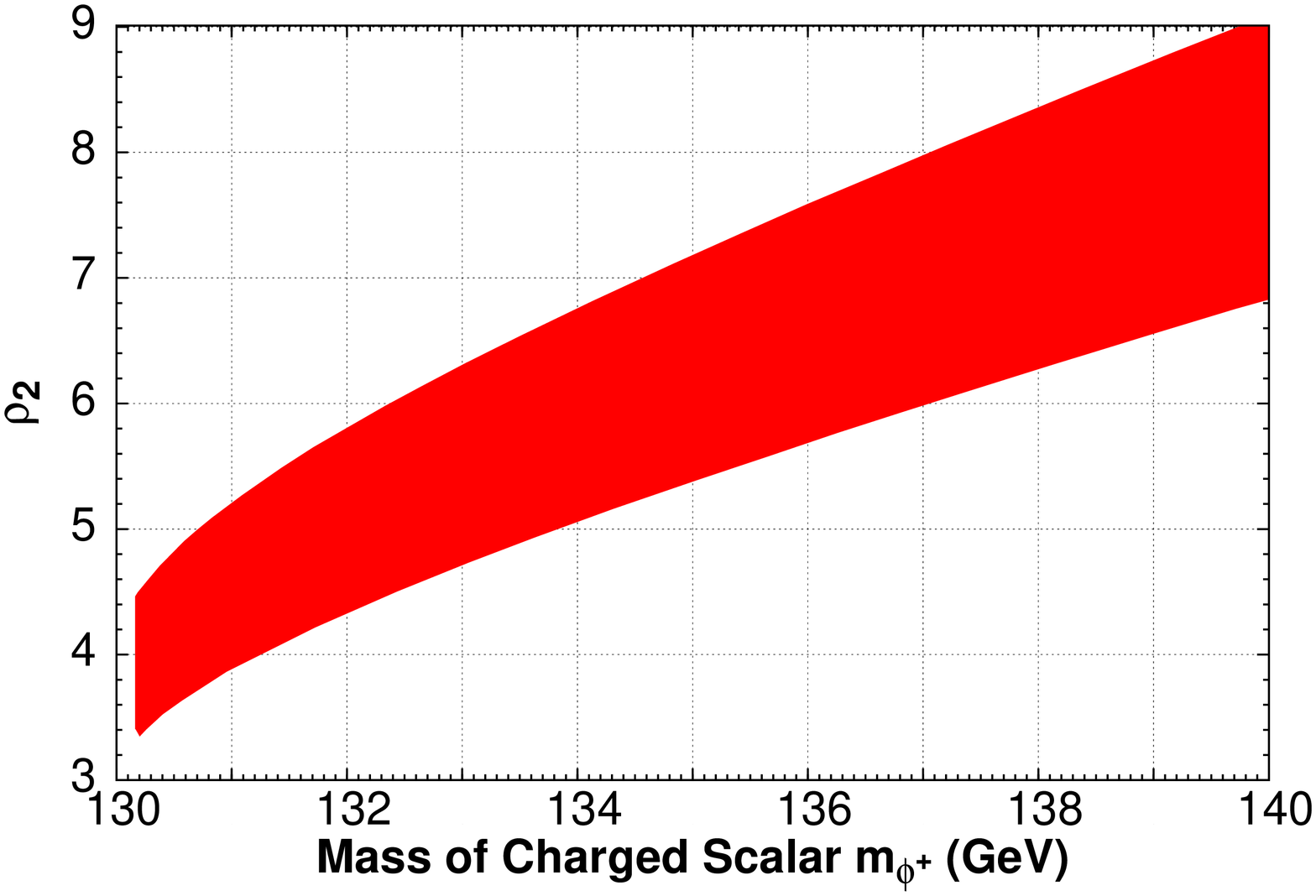}
\caption{Variations of $\rho_2$ with charged scalar mass $m_{\phi^+}$ for both
NFW (left panel) and Einasto (right panel) profile.}
\label{rho2-mcs-cont}
\end{figure}

Here we like to mention that we have checked the possibility that the continuum gamma-rays may
overshoot the monochromatic gamma-ray line. In Ref. \cite{Buchmuller},
the authors have shown that in order to distinguish the monochromatic gamma-line
(from DM DM $\rightarrow \gamma \gamma$ channel) from the continuum gamma-ray spectrum (produced by the 
secondary photons originating from the annihilation products of dark matter
e.g. gauge bosons, $q\bar{q}$, $f\bar{f}$), the branching ratio for the channel 
DM DM $\rightarrow \gamma \gamma$ must be greater than 1\% of total annihilation 
cross section (sum of annihilation cross sections for all possible channels). For the dark matter component
$\phi^0$ in our model we find that this ratio is nearly 1/70 (i.e. $>10^{-2}$) as seen from the left panel
of Fig. \ref{sigmaV} and Table \ref{tab4}.

A discussion on the choice of the value of charged scalar mass $m_{\phi^+}$ is in order.
In this work the value $m_{\phi^+} = 130.2$ GeV is adopted.
The viability of such a choice is demonstrated in Fig. \ref{fluxplot}.
In the left panel of Fig. \ref{fluxplot} we show the variations of
$\langle {\sigma {\rm v}}_{\phi^0 \phi^0 \rightarrow \gamma\gamma}\rangle$ for different
values of $m_{\phi^+}$. We have excluded the situation $m_{\phi^+} < m_{\phi^0}$
($m_{\phi^0}$ is fixed at 130 GeV), which can give rise to the 
possibility of having charged relic in the present scenario (with an 
unbroken ${\rm Z_2}$). The abundance of such a charged relic is severely 
constrained \cite{charged-relic} which therefore prompts us to analyse 
the $\langle {\sigma {\rm v}}_{\phi^0 \phi^0 \rightarrow \gamma\gamma}\rangle$  
for those values of $m_{\phi^+}$ which satisfy $m_{\phi^+} > m_{\phi^0}$.
We have checked with $m_{\phi^+} \sim 130.2$ GeV, $\phi^+$ can actually
decay\footnote{decay width turns out to be of the order of $\simeq 10^{-16}$ GeV.}
(e.g. $\phi^+ \longrightarrow \phi^0 + e^+ + \nu_{e}$) \cite{osland} before Big Bang Nucleosynthesis (BBN). 
We find that the annihilation cross section $\langle {\sigma {\rm v}}_{\phi^0 \phi^0 \rightarrow
\gamma\gamma}\rangle$ of $\phi^0$ to produce $\gamma$ decreases sharply with the increase of $m_{\phi^+}$ 
as is evident from the left panel of Fig. \ref{fluxplot}. The variations of the 
gamma-ray flux with $m_{\phi^+}$ are also displayed in the right panel of Fig. \ref{fluxplot}. 
Calculations for both the plots in Fig. \ref{fluxplot} are carried out for a 
chosen value of $\rho_2$ (see Table \ref{tab6}) such that the best fit value of
the gamma-ray flux from the Fermi-LAT data can be reproduced with $m_{\phi^+} = 130.2$ GeV.
Also in the right panel, we include the flux (the best fit value) of energy 130 GeV obtained
from Fermi-LAT data along with its error band of 95\% C.L. We therefore conclude that $m_{\phi^+}$ which  
would produce gamma-ray flux within 95\% C.L. of the experimental observation should lie in a very 
narrow interval (with a particular choice of other parameters involved),
in the vicinity of $m_{\phi^0}$ but not below it. 
For demonstrative purpose, we have shown two contour plots of the parameter
$\rho_2$ {\it vs} $m_{\phi^+}$ in both the panels of Fig. \ref{rho2-mcs-cont}.
These contours are drawn for both NFW and Einasto profile respectively
where each point within the contours produces gamma-ray flux which lies
within the 95\% C.L. of observed flux from Fermi-LAT data.
It is seen from both the panels of Fig. \ref {rho2-mcs-cont}
that a higher value of $m_{\phi^+}$ is also possible but at the expense of a
high value of the coupling $\rho_2$. However, we restrict ourselves with the
choice $m_{\phi^+} = 130.2$ GeV for the rest of our discussion. 
\section{Discussions and Conclusions}
\label{disc}
In the present work we propose a dark matter model which contains
two dark matter candidates. Such a two component dark
matter model can be obtained by adding a scalar singlet $S$ (singlet under SM gauge
group) and a doublet $\Phi$ (doublet under SM gauge group) to the
scalar sector of SM. We have introduced discrete symmetry ${\rm Z_2 \times
Z^{\prime}_2}$ under which only $S$ and $\Phi$ transform non-trivially.   
Both the scalar singlet $S$ and  doublet  $\Phi$ do
not produce any VEV. Consequently ${\rm Z_2 \times
Z^{\prime}_2}$ symmetry remains unbroken which ensure the stability 
of both the components ($S, \phi^0$) of the dark matter in the present model.
While the  component $\phi^0$ (neutral part of the doublet $\Phi$)
can produce the annihilation cross section required to obtain 130 GeV $\gamma$-line,
the value of the corresponding cross section for the scalar singlet component $S$
falls deficit by few orders of magnitude. However the component $\phi^0$ above,
having a mass of 130 GeV cannot solely account for the relic density predicted by
WMAP. This deficit in relic density is compensated by the
scalar singlet component $S$ such that the combined
relic density ($\Omega_{c}h^2$) for this two component dark matter model always lies
within the range given by WMAP. Combined relic density is the sum
of individual relic densities of both the components $S$ and $\phi^0$ which are obtained by
solving the coupled Boltzmann's equations numerically. We have found that the contribution of the component
$\phi^0$ to $\Omega_{c}h^2$ will be $\sim 62\%$ ($\sim 64\%$) when we consider $\alpha =
-0.037$ $(-0.045), \,\,|\lambda_6| \leq 3.2\times10^{-3}$
($ \leq 1.9\times10^{-3}$)\footnote{both $\alpha$ and $\lambda_6$
satisfy XENON 100 (2012) \cite{xenon2012} limit as well as limits from other dark
matter direct detection experiments
namely CDMS-II \cite{cdms09}, EDELWEISS-II \cite{edelweiss11} etc.},
$\Delta m = 0.5$ GeV, $\lambda_5 \sim 0.387$ ($\sim 0.426$)
and $m_{A^0} = 500$ GeV. Finally in the last section we have calculated the
annihilation cross section $\langle{\sigma {\rm{v}}}_{\phi^0\phi^0\rightarrow\gamma\gamma}\rangle$
for the channel $\phi^0\phi^0\rightarrow\gamma\gamma$ with the mass of
$\phi^0 \sim 130$ GeV. Using the expression of this annihilation cross section
($\langle{\sigma {\rm{v}}}_{\phi^0\phi^0\rightarrow\gamma\gamma}\rangle$)
we have computed the $\gamma-$ray flux of energy 130 GeV for two different
dark matter halo profiles namely the NFW profile and the Einasto profile.
The exact dark matter density at the galactic centre is unknown (e.g. Ref. \cite{Cirelli}
and references therein). This may produce an additional
uncertainty in the flux calculation. Depending on the value of the dark matter density at the galactic
centre, 130 GeV gamma-line may also be produced for a value of annihilation cross section
lower than the specified value of $\sim$ $10^{-27}$ cm$^3$/s. Fermi Collaboration placed an upper
limit \cite{fermix} $\langle{\sigma {\rm{v}}}_{\gamma \gamma}\rangle < 1.4\times 10^{-27}$ cm$^3$/s for 130 GeV
dark matter with an NFW profile and $\langle{\sigma {\rm{v}}}_{\gamma \gamma}\rangle
< 1.0\times 10^{-27}$ cm$^3$/s
with an Einasto profile. In the present work we indeed obtain ${\sigma {\rm{v}}}_{\gamma \gamma}$ in the
range of this upper limit ($\sim 10^{-27}$ cm$^3$/s).

\begin{figure}[h]
\centering 
\includegraphics[width=7.0cm,height=6.0cm]{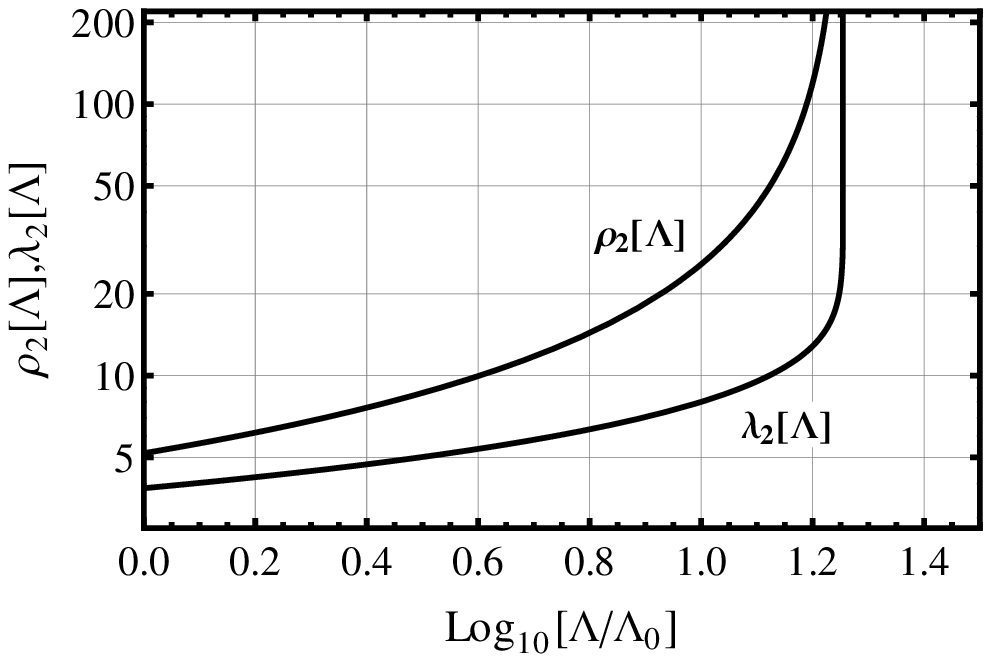}
\hspace{1cm}
\includegraphics[width=7.0cm,height=6.0cm]{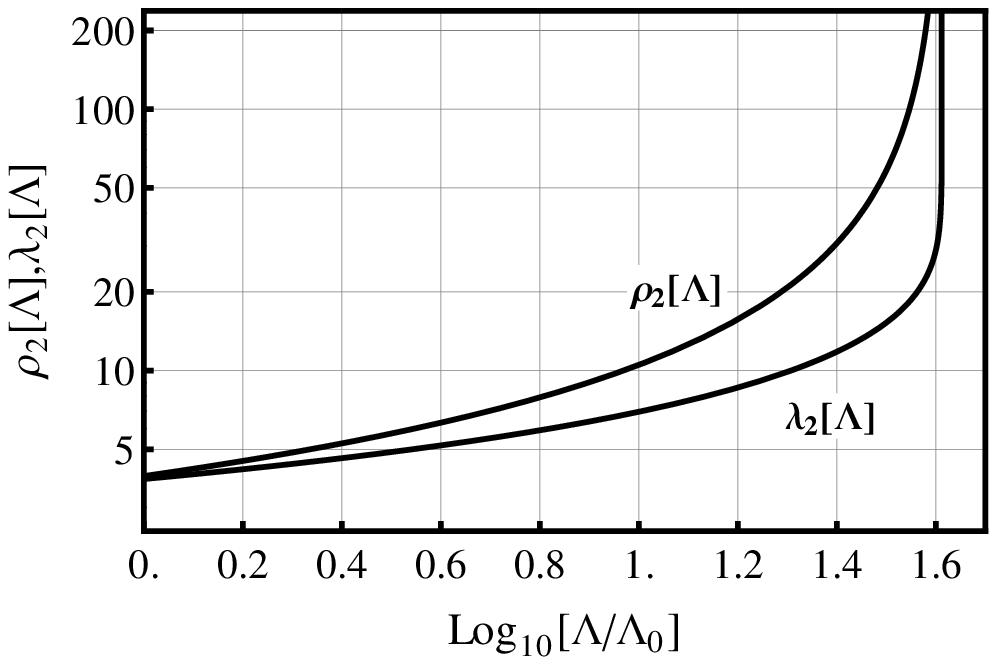}
\caption{Variation of the couplings $\rho_2$ and $\lambda_2$ with energy scale $\Lambda$ for both
NFW (left panel) and Einasto (right panel) profile.}
\label{rgrun}
\end{figure}

As a typical set of parameter space of the model under consideration, we have tabulated
values of all the parameters for specific choice of $\alpha = -0.045$ ($\alpha = -0.037$) in Table \ref{tab3.1}
(Table \ref{tab3}), which can contribute towards the DM relic density at an appropriate level as
well as produce the 130 GeV gamma-ray. It can be noticed that among all these parameters,
the quartic couplings $\rho_2$ (between $\phi^0$ and $\phi^{\pm}$, responsible for the production of 130 GeV 
$\gamma$ line, see Fig. {\ref{feyn_dia_gamma}}) and $\lambda_2$ (related to the
mass of $A^0$) are rather on the higher side. A comment on this particular choice of $m_{A^0}$ is relevant here.
A heavier $A^0$ indicates a larger choice of $\lambda_2$ as seen from
Table \ref{tab3}. Although a larger $m_{A^0}$ would maximise the contribution of $\phi^0$
towards DM relic density (i.e. to get a larger ratio $\left(\frac{\Omega_{\phi^0}h^2}{\Omega_{c}h^2}\right)$,
see Fig. \ref{result1}), we keep $m_{A^0}$ at 500 GeV so that the corresponding parameter $\lambda_2$ can have
a not-very-large value. As the couplings (particularly $\rho_2$ and $\lambda_2$) becomes stronger at high
energy scale, it would pose a threat to the validity of the model as the theory tends
to be non-perturbative at some high energy scale. In this scenario we have estimated
the Landau pole ($\Lambda_{\rm L}$) of the model. We have used the one loop beta functions
\cite{0907.1894, rg} appropriate for our inert Higgs doublet and singlet model.
Using the parameters in Table \ref{tab3.1}
as an initial choice at a energy scale $\Lambda_0 = m_{\phi^0} = 130$ GeV, we plot
the running of $\rho_2$ and $\lambda_2$ in both panels of Fig. \ref{rgrun}.
We find that for $\alpha = -0.045$,
$\Lambda_{\rm L} \sim 2.5$ TeV ($\sim$ 5 TeV) when NFW (Einasto) dark matter profile is considered.
Similar results are obtained for the case with $\alpha = -0.037$ (note that the mass $m_{A^0}$
of the only massive field $A^0$ is well within these limits).

Since the present two component dark matter consists of a singlet scalar
and an SU(2) inert doublet, they have different couplings with Higgs boson (Eqs. (\ref{RSDM}, \ref{IDM})).
The doublet component will also have additional interactions with gauge bosons. Hence both the
scattering cross section and the annihilation cross section for each of the
two components are different even though they have masses close to each other in the present
model. Therefore the rate of direct detection and its subsequent variations with the recoil
energies for the two components will be different. Also in the event that the scalar and 
inert doublet components of this two component dark matter are captured by the gravity of
the solar core and each component suffers subsequent annihilations in the core yielding
neutrinos as the final states then the spectra and the fluxes for such neutrinos will differ 
depending on which component of the dark matter (scalar singlet or inert doublet)
annihilates to produce them. Similar features would also be realised if these two dark
matter components annihilate to produces gamma-rays at a suitable dense site such as
galactic centre. If the measurements of such ``GeV neutrinos" from the solar core exhibit
two distinct natures for both the flux and spectrum and if such a difference can be
corroborated with the possible gamma-ray signal from the galactic centre then this may
indicate a probable indirect detection of such a two component dark matter. 
\vspace{1cm}

{\bf Acknowledgments} : A.B. would like to thank P. B. Pal, A. Ghosal and D. Adak
for useful suggestions and discussions. The authors thank J. Kopp, M. Raidal
and K. Kannike for very useful suggestions and comments. A.S acknowledges the
support from the Start Up grant from IIT Guwahati.

\begin{center}
------------------------
\end{center}   
\end{document}